# RETICOLO code for grating analysis

**Authors: J.P. Hugonin and P. Lalanne**
**arXiv:2101:00901**

**This technical note is composed of three independent parts:**

**Reticolo is a freeware operating under Matlab. To install it, copy the companion folder "reticolo_allege" and add the folder in the Matlab path. Reticolo can be downloaded here.**

The version V10 launched in 01/2025 features a few novelties:

- ✓ it includes a treatment of stacks of arbitrarily anisotropic multilayered thin-films.[1] Be aware that the substrate and superstrate cannot be anisotropic. This part is documented in Reticolo codes 1D-conical (or identically 2D).
- ✓ It features an option to visualize the Bloch modes.
- ✓ Diagonal anisotropy ($\boldsymbol{\varepsilon_{xx}} \neq \boldsymbol{\varepsilon_{yy}} \neq \boldsymbol{\varepsilon_{zz}}$) can be incorporated in structured grating layers and gratings with uniform layers having arbitrary anisotropy ($\boldsymbol{\varepsilon_{xy}} \neq \mathbf{0}$ ...) can also be handled)
- ✓ It is fully compatible with earlier versions.

**Last update: 01/2025**

**Technical contact: jean-paul.hugonin@institutoptique.fr**
**Contact: philippe.lalanne@institutoptique.fr**

[1] This is simply achieved by retaining a single Fourier harmonics coefficient in the expansion (nn=0). The extension is not optimal from numerical-efficiency perspectives, but it has been provided on demand of several users who additionally complained of mistakes in available freeware packages on thin films.

Note also that we have launched in 2023 the RETICOLOfilm-stack program, a MATLAB freeware that computes the reflection and transmission of arbitrary stacks of anisotropic thin films. RETICOLOfilm-stack is vectorialized and thus treats several wavelengths and incidences in a single instruction (see https://zenodo.org/record/7512710).

# I. RETICOLO CODE 1D: classical diffraction

**Reticolo code 1D is a free software for analyzing 1D gratings in classical mountings. It operates under Matlab. To install it, copy the companion folder "reticolo_allege" and add the folder in the Matlab path.**

## Outline

# 1. Generality

RETICOLO is a code written in the language MATLAB 9.0. It computes the diffraction efficiencies and the diffracted amplitudes of gratings composed of stacks of lamellar structures. It incorporates routines for the calculation and visualisation of the electromagnetic fields inside and outside the grating. With this version, 2D periodic (crossed) gratings cannot be analysed.

As free alternative to MATLAB, RETICOLO can also be run in GNU Octave with minimal code changes. For further information, please contact tina.mitteramskogler@profactor.at.

In brief, RETICOLO implements a frequency-domain modal method (known as the Rigorous Coupled wave Analysis/RCWA). To get an overview of the RCWA, the interested readers may refer to the following articles:

1D-classical and conical diffraction
M.G. Moharam et al., JOSAA **12**, 1068 (1995),
M.G. Moharam et al, JOSAA **12**, 1077 (1995),
P. Lalanne and G.M. Morris, JOSAA **13**, 779 (1996),
G. Granet and B. Guizal, JOSAA **13**, 1019 (1996),
L. Li, JOSAA **13**, 1870 (1996), see also C. Sauvan et al., Opt. Quantum Electronics **36**, 271-284 (2004) which simply explains the raison of the convergence-rate improvement of the Fourier-Factorization rules without requiring advanced mathematics on Fourier series and generalizes to other kinds of expansions.

2D-crossed gratings
L. Li, JOSAA **14**, 2758-2767 (1997),
E. Popov and M. Nevière, JOSAA **17**, 1773 (2000),
which describe the up-to-date formulation of the approach used in RETICOLO. Note that the formulation used in the last article (which proposes an improvement for analysing metallic gratings with continuous profiles like sinusoidal gratings) is not available in the RETICOLO version of the web. The RCWA relies on the computation of the eigenmodes in all the layers of the grating structure in a Fourier basis (plane-wave basis) and on a scattering matrix approach to recursively relate the mode amplitudes in the different layers.

**Eigenmode solver:** For conical diffraction analysis of 1D gratings, the Bloch eigenmode solver used in Reticolo is based on the article "P. Lalanne and G.M. Morris, JOSAA **13**, 779 (1996)".

**Scattering matrix approach:** The code incorporates many refinements that we have not published and that we do not plan to publish. For instance, although it is generally admitted that the S-matrix is inconditionnally stable, it is not always the case. We have developed an in-house transfer matrix method which is more stable and accurate. The new transfer matrix approach is also more general and can handle perfect metals. The essence of the method has been rapidly published in "J.-P. Hugonin, M. Besbes and P. Lalanne, Op. Lett. **33**, 1590 (2008)".

**Field calculation:** The calculation of the near-field electromagnetic fields everywhere in the grating is performed according to the method described in "P. Lalanne, M.P. Jurek, JMO **45**, 1357 (1998)" and to its generalization to crossed gratings (unpublished). Basically, no Gibbs phenomenon will be visible in the plots of the discontinuous electromagnetic quantities, but field singularities at corners will be correctly handled.

**Acknowledging the use of RETICOLO**: In publications and reports, acknowledgments have to be provided by referencing to J.P. Hugonin and P. Lalanne, RETICOLO software for grating analysis, Institut d'Optique, Orsay, France (2005), arXiv:2101:00901.

**In addition, one may fairly quote the following references in journal publications**:
-M.G. Moharam, E.B. Grann, D.A. Pommet and T.K. Gaylord, "Formulation for stable and efficient implementation of the rigorous coupled-wave analysis of binary gratings", J. Opt. Soc. Am. A **12**, 1068-1076 (1995), if TE-polarization efficiency calculations are provided
-P. Lalanne and G.M. Morris, "Highly improved convergence of the coupled-wave method for TM polarization", J. Opt. Soc. Am. A **13**, 779-789 (1996) and G. Granet and B. Guizal, "Efficient implementation of the coupled-wave method for metallic lamellar gratings in TM polarization", J. Opt. Soc. Am. A **13**, 1019-1023 (1996), if TM-polarization efficiency calculations are provided,
-P. Lalanne and M.P. Jurek, "Computation of the near-field pattern with the coupled-wave method for TM polarization", J. Mod. Opt.**45**, 1357-1374 (1998), if near-field electromagnetic-field distributions are shown.

# 2. The diffraction problem considered

In general terms, the code solves the diffraction problem by a grating defined by a stack of layers (in the z-direction) which have all identical periods in the x-direction and are invariant in the y direction, see Fig. 1. In the following, the (x,y) plane and the z-direction will be referred to as the transverse plane and the longitudinal direction, respectively. To define the grating structure, first we have to define a top and a bottom. This is rather arbitrary since the top or the bottom can be the substrate or the cover of a real structure. It is up to the user. Once the top and the bottom of the grating have been defined, the user can choose to illuminate the structure from the top or from the bottom. The z-axis is oriented from bottom to top.

RETICOLO is written with the $exp(-i\omega t)$ convention for the complex notation of the fields. So, if the materials are absorbant, one expects that all indices have a positive imaginary part. The Maxwell's equations are of the form

$$\nabla \times \mathbf{E} = \frac{2i\pi}{\lambda}\mathbf{H}\ (\varepsilon_0 = \mu_0 = c = 1)$$
$$\nabla \times \mathbf{H} = -\frac{2i\pi}{\lambda}\varepsilon\mathbf{E},$$

where $\varepsilon = n^2$ is the relative permittivity, a complex number, and $\lambda$ is the wavelength in a vacuum.
Two situations are considered in the following :
TE polarisation **E** is parallel to Oy,
TM polarisation **H** is parallel to Oy.

RETICOLO returns the diffraction efficiencies of the transmitted and reflected orders for a plane wave incident from the top and from the bottom with the same calculation. Of course, these two incident plane waves must have identical x-component of the parallel wave vector: $k_x^{inc}$. This possibility which is not mentioned in the literature to our knowledge is important in practice since the user may get, with the same computational loads, the diffraction efficiencies of the grating component illuminated from the substrate or from the cover.

RETICOLO-1D calculates the electric and magnetic fields diffracted by the grating for the following incident plane wave:

$$\boldsymbol{E}_{top}^{inc}\ exp\left(i\left(k_x^{inc}x + k_{z\,top}^{inc}(z-h)\right)\right)$$
$$\boldsymbol{H}_{top}^{inc}\ exp\left(i\left(k_x^{inc}x + k_{z\,top}^{inc}(z-h)\right)\right), \text{ if incident from the top layer,}$$
$$\text{where } k_{z\,top}^{inc} = -\sqrt{\left(2\pi n_{top}/\lambda\right)^2 - (k_x^{inc})^2}$$
$$\boldsymbol{E}_{bottom}^{inc}\ exp\left(i\left(k_x^{inc}x + k_{z\,bottom}^{inc}(z-h)\right)\right)$$
$$\boldsymbol{H}_{bottom}^{inc}\ exp\left(i\left(k_x^{inc}x + k_{z\,bottom}^{inc}(z-h)\right)\right), \text{ if incident from the bottom layer,}$$
$$\text{where } k_{z\,bottom}^{inc} = \sqrt{(2\pi n_{bottom}/\lambda)^2 - (k_x^{inc})^2}.$$

The $z$-component of the Poynting vector of the incident plane wave is ±0.5.

The Rayleigh-expansion of the diffracted electric fields are shown in the following figure.

$$\boldsymbol{E}_{top}^{dif} = \sum_m \boldsymbol{E}_{top}^m\ exp\left[i((k_x^{inc} + mK_x)x + k_{z\,top}^m(z-h)\right]$$
$$\boldsymbol{H}_{top}^{dif} = \sum_m \boldsymbol{H}_{top}^m\ exp\left[i((k_x^{inc} + mK_x)x + k_{z\,top}^m(z-h)\right]$$
$$\text{where } k_{z\,top}^m = \sqrt{\left(2\pi n_{top}/\lambda\right)^2 - (k_x^{inc} + mK_x)^2}$$
$$\boldsymbol{E}_{bottom}^{dif} = \sum_m \boldsymbol{E}_{bottom}^m\ exp[i((k_x^{inc} + mK_x)x + k_{z\,bottom}^m z]$$
$$\boldsymbol{H}_{bottom}^{dif} = \sum_m \mathbf{H}_{bottom}^m\ exp[i((k_x^{inc} + mK_x)x + k_{z\,bottom}^m z]$$
$$\text{where } k_{z\,bottom}^m = \sqrt{(2\pi n_{bottom}/\lambda)^2 - (k_x^{inc} + mK_x)^2}$$

They are shown in the following figure.

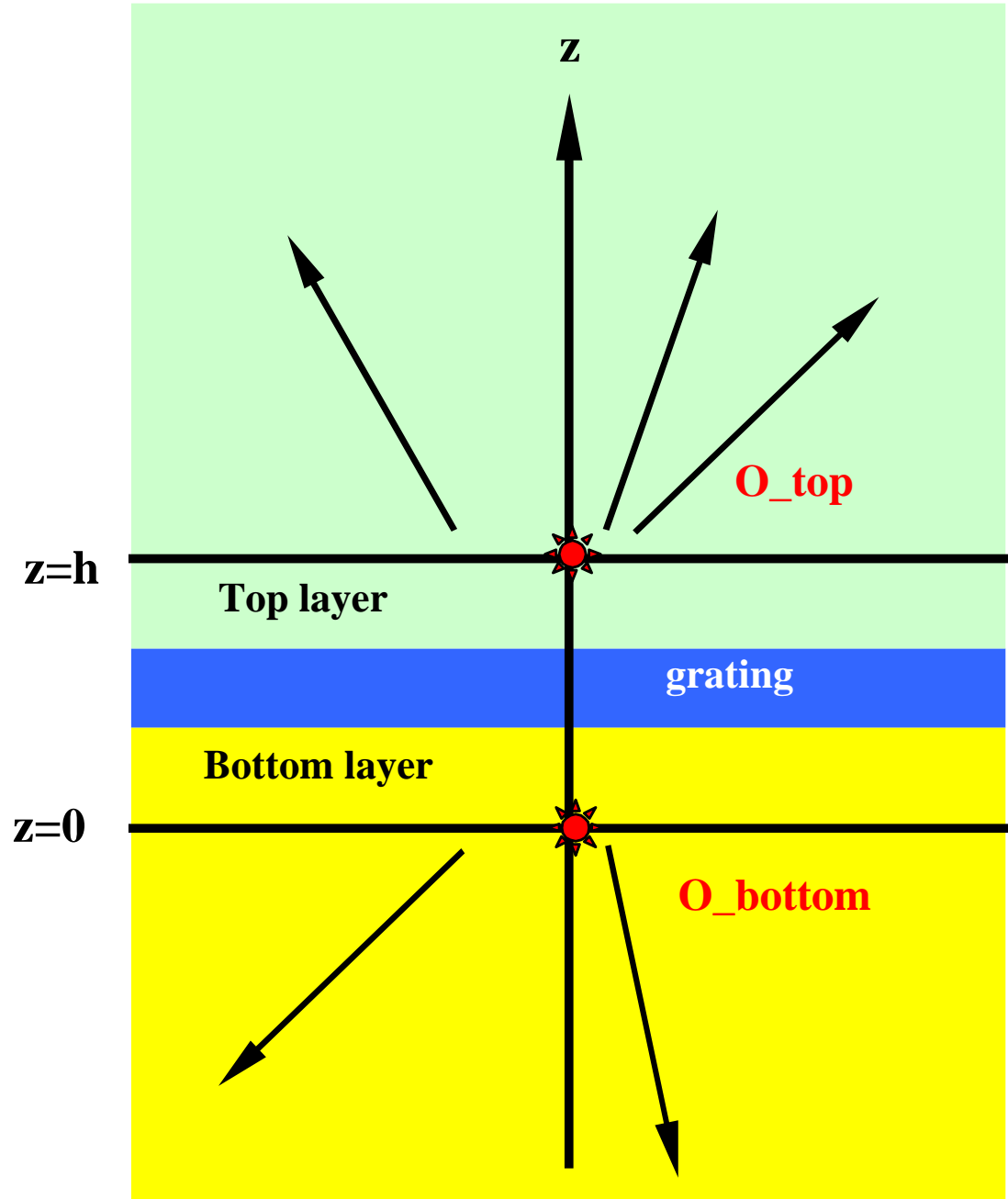

Fig. 1. Rayleigh expansion for the diffracted fields. $K_x = (2\pi)/period$. The m[th] order has a parallel momentum equal to $k_x^{inc} + mK_x$. We define two points $O_{top}$= (0,0,h) at the top of the grating, and $O_{bottom}$= (0,0,0) at the bottom of the grating.

The following is organized so that one can straightforwardly write a code using the software.

# 3. Preliminary input parameters

The name of the following parameters are given as examples. The user may define his own parameter vocabulary.

**wavelength** = **3**; % wavelength (λ) in a vacuum. It might be 3 nm or 3 µm. You do not need to specify the unit but all other dimensions are of course in the same unit as the wavelength.

**period** = in the x-direction.

**nn** = **20**; % this define the set of Fourier harmonics retained for the computation. More specifically, 2×nn+1 represent the number of Fourier harmonics retained from –nn to nn. This is a very important parameter ; for large n values, a high accuracy for the calculated data is achieved, but the computational time and memory is also large. If all the textures are homogeneous (case of a thin-film stack), we may set nn=1 and the period may be arbitrarily set to any value, 1 for example. NB: Because of our normalization (Poynting vector equal to 1), the computed reflected and transmitted amplitude coefficients are not identical to those provided by the classical Fresnel formulas found in textbooks.

**parm** = **res0(1)** for TE polarisation;
**parm** = **res0(-1)** for TM polarisation;
% res0.m is a function that set default values to all parameters used by the code and determine the polarisation.

**k_parallel** =$k_x^{inc}/(2\pi/\lambda)$ is the normalised parallel momentum of the incident plane wave.
If the grating is illuminated from the top region (or from the bottom region) under an incident angle θ, one has:
**k_parallel=n_inc*sin(θ)**,
where n_inc is the refractive index of the top (or bottom) layer. One expects that it is a positive real number and that the texture (see Section 4.1) associated to the top (or the bottom) layer has a background with a uniform refractive index “n_inc”.
(Note that the “k_parallel” variable is defined **without** the factor $\boldsymbol{2\pi/\lambda}$.)

It is very important to keep in mind that wether one defines the incident plane wave in the top layer or in the bottom layer, the calculation will be done for both an incident wave from the top and an incident wave from the bottom, with an identical parallel momentum **k_parallel**.

These 5 parameters ("wavelength, nn, parm and k_parallel) are required by the code. Some other parameters can additionally be defined. For example, the default parameters do not take the symmetry of the problem into account. So if one wants to use symmetries, a new parameter has to be defined: "**parm.sym.x**", (see section 7). If one wants to calculate accurately the electromagnetics fields, one has to define: " **parm.res1.champ=1",** but this increases the calculation time and memory loads (see section 8).

# 4. Structure definition (grating parameters)

The grating encompasses a uniform upperstrate, called the top in the following, a uniform substrate, called the bottom in the following, and many layers which define the grating, which is defined by a stack of layers. Every layer is defined by a "texture" and by its thickness. Two different layers may be identical (identical texture and thickness), may have different thicknesses with identical texture, may have different thicknesses and textures. To define the diffraction geometry, we need to define the different textures and then the different layers.

## 4.1. How to define a texture?

Every texture is defined by a cell-array composed of two line-vectors of identical length. The first vector, let us say $[x_1\ x_2\ ...\ x_p\ ...x_N]$, contains all the x-values of the discontinuities. One *must* have :

$N>1$,

$x_p<x_{p+1}$ for any p,

and $x_N - x_1<\text{period}$.

The second line-vector $[n_1\ n_2\ ...\ n_p\ ...\ n_N]$ contains the refractive indices of the material between the discontinuities. More explicitly, we have a refractive index $n_p$ for $x_{p-1}<x<x_p$. Because of periodicity, note that the refractive index for $x_N<x<x_1+\text{period}$ is equal to $n_1$.
The specific case of a uniform texture with a refractive index n is easily defined by texture{1}={n}. In that specific case, no need of a second vector since there is no discontinuity.

The textures have all to be to be packed together in a cell array textures={textures{1}, textures{2}, textures{3}} prior calling subroutine **res1.m.**

Example :
period=17;
textures =cell(1,2);
textures{1}={1.5}; %uniform texture
textures{2}={[-5,-3,1,6],[2,1.3,1.5,3]}; %texture composed of 4 different refractive indices

The following figure shows the refractive indices of the two textures.

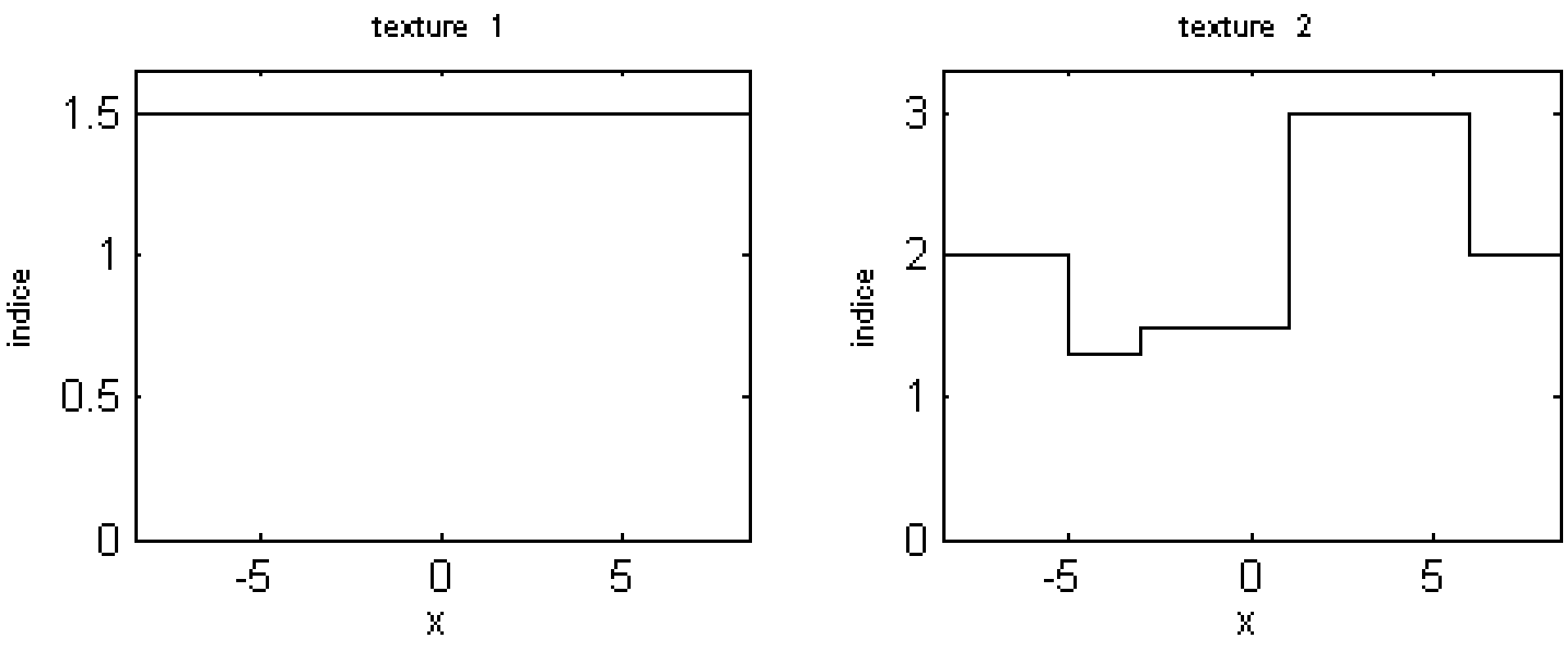

Fig. 2. Textures{1} and {2}.

Slits in perfectly-conducting metallic textures:
Mixing perfectly-conducting metallic textures and dielectric textures in the same grating structure is possible. We have first to define a background by its refractive index “inf” (for infinity). In this uniform background, we can incorporate strip inclusions with a complex or real refractive index “ninclusion” defined by the position c of its center and its x-width L. The inclusions cannot overlap.
For example:

textures {3}= {inf, [c1,L1,ninclusion1],[c2,L2, ninclusion2]}

Anisotropic layers:
Grating layers (not the substrate nor the superstrate) can be anisotropic with diagonal tensors ($\varepsilon_{xy} = \varepsilon_{xz} ... = 0$).
To implement diagonal anisotropy

parm.res1.change_index={[$n_{prov}^1$, $n_x^1$, $n_y^1$, $n_z^1$] , [$n_{prov}^2$, $n_x^2$, $n_y^2$, $n_z^2$]}; % $n_{prov}^1 \neq n_{prov}^2$

The refractive index $n_{prov}^1$ is then replaced **in all textures** by epsilon=diag([$(n_x^1)^2$, $(n_y^1)^2$, $(n_z^1)^2$]). Beware if the superstate (or substrate) has a refractive index $n_{prov}^1$, it will also be replaced and this is not allowed. Thus we recommend using an unusual value for $n_{prov}^1$ (e.g. 89.99999 or rand(1)).
The user may also diagonal permeability tensors

parm.res1.change_index={ [$n_{prov}^1$, $n_x^1$, $n_y^1$, $n_z^1$ , $m_x^1$, $m_y^1$, $m_z^1$ ] , [$n_{prov}^2$, $n_x^2$, $n_y^2$, $n_z^2$] };

The refractive index $n_{prov}^1$ is then replaced **in all textures** by epsilon=diag( [$(n_x^1)^2$, $(n_y^1)^2$, $(n_z^1)^2$] ), mu=diag( [$(m_x^1)^2$, $(m_y^1)^2$, $(m_z^1)^2$] ).
For slits in perfectly-conducting metallic textures, anisotropy cannot be implemented.

In order to check if the set of textures is correctly set up, the user can set the variable parm.res1.trace equal to 1: “parm.res1.trace = 1;”. Then a Matlab figure will show up the refractive-index distribution of all textures. Each texture is represented with the coordinate x varying from –period/2 to period/2.

## 4.2. How to define the layers?

This is performed by defining the “Profile” variable which contains, starting from the top layer and finishing by the bottom layer, the successive information (thickness and texture-label) relative to every layer. Here is an example that illustrates how to set up the “Profile” variable:

**Profile** = {[0,1,0.5,0.5,1,0.5,0.5,2,0],[1,3,2,4,3,2,4,6,2]}; (1)

It means that from the top to the bottom we have: the top layer is formed by a thickness 0 of texture 1, then we have twice textures 3, 2 and 4 with depth 1, 0.5 and 0.5 respectively, texture 6 with depth 2, and finally the bottom layer (formed by texture 2) with null thickness. Since textures 1 and 2 correspond to the top and bottom layers, they must be uniform. In this example, the top and bottom layers have a null thickness. However, one may set an arbitrary thickness. Especially, if one needs to plot the electromagnetic fields in the bottom and top layers, the thicknesses $h_b$ and $h_h$ (see Fig. 4) over which the fields have to be visualized has to be specified. For $h_b$=$h_h$=0, the Rayleigh expansions of the fields in the top and bottom layers are not plotted.

In this particular Profile, the structure formed by texture 3 with thickness 1, texture 2 with thickness 0.5 and texture 4 with thickness 0.5 is repeated twice. It is possible to simplify the instruction defining the “Profile” variable in order to take into account the repetitions:

**Profile** = {{0,1},{[1,0.5,0.5], [3,2,4], 2},{[2,0],[6,2]}}; (2)

If a structure is repeated many times, the above “factorized” instruction of Eq. 2 is better than the “expanded” one of Eq. 1, in terms off computational speed, because the calculation will take into account the repetitions.

The profile is shown below.

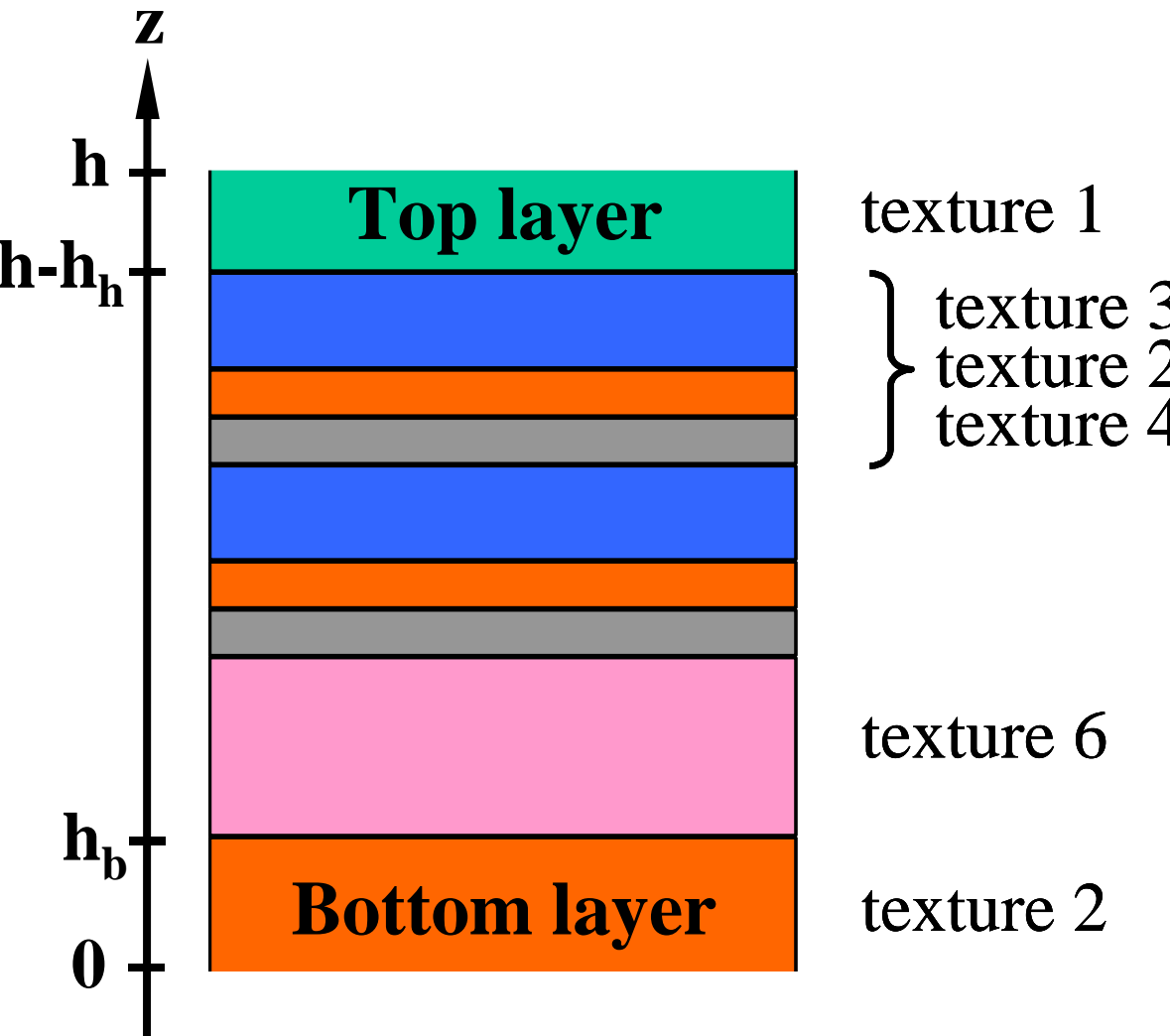

Fig. 3. Texture stacks. The example corresponds to a profile defined by **Profile** = {[$h_h$,1,0.5,0.5,1,0.5,0.5,2, $h_b$],[1,3,2,4,3,2,4,6,2]}; . The top and bottom layers have uniform textures.

## 5. Solving the eigenmode problem for every texture

The first computation with the RCWA consists in calculating the eigenmodes associated to all textures. This is done by the subroutine “res1.m”, following the instruction:

**aa = res1(wavelength,period,textures,nn,k_parallel,parm)**;

This subroutine has 6 input arguments: the wavelength “**wavelength**”, the period of the grating “**period**”, the “**textures**” variable, the number of Fourier harmonics “**nn**”, the normalized parallel incident wave vector “**k_parallel**, and the “**parm**” variable containing the values of all parameters used by the code and the selected the polarisation. If one has to study the diffraction by different gratings composed of the same textures, one needs to compute only once the eigenmodes. It is possible to save the “aa” variable in a “.mat” file and to reload it for the computation of the diffracted waves, see an example in Annex 10.3.

## 6. Computing the diffracted waves

This is the second step of the computation. This is done by the subroutine “**res2.m**”, following the instruction:

**result = res2(aa, Profile)**;

This subroutine has 2 input arguments: the output “**aa**” of the subroutine “res1.m” and the “**Profile**” variable. The output argument “**result**” contains all the information on the diffracted fields. “**result**” is an object of class

'reticolo' that can be indexed as an usual structure with parentheses, or with the labels of the considered orders between curly braces. Examples will be given in the following.
This information is divided into the following sub-structures fields :

- "**result. inc_top**"
- "**result. inc_top_reflected**"
- "**result. inc_top_transmitted**"

- "**result. inc_bottom**"
- "**result.inc_bottom_reflected**"
- "**result. inc_bottom_transmitted**"

The sub-structure "**result.inc_top_reflected**" contains all the information concerning the propagative *reflected* waves *for an incident wave from the top layer* of the grating. The incident wave is described in the sub-structure "**result.inc_top**".

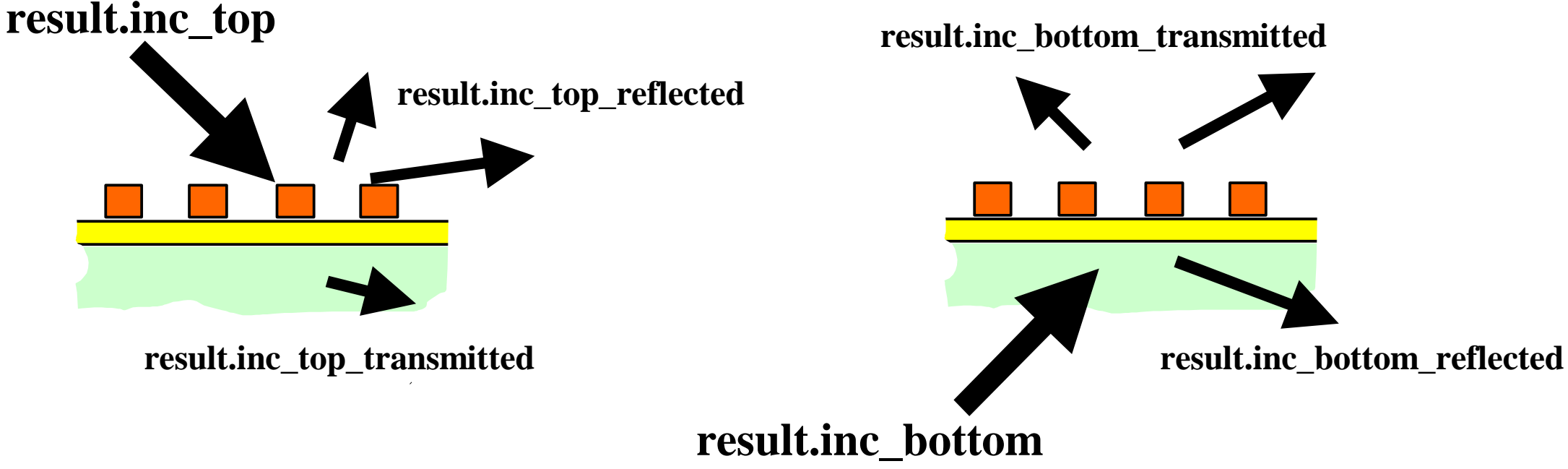

Fig. 4. The two obtained solutions.

Each sub-structure of **result** is composed of the several fields. Each field is a Matlab column vector or matrix having the same number N of lines. N is the number of propagative orders considered and can be 0.

| **Field name** | **Signification** | size |
|---|---|---|
| order | orders of the diffracted propagative plane waves | N, 1 |
| theta | angle $\theta_m$ of each diffracted order | N, 1 |
| **K** | normalised wave vector | N, 3 |
| efficiency | efficiency of each diffracted order | N, 1 |
| amplitude | complexe amplitude in TE polarization of every order | N, 1 |
| **E** | electric field $(E_x, E_y, E_z)$ of the diffracted orders at O_top or O_bottom when the amplitude of the incident plane wave is one. | N, 3 |
| **H** | magnetic field $(H_x, H_y, H_z)$ of the diffracted orders at O_top or O_bottom when the amplitude of the incident plane wave is one. | N, 3 |
| **PlaneWave_E** | E-vector components of the $\overrightarrow{PW}$ 's (in the Oxyz basis) | N, 3 |
| **PlaneWave_H** | H-vector components of the $\overrightarrow{PW}$ 's (in the Oxyz basis) | N, 3 |

(To use the same notations as in the conical code or in the crossed-grating code, set parm.res1.result=-1 before calling res1.m).

## 6.1. Efficiencies

For a given diffraction order n, the diffraction efficiency is defined as the ratio between the flux of the diffracted Poynting vector and the flux of the incident Poynting vector (flux through a period of the grating).

The efficiencies of all propagative reflected and transmitted waves for an incident wave from the top of the grating are given by the two vectors "**result.inc_top_reflected.efficiency**" and "**result.inc_top_transmitted.efficiency**". If all refractive indices are real, the sum of all elements of these two vectors is equal to one because of the energy conservation. The labels n of the corresponding orders are in "**result.inc_top_reflected.order**" (see below for a description of the other fields of this sub_structure).

<u>Some examples</u>

1) The efficiency of the reflected order -2 ( $k_{//}=k_x^{inc}-2K_x$ ) when the grating is illuminated from the top is equal to **result. inc_top_reflected.efficency{-2}**. If this order is evanescent, the efficiency is 0.

It is important to have in mind the difference between :
**result.inc_top_reflected.efficiency{-2}** : efficiency of order 2
**result.inc_top_reflected.efficiency(-2)** : gives an error !
**result.inc_top_reflected.efficiency{2}** : efficiency of order 2
**result.inc_top_reflected.efficiency(2)** : efficiency in order **result. inc_top_reflected.order(2)**;

2) The orders of all the transmitted-propagative plane waves for an incident wave from the top of the grating are given by the vector "**result.inc_top_transmitted.order**".

3) The efficiencies of all propagative reflected waves for an incident wave from the bottom in TM polarization are given by the vector "**result.inc_bottom_reflected.efficiency**".

## 6.2. Rayleigh expansion for propagatives modes

The coefficients of the Rayleigh expansion of Fig. 1 can be obtained from the structure **result**. For instance, when the grating is illuminated from the bottom with a TE polarised mode, we have :

$\mathbf{E}^{m}_{\mathrm{bottom}}$=result.inc_bottom_reflected.E{m} (3 components in Oxyz)

$\mathbf{H}^{m}_{\mathrm{bottom}}$=result inc_bottom_reflected.H{m} (3 components in Oxyz)

$\mathbf{E}^{m}_{\mathrm{top}}$ =result.inc_bottom_transmitted.E{m} (3 components in Oxyz)

$\mathbf{H}^{m}_{\mathrm{top}}$=result.inc_bottom_ transmitted.H{m} (3 components in Oxyz)

and the incident plane wave defined in page 4 is given by :

$\mathbf{E}^{\mathrm{inc}}_{\mathrm{bottom}}$=result inc_bottom.E (3 components in Oxyz)

$\mathbf{H}^{\mathrm{inc}}_{\mathrm{bottom}}$=result.inc_bottom.H (3 components in Oxyz).

## 6.3. Amplitude of diffracted propagative waves

### 6.3.1 Angle $\theta_m$

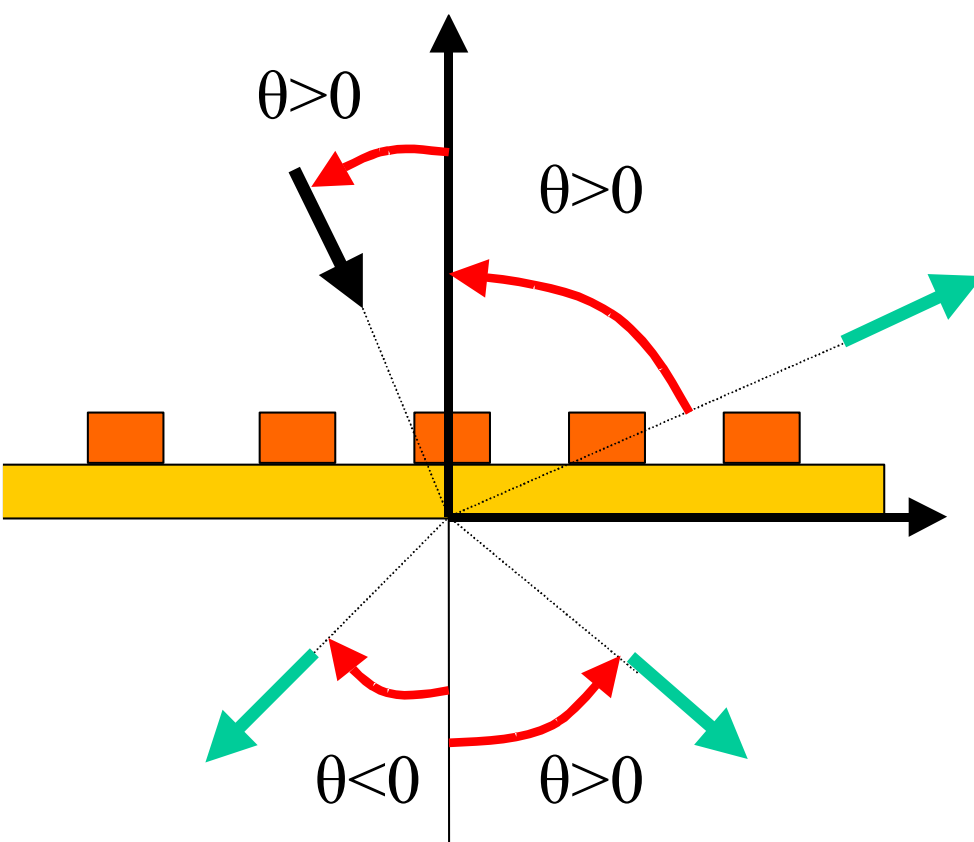

Fig. 5 $\theta_m$ angle orientations.

The angle $\theta_m$ related to order m is varying between –90 and 90. It is oriented in such a way that the k-parallel momentum of the corresponding wave vector (incident or diffracted) is
$k_x^{inc}+mk_x = (2\pi/\lambda)\ n_{_top}\ \sin(\theta_m)$ or $(2\pi/\lambda)\ n_{_bottom_}\sin(\theta_m)$.

### 6.3.2 $O_{top}$ and $O_{bottom}$ points

$O_{top}$ and $O_{bottom}$ are 2 important points (see Fig. 1). In the Cartesian coordinates system Oxyz , they are defined by: $O_{top}$=(0,0,h) at the top of the grating, and $O_{bottom}$=(0,0,0) at the bottom of the grating.

In addition, let us consider an arbitrary point M=(x,y,z) in the 3D space in Oxyz. Associated to this point, we define the two vectors :
$\mathbf{r}_{top}= \overrightarrow{O_{top}M}$ , and
$\mathbf{r}_{bottom}= \overrightarrow{O_{bottom}M}$ .

### 6.3.3 Jones' coefficient

Let us assume that the grating is illuminated from the top layer and let us consider a diffracted order m in the bottom layer. Any other diffraction situation is straighforwardly deduced.
Let α be a given complex number. The incident electromagnetic field (6 components of **E** and **H** in every points of the 3D space) can be written :

$$\mathbf{W}^{inc}=\alpha\,\overrightarrow{PW}\ ,$$

where $\overrightarrow{PW}$ is a plane wave defined in every point by $\overrightarrow{PW}=\mathbf{A}\exp\left(i\mathbf{k}_{top}^{inc}\mathbf{r}_{top}\right)$, **A** being the electromagnetic fields (6 components) of the plane wave at M=$O_{top}$, and $\mathbf{k}_{top}^{inc}$ is the incident wave vector. **A** and $\mathbf{K}=\mathbf{k}_{top}^{inc}/\left|\mathbf{k}_{top}^{inc}\right|$ are given by the structure “**result“** as will be defined later.

Similarly, the diffracted electromagnetic field in the m bottom order can be written :

$$\mathbf{W}_m^{dif}=\gamma\ \overrightarrow{PW^m}\ \ ,$$

where γ is a complex number, $\overrightarrow{PW^m}$ is a plane wave defined in every point by $\overrightarrow{PW^m}=\mathbf{A}^m\ \exp\left(i\mathbf{k}_{bottom}^m\ \mathbf{r}_{bottom}\right)$, $\mathbf{A}^m$ is the electromagnetic fields (6 components) of the plane wave at M=$O_{bottom}$, and $\mathbf{k}_{bottom}^m$ is the wave vector of the mth transmitted order. $\mathbf{A}^m$ and, $\mathbf{K}^m=\mathbf{k}_{bottom}^m/\left|\mathbf{k}_{bottom}^m\right|$ are given by the structure “**result“** as will be defined later.

We define the Jones'coefficient J, associated to the order m by
$\gamma = J\ \alpha$
**A** and $\mathbf{A}^m$ are normalized so that the $|J|^2$ is the diffraction efficiency. For instance, $|J|^2$ = **result. inc_top_transmitted.efficency {m }**.

We now define all these data from the “result” structure :
**K** = result. inc_top.K.
$\mathbf{K}^m$ = result. inc_top_transmitted.K{m}.

In the Cartesian coordinate system Oxyz :

$$\mathbf{A} = \begin{pmatrix}\text{result.inc_top.PlaneWave_E}\\ \text{result.inc_top.PlaneWave_H}\end{pmatrix}$$

$$\mathbf{A}^m = \begin{pmatrix}\text{result.inc_top_transmitted.PlaneWave_E\{m\}}\\ \text{result.inc_top_transmitted.PlaneWave_H\{m\}}\end{pmatrix}.$$

The Jones' coefficients is:
J = result.Jones.inc_top_transmitted {m} (=result.inc_top_transmitted.amplitude {m}).

# 7. Using symmetries to accelerate the computational speed

When the grating possesses some mirror symmetry for the plane $x=x_0$, one may define "**parm.sym.x=** $x_0$. Then when k_parallel =0, the code will use the symmetry property for speeding up the calculation.
Note that the code does not verify if the symmetries of the grating defined by the user are in agreement with the "textures" parameters. It is up to the user to define carefully the parameters parm.sym.x. All textures used in the calculation must possess the same symmetry.

# 8. Plotting the electromagnetic field and calculating the absorption loss

## 8.1. Computation of the electromagnetic fields

Once the eigenmodes associated to all textures are known, the calculation of the electromagnetic fields everywhere in the grating can be performed. This calculation is done by the function "**res3.m**", following the instruction:

**[e,z,index] = res3(x,aa,Profile, inc,parm);**

The function"res3.m" can be called without calling "res2.m". This subroutine has 5 input arguments:
-the "**x**" variable is a vector containing the locations where the fields will be calculated in the x-direction [for instance we may set **x = linspace(-period_x/2, period_x/2, 51);** for allocating 51 sampling points in the x-direction]**,**
the "**aa**" variable contains all the information on the eigenmodes of all textures and is computed by the subroutine res1.m,
-the variable "**Profile**" is defined in Section 4.2; note that it can be redefined,
-the variable "**inc**" defines the y component of the complex amplitude of the incident electric (in TE polarisation) or magnetic field (in TM polarisation) field at O_top or O_bottom .
For illuminating the grating exactly by the TE-polarized incident $\overrightarrow{PW}$ defined above, one should set:
einc= **result.inc_top PlaneWave_E(2)** for TE polarisation; einc= **result.inc_top PlaneWave_H(2)** for TM polarisation.

-the "**parm**" variable, already mentioned is discussed in the following.

There are three possible output arguments for the subroutine "**res3.m**". The variable "**e**" contains all the electromagnetic field quantities:

**$E_y$=e(:,:,1); $H_x$=e(:,:,2); $H_z$=e(:,:,3)**; in TE polarization.
**$H_y$=e(:,:,1); $E_x$=e(:,:,2); $E_z$=e(:,:,3)**; in TM polarization.

The second variable "**z**" is the vector containing the z-coordinate of the sampling points. Note that in the matrix $E_x$=e(:,:,1), the first index refer to the z coordinate, and the second to the x-coordinate. Thus $E_x$(i,j) is the $E_x$ field component at the location {z(i), x(j)}. The third variable "**index**" is the complex refractive index of the considered grating. index(i,j) is the refractive index at the location {z(i), x(j)}. It can be useful to test the profile of the grating.

Some **important** comments on the **parm**" variable :
1. For calculating precisely the electromagnetics fields, one has to set : "**parm.res1.champ=1"** before calling **res1.m.** This increases the calculation time and memory load, but it is highly recommended. If not, the computation of the field will be correct only in homogenous textures (for example in the top layer and in the bottom layer).
2. Illuminating the grating from the top or the bottom layer : As mentioned earlier, the code compute the diffraction efficiencies of the transmitted and reflected orders for an incident plane wave from the top and for an incident plane wave from the bottom at the same time. When plotting the field, the user must specify the direction of the incident plane wave. This is specified with variable **parm.res3.sens**. For **parm.res3.sens=1**, the grating is illuminated from the top and for **parm.res3.sens=-1**, the grating is illuminated from the bottom (default is **parm.res3.sens=1**).

3. Specifying the z locations of the computed fields: This is provided by the variable **parm.res3.npts**. **parm.res3.npts** is a vector whose length is equal to the number of layers. For instance let us imagine, a grating defined by **Profile** = {[0.5,1,2,0.6],[1,2,3,4]}. Setting **parm.res3.npts=[2,3,4,5]** implies that the field will be computed in two z=constant plans in the top layer, in three z=constant plans in the first layer (texture 2), in four z=constant plans in the second layer (texture 3), and in five z=constant plans in the bottom layer. Default for **parm.res3.npts** is 10 z=constant plans per layer.

**VERY IMPORTANT** : where is the z=0 plan and what are the z-coordinates of the z=constant plan? The z=0 plan is defined at the bottom of the bottom layer. Thus, the field calculation is performed only for z>0 values. For the example **Profile** = {[0.5,1,2,0.6],[1,2,3,4]}, and if we refer to texture 4 as the substrate, the z=0 plan is located in the substrate at a distance 0.6 under the grating. The z=constant plans are located by an equidistant sampling in every layer. Always referring to the previous example, it implies that the five z=constant plans in the substrate are located at coordinate z=(p-0.5) 0.6/5, where p=1,2,…5. Note that the z coordinates for the z=constant plans are always given by the second output variable of res3.m.

4. How can one specify a given z=constant plan? First, one has to redefine the variable **Profile**. For the grating example with the two layers discussed above, let us imagine that one wants to plot the field at z=z0+0.6+0.2 in layer 2. Then one has to set: **Profile** = {[0.5,1-z0,0,z0,0.2,0.6],[1,2,2,2,3,4]} and set **parm.res3.npts=[0,0,1,0,0,0]**. Note that it is not necessary to redefine the variable **Profile** at the beginning of the program. One just needs to redefine this variable before calling subroutine res3.m.

5. Automatic plots: an automatic plot (showing all the components of the electromagnetic fields and the grating refractive index distribution) is provided by setting **parm.res3.trace**=1. If one wants to plot only some components of the fields, one can set for instance in TE polarization: **parm.res3.champs**=[1,0] to plot $E_y$ and the objet, **parm.res3.champs**=[2] to plot only $H_x$.

## 8.2. Computation of the absorption loss

Loss computation is performed with the subroutine "**res3.m**".

<u>First approach based on integrals (not valid for homogeneous layers with non-diagonal anisotropy):</u>
The absorption loss in a surface $S$ is given by:
$L = \frac{\pi}{\lambda}\int_S Im\ \varepsilon(M)\ \left|E_y(M)\right|^2\ dS$ for TE polarization.
$L = \frac{\pi}{\lambda}\int_S Im\ (\varepsilon_{XX}(M)|E_X(M)|^2 + \varepsilon_{ZZ}(M)|E_z(M)|^2)\ dS$ for TM polarization.
These integrals can be computed with the following instruction

[**e, Z, index, wZ, loss_per_layer, loss_of_Z, loss_of_Z_X, X, wX**] = res3(**x,aa,Profile,einc,parm**);

The important ouput arguments are:
**loss_per_layer**: the loss in every layer defined by **Profile**, **loss_per_layer**(1) is the loss in the top layer, **loss_per_layer**(2) is the loss in layer 2, ... and **loss_per_layer**(end) is the loss in the bottom layer
**loss_of_Z**: the absorption loss density (integrated over **X**) as a function of **Z** (like for **X**, the sampling points **Z** are not equidistant. You may plot this loss density as follows: plot(**Z**, **loss_of_Z**), xlabel('Z'), ylabel('absorption')
**loss_of_Z_X**(**Z**,**X**) = π/λ Im(**index**(**Z**,**X**).^2) |**e**(**Z**,**X**,**1**)|$^2$ in TE polarization
**loss_of_Z_X**(**Z**,**X**) = π/λ Im(**index**(**Z**,**X**).^2) ( **e**(**Z**,**X**,**2**)|$^2$+|**e**(**Z**,**X**,**3**)|$^2$) in TM polarization
**index**: **index**(i,j) is the complex refractive index at the location {**z**(i), **x**(j)}.

<u>Second approach based on Poynting theorem (always valid, even for homogeneous layers with non-diagonal anisotropy):</u>
An alternative approach to compute the losses in the layers consists in calculating the difference in the flux of the incoming and outgoing Poynting vectors. This approach is faster, but in some cases, the computation of the integral can be more accurate. In homogeneous layers with non-diagonal anisotropy, only this approach is possible.
To specify which approach used per layer, we define a vector

**parm.res3.pertes_poynting** = [0,0,0,1,0]; % for instance for a 5-layer grating with "0", the integral approach is used (default option) and with "1", the Poynting approach is used. The length of **parm.res3.pertes_poynting** is equal to the number of layers. We may set **parm.res3.pertes_poynting** = 0 or 1; the scalar is then repeated for all layers.

We may then compute the flux of the Poynting vector in the layer-boundary planes

[**e, Z, index, wZ,loss_per_layer,loss_of_Z,loss_of_Z_X,X,wX,Flux_Poynting**] = res3(**x,aa,Profile,einc,parm**);

**Flux_Poynting** is a vector. **Flux_Poynting(1)** corresponds to the upper interface of the top layer. The flux is computed for a normal vector equal to the $\hat{\mathbf{z}}$ vector. If **Flux_Poynting(p)** > 0, the energy flows toward the top and if it is negative, the energy flows toward the bottom.

For an illumination from the top and a lossy substrate, the substrate absorption is **−Flux_Poynting (end)/(0.5*period)**. For an illumination from the bottom and a lossy superstrate, the superstrate absorption is **Flux_Poynting (1)/(0.5*period)**.

Note on the computation accuracy of the integral approach:
To compute integrals like the loss or the electromagnetic energy, RETICOLO uses a Gauss-Legendre integration method. This method, which is very powerful for 'regular' functions, becomes inaccurate for discontinuous functions. Thus, the integration domain should be divided into subdomains where the electric field **E** is continuous. For the integration in **X**, this difficult task is performed by the program, so that the user should only define the limits of integration: the input "**x**" argument is now a vector of length 2, which represent the limits of the x interval (to compute the loss over the entire period, we may take **x**(2)=**x**(1)+**period**. The integration domain is then divided into subintervals where the permittivity is continuous, each subinterval having a length less than $\lambda/(2\pi)$. For every subinterval, a Gauss-Legendre integration method of degree 10 is used. This default value can be changed by setting **parm.res3.gauss_x**=.... The actual points where the field is computed are returned in the output argument **X**.

For the z integration, the discontinuity points are more easily determined by the variable **'Profile'**. The user may choose the number of subintervals and the degree in every layer using the parameter **parm.res3.npts**, which is now an array with two lines (in subsection 8.1 this variable is a line vector): the first line defines the degree and the second line the numbers of subintervals of every layer. For example: **parm.res3.npts** = [ [10,0,12] ; [3,1,5] ]; means that 3 subintervals with 10-degree points are used in the first layer, 1 subintervals with 0 point in the second layer, 5 subintervals with 12degree points in the third layer.

The actual z-points of computation of the field are returned in the output variable **Z**, and the vector **wZ** represents the weights and we have sum(**loss_of_Z**.***wZ**)=sum(**loss_per_layer**). Although the maximum degree that can be handled by reticolo is 1000, it is recommended to limit the degree values to modest numbers (10-30 maximum) and to increase the number of subintervals (the larger the degree, the denser the sampling points in the vicinity of the subinterval boundaries).

Note that if **einc**= **result. inc_top PlaneWave_E(2)**, in TE ploarization, or **einc**= **result. inc_top PlaneWave_H(2)**, in TE ploarization **,** the energie conservation test for an incident plane wave from the top is

sum(**result. inc_top_reflected.efficiency**)+
sum(**result. inc_top_transmitted.efficiency**)+
sum(**loss_per_layer**) / (.5***period**) = 1.

Usually, this equality is achieved with an absolute error of $<10^{-5}$.

For specialists:
-**loss_of_Z_X** =pi/ **wavelength***imag(**index**.^2).* abs(**e**(:,:,1)).^2; in TE polarization
-**loss_of_Z_X** =pi/ **wavelength***imag(**index**.^2).*sum(abs(**e**(:,:,2:3)).^2,3); in TM polarization
-**loss_of_Z** =(**loss_of_Z_X***wX**(:)).';
-by setting **index**(**index** ~= index_chosen)=0 in the previous formulas, one may calculate the absorption loss in the medium of refractive index index_chosen.

# 9. Bloch-mode effective indices

RETICOLO gives access to another output: the Bloch mode associated to all textures. The Bloch mode *k* of the texture *l* can be written

$$\left|\Phi_k{}^l\right\rangle = \sum_m a_m^{k,l} \, exp\left[i(k_x^{inc} + mK_x)x\right] \, exp\left(i\frac{2\pi}{\lambda} n_{eff}^{k,l} z\right),$$

where $n_{eff}^{k,l}$ is the effective index of the Bloch mode *k* of the texture *l*.

Instruction:
**[aa, n_eff] = res1(wavelength,period,textures,nn,kparallel, parm)**;
Note that the "n_eff" variable is a Matlab cell array: "**n_eff{ii}**" is a column vector containing all the Bloch-mode effective indices associated to the texture "**textures{ii}**". The element number 5 of this vector, for example, is called by the instruction "**n_eff{ii}(5)**;". An attenuated Bloch-mode has a complex effective index.

Bloch mode profile visualization:

To plot the profile of Bloch mode Num_mode of the texture Num_texture:
**res1(aa, neff, Num_texture, Num_mode)**;
To obtain the profile datas in the format given by res3:
**[e,o,x] = res1(aa, neff, Num_texture, Num_mode)**;  % by default, |x| < period/2
**[e,o] = res1(aa, neff, Num_texture, Num_mode, x)**; % by specifying the x vector, x=linspace(0, 3*period(1),100) for example.

# 10. Annex

## 10.1. Checking that the textures are correctly set up

Setting "**parm.res1.trace = 1**;" generates a Matlab figure which represents the refractive-index distribution of all the textures.

## 10.2. The "retio" & "retefface" instructions

RETICOLO automatically creates temporary files in order to save memory. These temporary files are of the form "abcd0.mat", "abcd1.mat" … with abcd are randomly chosen) .They are created in the current directory. In general RETICOLO automatically erases these files when they are no longer needed, but it is recommended to finish all programs by the instruction "**retio**;", which erases all temporary files. Also, if a program anormally stopsone may execute the instruction "**retio**" before restarting the program.
The "retefface" instruction allows to know all the "abcd0.mat" files and to erase them if wanted.

If we are not limited by memory (this is often the case with modern computers), we can prevent the writing of intermediate files on the hard disk by the setting

parm.not_io = 1;

before the call to res1. Then it is no longer necessary to use the retio instruction at the end of the programs to erase the files.
IMPORTANT: to use parfor loops, it is imperative to take the option parm.not_io = 1.

## 10.3. How to save and to reload the "aa" variable

To save the "**aa**" variable in a ".mat" file, the user has to define a new parameter containing the name of the file he or she wants to create : "**parm.res1.fperm = 'file_name'**;". field_name is a char string with at least one letter. The program will automatically save "**aa**" in the file "**file_name.mat**". In a new utilisation it is sufficient to write aa== **'file_name'**;.

<u>Example of a program which calculates and saves the "aa" variable</u>
[...]   % Definition of the input parameters, see Section 3
**parm.res1.fperm = 'toto'**;
[...]   % Definition of the textures, see Section 4.1
**aa = res1(wavelength,period,textures,nn,k_parallel,parm)**;
<u>Example of a program which uses the "aa" variable and then calculates the diffracted waves</u>
[...]   % Definition of the profile, see Section 4.2. Note that the textures used to define the profile argument have to correspond to the textures defined in the program which has previously calculated the "aa" variable.
**aa='toto'**;
**result = res2(aa,Profile)**;
**retio**;

## 10.4. Asymmetry of the Fourier harmonics retained in the computation

**nn = [-15;20]**;   % this defines the set of non-symmetric Fourier harmonics retained for the computation. In this case, the Fourier harmonics from –15 to +20 are retained.
The instructions "**nn = 10**;" and "**nn = [-10;10]**;" are equivalent.

Take care that the use of symmetry imposes symmetric Fourier harmonics if not the computation will be done without any symmetry consideration.

# 11. Summary

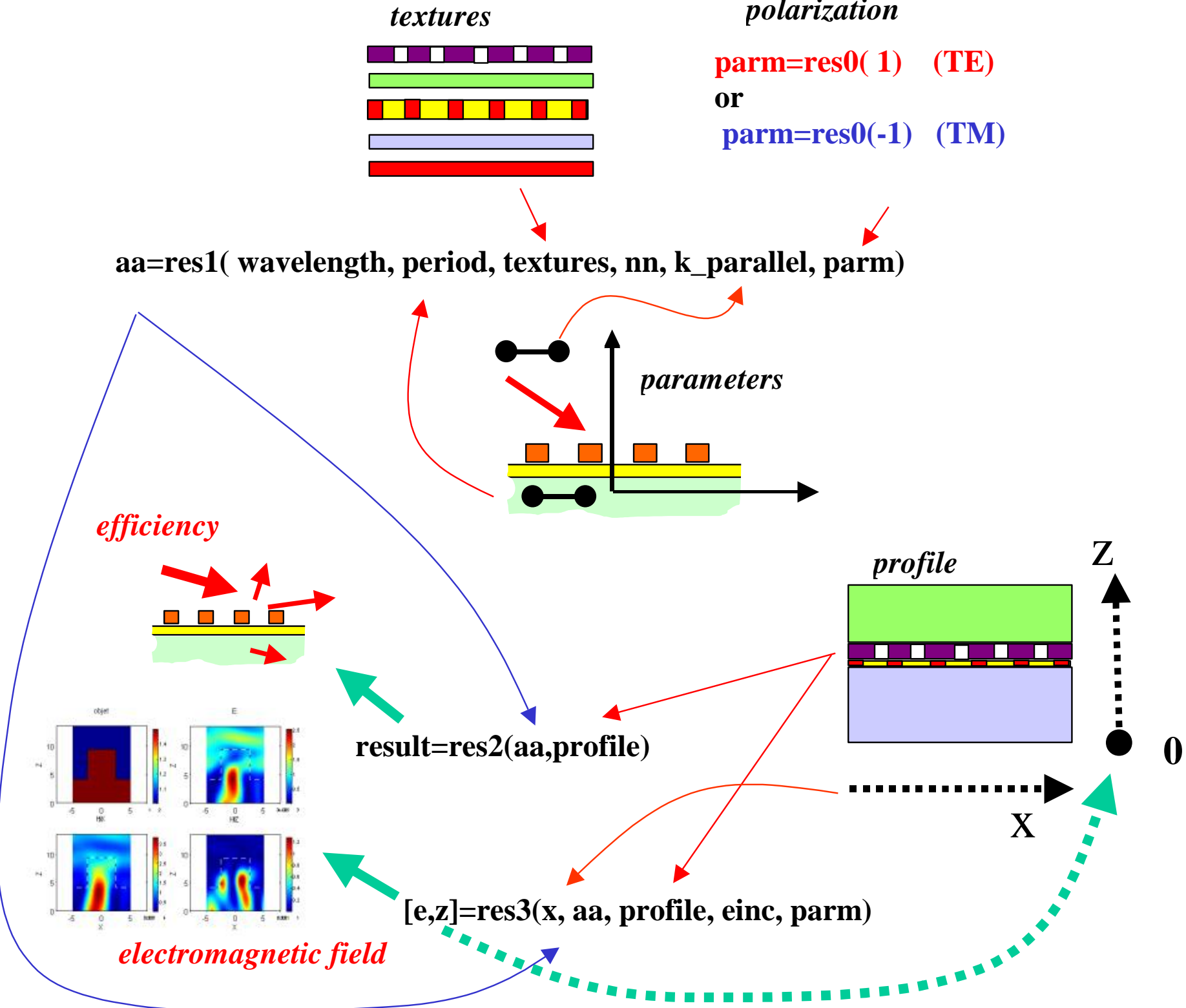

Fig. 6 Summary.

**parm** = **res0( 1)** for TE polarisation;
**parm** = **res0(-1)** for TM polarisation;

**aa** = **res1(wavelength,period,textures,nn,k_parallel,parm)**;
**result** = **res2(aa,Profile)**;
**J** = **result.Jones.inc_top_transmitted {m}**
**[e,z,o]** = **res3(x,aa,Profile, inc,parm)**;

# 12. Examples

The following example can be copied and executed in Matlab.

```
%%%%%%%%%%%%%%%%%%%%%%%%%
% EXAMPLE 1D (TE or TM) %
%%%%%%%%%%%%%%%%%%%%%%%%%
wavelength=8;
period=10;% same unit as wavelength
n_incident_medium=1;% refractive index of the top layer
n_transmitted_medium=1.5;% refractive index of the bottom layer

angle_theta0=-10;k_parallel=n_incident_medium*sin(angle_theta0*pi/180);

parm=res0(1);% TE polarization. For TM : parm=res0(-1)
parm.res1.champ=1;% the electromagnetic field is calculated accurately

nn=40;% Fourier harmonics run from [-40,40]
```

```
% textures for all layers including the top and bottom layers
texture=cell(1,3);
textures{1}= n_incident_medium;                    % uniform texture
textures{2}= n_transmitted_medium;                 % uniform texture
textures{3}={[-2.5,2.5],[n_incident_medium,n_transmitted_medium] };

aa=res1(wavelength,period,textures,nn,k_parallel,parm);

Profile={[4.1,5.2,4.1],[1,3,2]};

one_D_TE=res2(aa,Profile)
eff=one_D_TE.inc_top_reflected.efficiency{-1}
J=one_D_TE.Jones.inc_top_reflected{-1};% Jones'coefficients
abs(J)^2 % first order efficiency for an illumination from the top layer

% field calculation
x=linspace(-period/2,period/2,51);% x coordinates(z-coordinates are determined by
res3.m)
einc=1;
parm.res3.trace=1; % plotting automatically
parm.res3.npts=[50,50,50];
[e,z,index]=res3(x,aa,Profile,einc,parm);
figure;pcolor(x,z,real(squeeze(e(:,:,1)))); % user plotting
shading flat;xlabel('x');ylabel('y');axis equal;title('Real(Ey)');

% Loss calculation
textures{3}={[-2.5,2.5],[n_incident_medium,.1+5i] };
aa_loss=res1(wavelength,period,textures,nn,k_parallel,parm);
one_D_loss=res2(aa_loss,Profile)
parm.res3.npts=[[0,10,0];[1,3,1]];
einc=one_D_loss.inc_top.PlaneWave_E(2);
[e,z,index,wZ,loss_per_layer,loss_of_Z,loss_of_Z_X,X,wX]=res3([-
period/2,period/2],aa_loss,Profile,einc,parm);
Energie_conservation=sum(one_D_loss.inc_top_reflected.efficiency)+sum(one_D_loss.in
c_top_transmitted.efficiency)+sum(loss_per_layer)/(.5* period)-1

retio % erase temporary files
```

# II. RETICOLO CODE 1D: conical diffraction

**Reticolo code 1D-conical is a free software for analyzing 1D gratings in classical and conical mountings. It operates under Matlab. To install it, copy the companion folder "reticolo_allege" and add the folder in the Matlab path.**

## Outline

# 1. Generality

RETICOLO is a code written in the language MATLAB 7.0. (or 6.1) It computes the diffraction efficiencies and the diffracted amplitudes of gratings composed of stacks of lamellar structures. It incorporates routines for the calculation and visualisation of the electromagnetic fields inside and outside the grating. With this freeware version, 2D periodic (crossed) gratings cannot be analysed.

As free alternative to MATLAB, Reticolo can also be run in GNU Octave with minimal code changes. For further information, please contact tina.mitteramskogler@profactor.at.

In brief, RETICOLO implements a frequency-domain modal method (known as the Rigorous Coupled wave Analysis/RCWA). To get an overview of the RCWA, the interested readers may refer to the following articles :
M.G. Moharam et al., JOSAA **12**, 1068 (1995),
M.G. Moharam et al, JOSAA **12**, 1077 (1995),
P. Lalanne and G.M. Morris, JOSAA **13**, 779 (1996),
G. Granet and B. Guizal, JOSAA **13**, 1019 (1996),
L. Li, JOSAA **13**, 1870 (1996), see also L. Li, in *Mathematical Modeling in Optical Science*, (SIAM, Philadelphia), Eds. G. Bao, L. Cowsar, and W. Masters, pp. 111-139, 2001,
L. Li, JOSAA **14**, 2758-2767 (1997),
E. Popov and M. Nevière, JOSAA **17**, 1773 (2000),
which describe the up-to-date formulation of the approach used in Reticolo. Note that the formulation used in the last article (which proposes an improvement for analysing metallic gratings with continuous profiles like sinusoidal gratings) is not implemented in Reticolo. The RCWA relies on the computation of the eigenmodes in all the layers of the grating structure in a Fourier basis (plane-wave basis) and on a scattering matrix approach to recursively relate the mode amplitudes in the different layers.

**Eigenmode solver:** For conical diffraction analysis of 1D gratings, the Bloch eigenmode solver used in Reticolo is based on the article "P. Lalanne and G.M. Morris, JOSAA **13**, 779 (1996)".

**Scattering matrix approach:** The code incorporates many refinements that we have not published and that we do not plan to publish. For instance, although it is generally admitted that the S-matrix is inconditionally stable, it is not always the case. We have developed an in-house transfer matrix method which is more stable and accurate. The new transfer matrix approach is also more general and can handle perfect metals.

**Field calculation:** The calculation of the near-field electromagnetic fields everywhere in the grating is performed according to the method described in "P. Lalanne, M.P. Jurek, JMO **45**, 1357 (1998)" and to its generalization to crossed gratings (unpublished). Basically, no Gibbs phenomenon will be visible in the plots of the discontinuous electromagnetic quantities, but field singularities at corners will be correctly handled.

**Acknowledging the use of Reticolo**: In publications and reports, acknowledgments have to be provided by referencing to J.P. Hugonin and P. Lalanne, *Reticolo software for grating analysis*, Institut d'Optique, Orsay, France (2005).

In journal publications and in addition, one may fairly quote the following references:
-M.G. Moharam, E.B. Grann, D.A. Pommet and T.K. Gaylord, "Formulation for stable and efficient implementation of the rigorous coupled-wave analysis of binary gratings", J. Opt. Soc. Am. A **12**, 1068-1076 (1995),
-P. Lalanne and G.M. Morris, "Highly improved convergence of the coupled-wave method for TM polarization", J. Opt. Soc. Am. A **13**, 779-789 (1996),
-P. Lalanne and M.P. Jurek, "Computation of the near-field pattern with the coupled-wave method for TM polarization", J. Mod. Opt.**45**, 1357-1374 (1998), if near-field electromagnetic-field distributions are shown.

# 2. The diffraction problem considered

In general terms, the code solve the diffraction problem by a grating defined by a stack of layers which have all identical periods in the x- directions and are invariant in the y direction see the following figure. In the following, the (x,y) plane and the z-direction will be referred to as the transverse plane and the longitudinal direction, respectively. To define the grating structure, first we have to define a top and a bottom. This is rather arbitrary since the top or the bottom can be the substrate or the cover of a real structure. It is up to the user. Once the top and the bottom of the grating have been defined, the user can choose to illuminate the structure from the top or from the bottom. The z-axis is oriented from bottom to top.

RETICOLO is written with the exp(-iωt) convention for the complex notation of the fields. So, if the materials are absorbant, one expects that all indices have a positive imaginary part. The Maxwell's equations are of the form :

**rot**(**E**) = (2iπ/λ)* **H**    ( $\varepsilon_0 = \mu_0 = c = 1$)
**rot**(**H**) = - (2iπ/λ)  ε **E**,

where $\varepsilon = n^2$ is the relative permittivity, a complex number, and λ is the wavelength in a vacuum.

RETICOLO-1D returns the diffraction efficiencies of the transmitted and reflected orders for an incident plane wave from the top and for an incident plane wave from the bottom, both for TM and TE polarizations. The four results are obtained by the same calculation (incident TE wave from the top, incident TM wave from the top, incident TE wave from the bottom and incident TM wave from the bottom). Of course, the two incident plane waves must have identical parallel wave vector in the transverse plane [ $k_x^{inc}, k_y^{inc}$]. This possibility which is not mentioned in the literature to our knowledge is important in practice since the user may get, for the same computational loads, the grating diffraction efficiencies for an illumination from the substrate or from the cover.

RETICOLO-1D calculates the electric and magnetic fields diffracted by the grating for the following incident plane wave :

$$\mathbf{E}_{top}^{inc} \exp\left[i\left(k_x^{inc}x + k_y^{inc}y + k_{z\,top}^{inc}(z-h)\right)\right]$$
$$\mathbf{H}_{top}^{inc} \exp\left[i\left(k_x^{inc}x + k_y^{inc}y + k_{z\,top}^{inc}(z-h)\right)\right]$$, if incident from the top layer,

where $k_{z\,top}^{inc} = -\sqrt{\left(2\pi n_{top}/\lambda\right)^2 - \left(k_x^{inc}\right)^2 - \left(k_y^{inc}\right)^2}$ ,

$$\mathbf{E}_{bottom}^{inc} \exp\left[i\left(k_x^{inc}x + k_y^{inc}y + k_{z\,bottom}^{inc}z\right)\right]$$
$$\mathbf{H}_{bottom}^{inc} \exp\left[i\left(k_x^{inc}x + k_y^{inc}y + k_{z\,bottom}^{inc}z\right)\right]$$, if incident from the bottom layer,

where $k_{z\,bottom}^{inc} = \sqrt{\left(2\pi n_{bottom}/\lambda\right)^2 - \left(k_x^{inc}\right)^2 - \left(k_y^{inc}\right)^2}$ ,

The z componant of the Poynting vector of the incident field is ±0.5
The Rayleigh-expansion of the diffracted electric fields are shown in the following figure.

$$\mathbf{E}_{top}^{dif} = \sum_m \mathbf{E}_{top}^m \exp\left[i((k_x^{inc} + mK_x)x + k_y^{inc}y + k_{z\,top}^m(z-h)\right]$$

$$\mathbf{H}_{top}^{dif} = \sum_m \mathbf{H}_{top}^m \exp\left[i((k_x^{inc} + mK_x)x + k_y^{inc}y + k_{z\,top}^m(z-h)\right]$$

$$\text{where: } k_{z\,top}^m = \sqrt{\left(2\pi n_{top}/\lambda\right)^2 - \left(k_x^{inc} + mK_x\right)^2 - k_y^{inc\,2}}$$

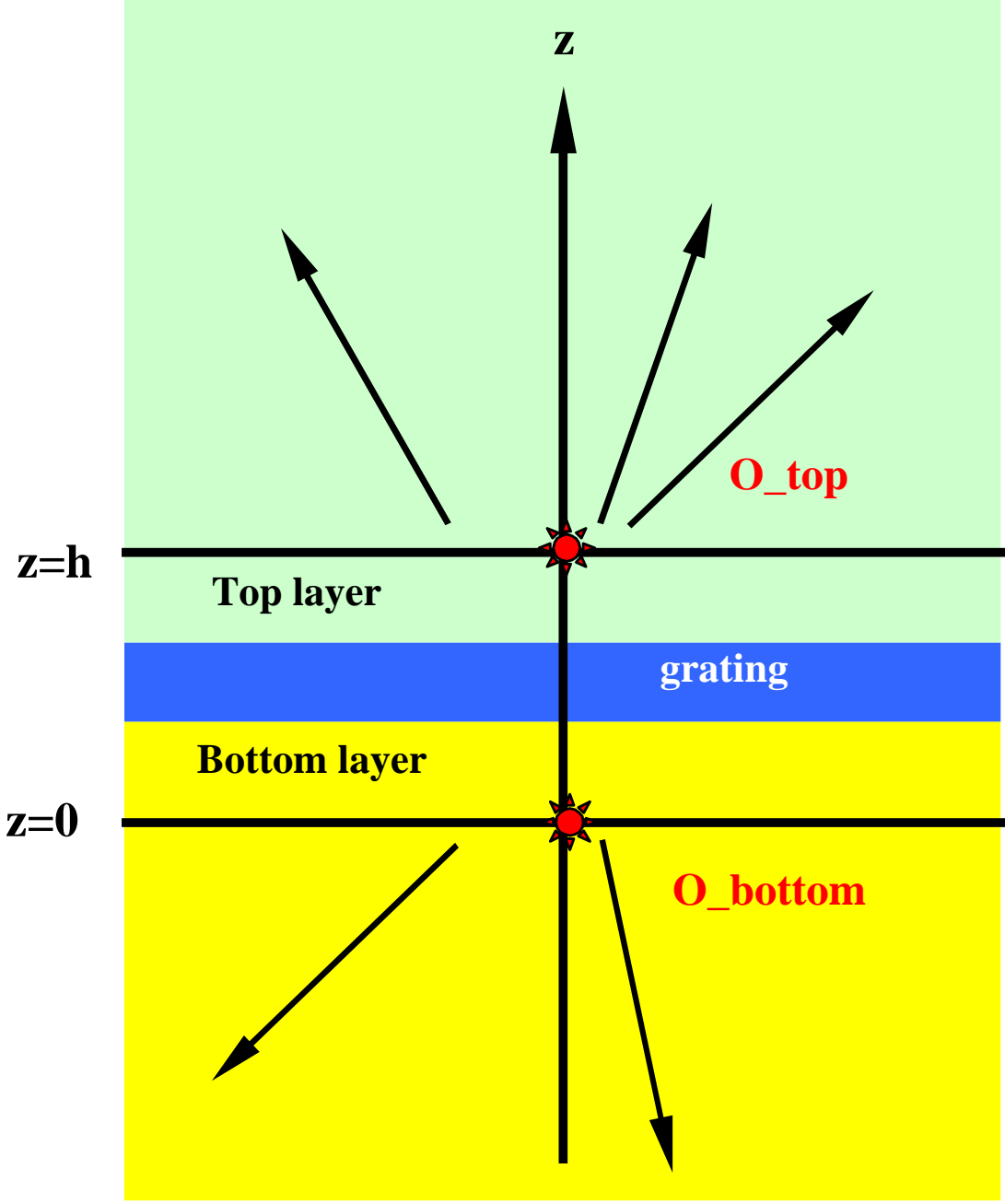

$$\mathbf{E}^{dif}_{bottom}=\sum_m \mathbf{E}^m_{bottom}\exp\left[i((k_x^{inc}+mK_x)x+k_y^{inc}y+k^m_{z\,bottom}\,z\right]$$

$$\mathbf{H}^{dif}_{bottom}=\sum_m \mathbf{H}^m_{bottom}\exp\left[i((k_x^{inc}+mK_x)x+k_y^{inc}y+k^m_{z\,bottom}z\right]$$

$$\text{where}: k^m_{z\,bottom}=-\sqrt{\left(2\pi n_{bottom}/\lambda\right)^2-\left(k_x^{inc}+mK_x\right)^2-k_y^{inc\,2}}$$

Fig. 1. Rayleigh expansion for the diffracted fields. $K_x=(2\pi)/\text{period}$. The $m^{th}$ order has a parallel momentum equal to $\left(k_x^{inc}+mK_x\right)\vec{x}+k_y^{inc}\vec{y}$. We define two points $O_{top}$= (0,0,h) at the top of the grating, and $O_{bottom}$= (0,0,0) at the bottom of the grating.

The following is organized so that one can straightforwardly write a code using the software.

# 3. Preliminary input parameters

The name of the following parameters are given as examples. The user may define his own parameter vocabulary.

**wavelength = 3**; Wavelength (λ) in a vacuum. It might be 3 nm or 3 µm. You do not need to specify the unit but all other dimensions are of course in the same unit as the wavelength.

**period** % in the x-direction.

**nn = 20**; This defines the set of Fourier harmonics retained for the computation. More specifically, 2*nn+1 represent the number of Fourier harmonics retained from –nn to nn. This is a very important parameter ; for large nn values, a high accuracy for the calculated data is achieved, but the computational time and memory is also large.

**angle_delta = 30**; In degrees, see the following figure for a definition of "angle_delta" for the incident plane wave. This angle is varying between 0° and 360°. This angle has to be defined in the incident medium.

**k_parallel = n_incident_medium*sin(angle_theta*pi/180)**;
The parameters "**angle_delta**" and "**angle_theta**" which are used to specify **the plane of incidence and the angle of incidence** are denoted by δ and $\theta_{inc}$ in Fig. 2. The angle δ defines the plane of incidence. This plane allows to define the polarization of the incident plane wave : if the electric field of the incident plane wave is perpendicular to this plane, the incident wave is TE polarized, and if it is parallel to this plane, the incident wave is TM polarized. The incident wave vector is :

$\mathbf{k}_{inc}=(2\pi/\lambda)\ n_{inc}\ \mathbf{K}_{inc}$

with $\mathbf{K}_{inc}$ = [sin(θ)cos(δ), sin(θ)sin(δ), -cos(θ) ].
$n_{inc}$ is the refractive index of the top (or bottom) layer. One expects that it is a positive real number and that the texture (see Section 4.1) associated to the top (or the bottom) layer has a background with a uniform refractive index "$n_{inc}$".
(Note that the "k_parallel" variable is defined **without** the factor 2π/λ.)

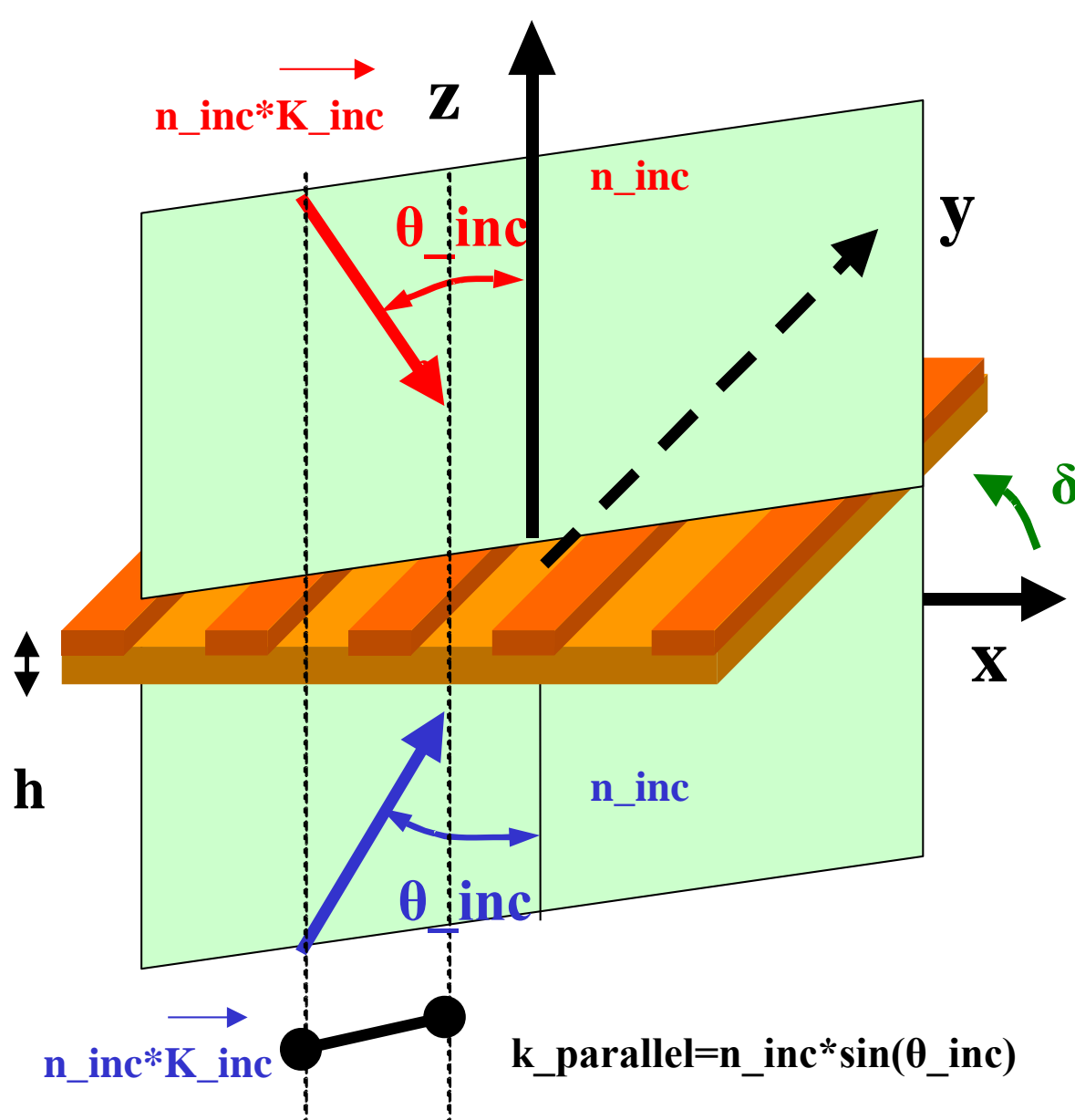

Fig. 2. Definition of θ_inc, δ, k_parallel.

In general, the user has in mind to illuminate the grating from the substrate or from the upperstrate (air in general). "n_incident_medium" (denoted also $n_{inc}$) is the refractive index of the incident medium. One expects that it is a positive real number and that the texture (see Section 4.1) associated to the top or the bottom layer has a background with a refractive index "n_incident_medium".

It is very important to keep in mind that wether the user defines the incident plane wave in the top layer or in the bottom layer, the calculation will be done for both an incident wave from the top and an incident wave from the bottom, with an identical parallel wave vector, i.e. for a specified $\left[k_x^{inc}, k_y^{inc}\right]$ which is the same in the bottom and top layers.

# 4. Structure definition (grating parameters)

The grating encompasses a uniform upperstrate, called the top in the following, a uniform substrate, called the bottom in the following, and many layers which define the grating, which is defined by a stack of layers. Every layer is defined by a "texture" and by its thickness. Two different layers may be identical (identical texture and thickness), may have different thicknesses with identical texture, may have different thicknesses and textures. To define the diffraction geometry, one needs to define the different textures and then the different layers.

## 4.1. How to define a texture?

Every texture is defined by a cell-array composed of two line-vectors of identical length. The first vector, let us say [$x_1$ $x_2$ ... $x_p$ ...$x_N$], contains all the x-values of the discontinuities. One *must* have :

N>1,

$x_p < x_{p+1}$ for any p,

and $x_N - x_1 < \text{period}$.

The second line-vector $[n_1\ n_2\ \ldots\ n_p\ \ldots\ n_N]$ contains the refractive indices of the material between the discontinuities. More explicitly, we have a refractive index $n_p$ for $x_{p-1} < x < x_p$. Because of periodicity, note that the refractive index for $x_N < x < x_1+\text{period}$ is equal to $n_1$.
The specific case of a uniform texture with a refractive index n is easily defined by texture{1}={n}. In that specific case, no need of a second vector since there is no discontinuity.

The textures have all to be to be packed together in a cell array textures={textures{1}, textures{2}, textures{3}} prior calling subroutine **res1.m.**

<u>Example</u>
period=17;
textures =cell(1,2);
textures{1}={ 1.5}; %uniform texture
textures{2}={[-5,-3,1,6],[2,1.3,1.5,3]}; %texture composed of 4 different refractive indices

The following figure shows the refractive indices of the two textures.

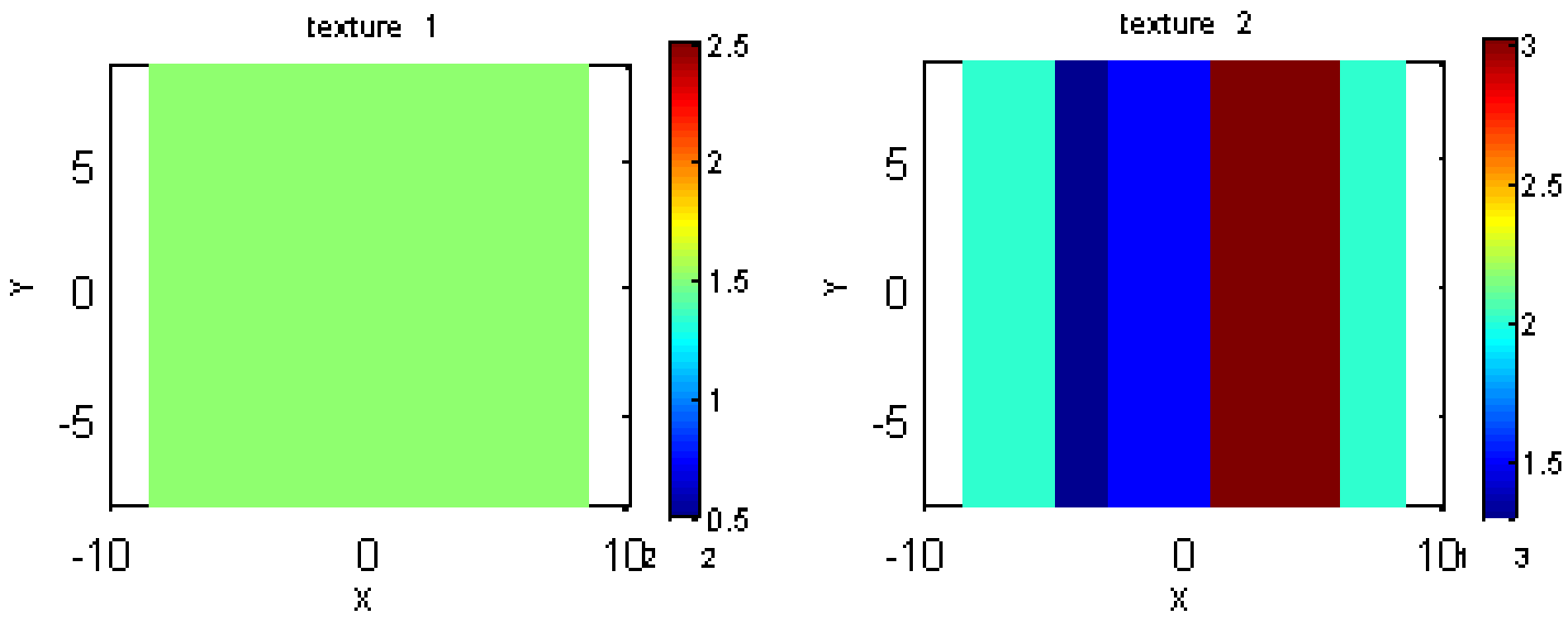

Fig. 3. Textures{1} and {2}.

<u>Slits in perfectly-conducting metallic textures :</u>
We have first to define a background by its refractive index “inf”. In this uniform background, we can incorporate strip inclusions with a complex or real refractive index “ninclusion” defined by the position c of its center and its x-width L. The inclusions cannot overlap.
For example:
textures {3}= {inf, [c1,L1,ninclusion1],[c2,L2, ninclusion2]}

In order to check if the set of textures is correctly set up, the user can set the variable parm.res1.trace equal to 1 : “parm.res1.trace = 1;”. Then a Matlab figure will show up the refractive-index distribution of all textures. Each texture is represented with the coordinate x varying from –period/2 to period/2.

## 4.2. How to define the layers?

This is performed by defining the “profile” variable which contains, starting from the top layer and finishing by the bottom layer, the successive information (thickness and texture-label) relative to every layer. Here is an example that illustrates how to set up the “profile” variable:

**profile** = {[0,1,0.5,0.5,1,0.5,0.5,2,0],[1,3,2,4,3,2,4,6,2]}; (1)

It means that from the top to the bottom we have: the top layer is formed by a thickness 0 of texture 1, then we have twice textures 3, 2 and 4 with depth 1, 0.5 and 0.5 respectively, texture 6 with depth 2, and finally the bottom layer (formed by texture 2) with null thickness. Since textures 1 and 2 correspond to the top and bottom layers, they must be uniform. In this example, the top and bottom layers have a null thickness. However, one may set an arbitrary thickness. Especially, if one needs to plot the electromagnetic fields in the bottom and top layers, the thicknesses $h_b$ and $h_h$ (see Fig. 4) over which the fields have to be visualized has to be specified. For $h_b$=$h_h$=0, the Rayleigh expansions of the fields in the top and bottom layers are not plotted.

In this particular profile, the structure formed by texture 3 with thickness 1, texture 2 with thickness 0.5 and texture 4 with thickness 0.5 is repeated twice. It is possible to simplify the instruction defining the “profile” variable in order to take into account the repetitions:

**profile** = {{0,1},{[1,0.5,0.5], [3,2,4], 2},{[2,0],[6,2]}}; (2)

If a structure is repeated many times, the above “factorized” instruction of Eq. 2 is better than the “expanded” one of Eq. 1, in terms of computational speed, because the calculation will take into account the repetitions.

The profile is shown below.

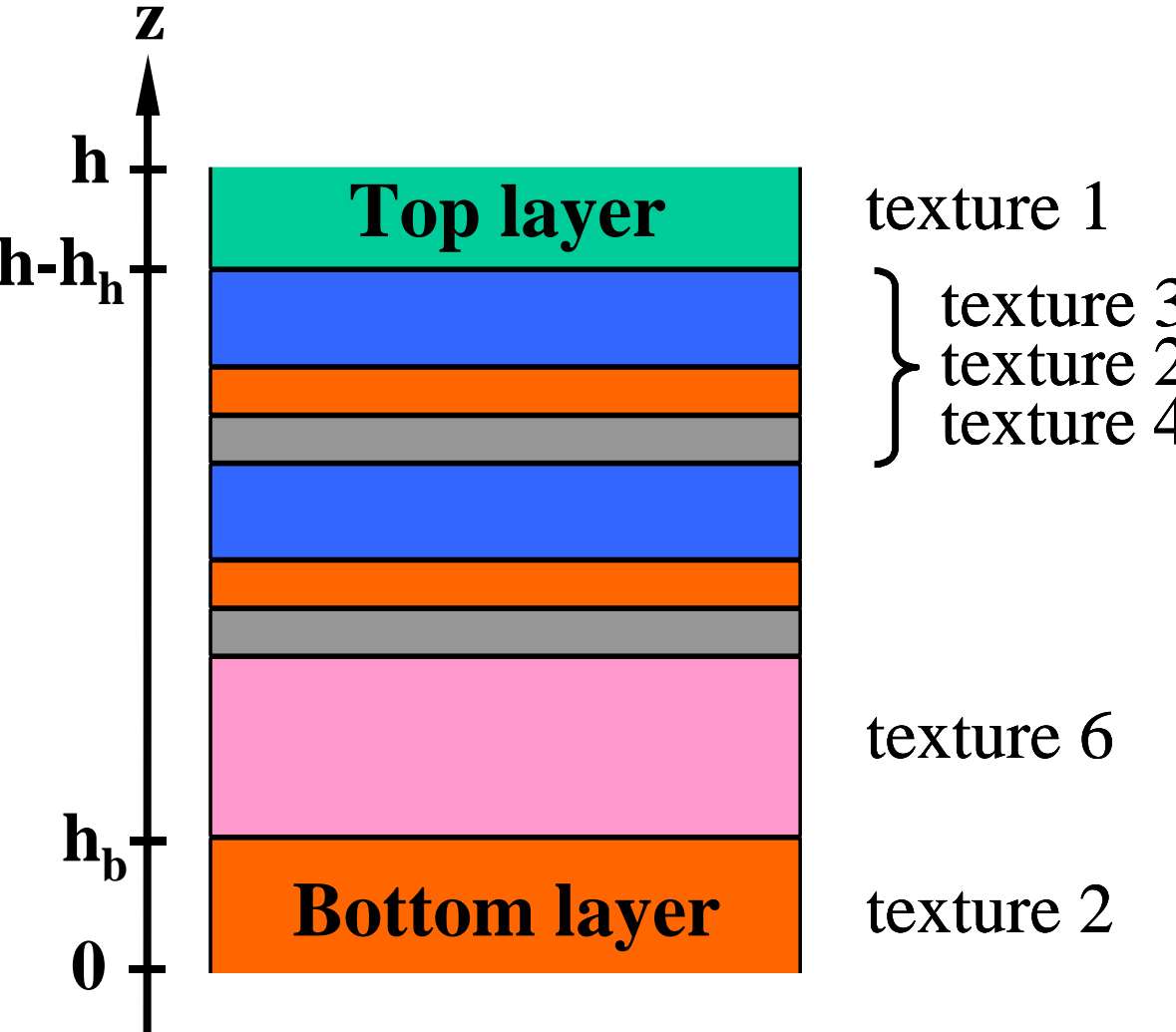

Fig. 4. Texture stacks. The example illustrates a profile defined by **profile** = {[$h_h$,1,0.5,0.5,1,0.5,0.5,2, $h_b$],[1,3,2,4,3,2,4,6,2]}. The top and bottom layers have uniform textures.

# 5. Solving the eigenmode problem for every texture

The first computation with the RCWA consists in calculating the eigenmodes associated to all textures. This is done by the subroutine “res1.m”, following the instruction:

**aa = res1(wavelength,period,textures,nn,k_parallel,angle_delta,parm)**;

The first-six input parameters are absolutely required by the code : the wavelength “**wavelength**”, the period of the grating “**period**”, the “**textures**” variable, the number of Fourier harmonics “**nn**”, the norm of the parallel incident wave vector “**k_parallel**”, the angle that defines the plane of incidence “**angle_delta**.
Some other additional parameters can be defined. For example, the default parameters do not take the symmetry of the problem into account. So if the user wants to use symmetries, new parameters have to be defined : “**parm.sym.x**”, “**parm.sym.y**”, and “**parm.sym.pol**”. These parameters are defined in Section 7.

**parm = res0**;
res0.m is a function that changes the default values. This instruction has to be executed before res1.m, if one wants to modify the default values (for instance to use symmetry).

It is very important to note that if one has to study the diffraction by many different gratings composed of the same textures, one needs to compute only once the eigenmodes. It is possible to save the “aa” variable in a “.mat” file and to reload it for the computation of the diffracted waves, see an example in Annex 9.3.

# 6. Computing the diffracted waves

This is the second step of the computation. This is done by the subroutine “res2.m”, following the instruction:

**result = res2(aa,profile)**;

This subroutine has 2 input arguments: the output “**aa**” of the subroutine “res1.m” and the “**profile**” variable. The output argument “**result**” contains all the information on the diffracted fields. “**result**” is an object of class ‘reticolo’ that can be indexed as an usual structure with parentheses, or with the labels of the considered orders between curly braces. Examples will be given in the following.
This information is divided into the following sub-structures fields :

- “**result.TEinc_top**”
- “**result.TEinc_top_reflected**”
- “**result.TEinc_top_transmitted**”

- “**result.TEinc_bottom**”
- “**result.TEinc_bottom_reflected**”
- “**result.TEinc_bottom_transmitted**”

- “**result.TMinc_top**”
- “**result.TMinc_top_reflected**”
- “**result.TMinc_top_transmitted**”

- “**result.TMinc_bottom**”
- “**result.TMinc_bottom_reflected**”
- “**result.TMinc_bottom_transmitted**”

The sub-structure “**result.TEinc_top_reflected**” contains all the information concerning the propagative *reflected* waves for *the incident wave from the top* of the grating in *TE polarization* which is described in the sub-structure “**result.TEinc_top”**
The sub-structure “**result.TMinc_bottom_transmitted**” contains all the information concerning the propagative *transmitted* waves for *the incident wave from the bottom* of the grating *in TM polarization* which is described in the sub-structure “**result.TMinc_ bottom”**. And so on.

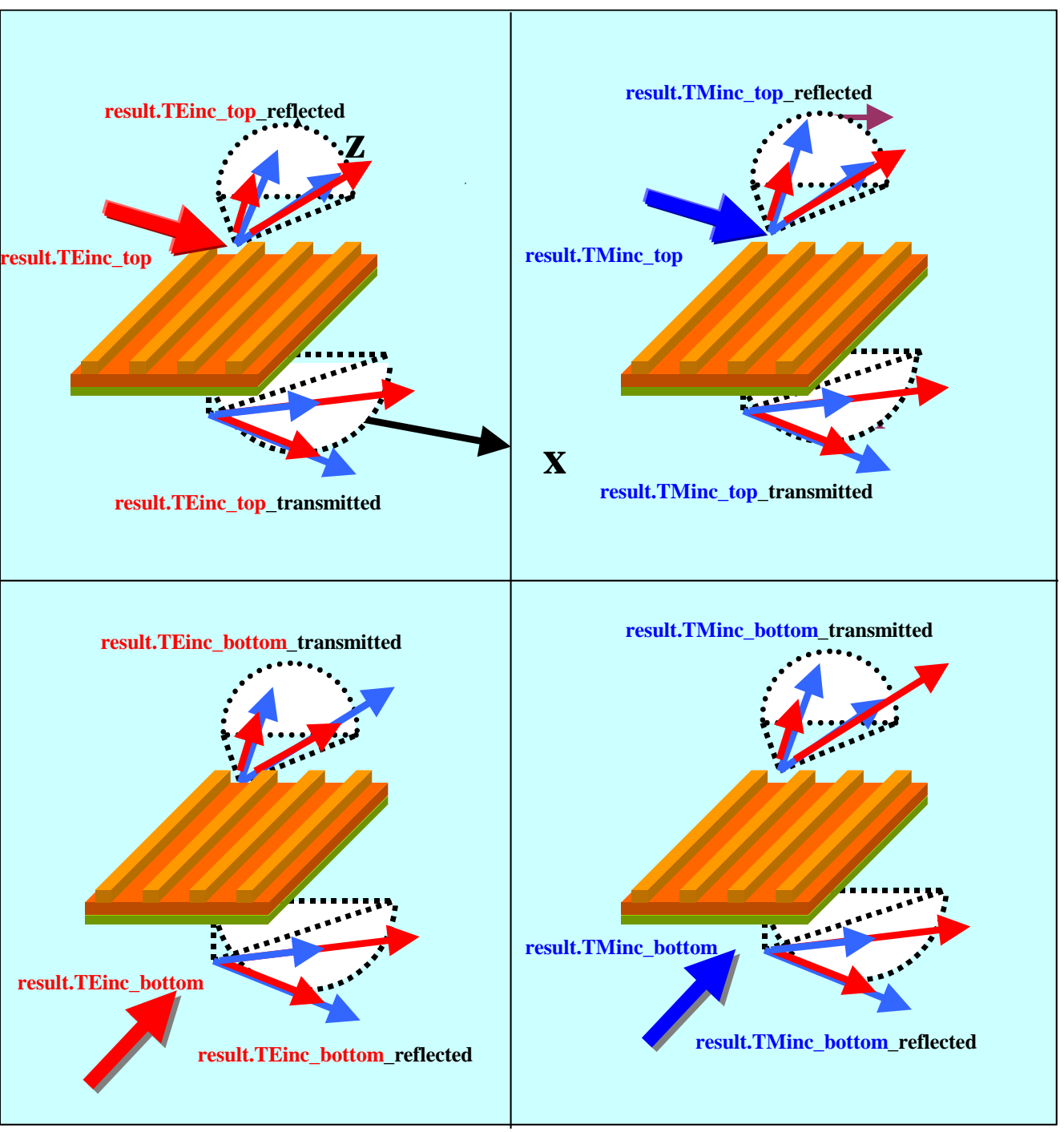

Fig. 5. The 4 solutions obtained.

Each sub-structure of **result** is composed of the following fields. Each field is a Matlab column vector.or matrix having the same number N of lines. N is the number of propagative orders considered and can be 0.

| **Field name** | **signification** | size |
|---|---|---|
| order | orders of the diffracted propagative plane waves | N, 1 |
| theta | angle $\theta_m$ of every diffracted order | N, 1 |
| delta | angle $\delta_m$ of every diffracted order | N, 1 |
| **K** | normalized wave vector | N, 3 |
| efficiency | efficiency in each order | N, 1 |
| efficiency_TE | efficiency in TE polarization in every order | N, 1 |
| efficiency_TM | efficiency in TM polarization in every order | N, 1 |
| amplitude_TE | complexe amplitude in TE polarization in every order | N, 1 |
| amplitude_TM | complexe amplitude in TM polarization in every order | N, 1 |
| **E** | electric field ($E_x$,$E_y$,$E_z$) of the diffracted orders at O_top or O_bottom when the amplitude of the incident plane wave is one. | N, 3 |
| **H** | magnetic field ($H_x$,$H_y$,$H_z$) of the diffracted orders at O_top or O_bottom when the amplitude of the incident plane wave is one. | N, 3 |
| **PlaneWave_TE_E** | **E**-vector components of the TE-polarized $\overrightarrow{PW}$ 's (in the Oxyz basis) | N, 3 |
| **PlaneWave_TE_H** | **H**-vector components of the TE-polarized $\overrightarrow{PW}$ 's (in the Oxyz basis) | N, 3 |
| **PlaneWave_TE_Eu** | **E**-vector components of the TE-polarized $\overrightarrow{PW}$ 's (in the $\mathbf{u}_{TM}$ $\mathbf{u}_{TE}$ basis) | N, 2 |
| **PlaneWave_TE_Hu** | **H**-vector components of the TE-polarized $\overrightarrow{PW}$ 's (in the $\mathbf{u}_{TM}$ $\mathbf{u}_{TE}$ basis) | N, 2 |
| **PlaneWave_TM_E** | **E**-vector components of the TM-polarized $\overrightarrow{PW}$ 's (in the Oxyz basis) | N, 3 |
| **PlaneWave_TM_H** | **H**-vector components of the TM-polarized $\overrightarrow{PW}$ 's (in the Oxyz basis) | N, 3 |
| **PlaneWave_TM_Eu** | **E**-vector components of the TM-polarized $\overrightarrow{PW}$ 's (in the $\mathbf{u}_{TM}$ $\mathbf{u}_{TE}$ basis) | N, 2 |
| **PlaneWave_TM_Hu** | **H**-vector components of the TM-polarized $\overrightarrow{PW}$ 's (in the $\mathbf{u}_{TM}$ $\mathbf{u}_{TE}$ basis) | N, 2 |

## 6.1. Efficiency

For a given diffraction order n, the diffraction efficiency is defined as the ratio between the flux of the diffracted Poynting vector and the flux of the incident Poynting vector (flux through a period of the grating).

The efficiencies of all propagative reflected and transmitted plane waves (for a TE-polarized plane wave incident from the top of the grating) are given by the two vectors "**result.TEinc_top_reflected.efficiency**" and "**result.TEinc_top_transmitted.efficiency**". If all refractive indices are real, the sum of all elements of these two vectors is equal to one because of the energy conservation. The label "m" of the corresponding orders are found in "**result.TEinc_top_reflected.order**" (see below for a description of the other fields of this sub-structure).

<u>Some examples</u>

1) The TE-efficiency of the reflected order -2 ( $k_x^{inc} - 2K_x$ ) of the grating illuminated from the top by a TM-polarized plane wave is equal to **result.TMinc_top_reflected.efficency_TE{-2}**. If this order is evanescent the efficiency is equal to zero.
The total efficiency (TE+TM) in this order is **result.TMinc_top_reflected.efficency{-2}**.

It is important to have in mind the difference between :
**result.TMinc_top_reflected.efficiency{-2}** : efficiency of order 2
**result.TMinc_top_reflected.efficiency(-2)** : gives an error !
**result.TMinc_top_reflected.efficiency{2}** : efficiency of order 2
**result.TMinc_top_reflected.efficiency(2)** : efficiency in order **result. inc_top_reflected.order(2)**;

2) The orders of the transmitted waves for an incident wave from the top of the grating in TE polarization are given by the vector "**result.TEinc_top_transmitted.order**".

3) The efficiencies of all propagative reflected waves for an incident wave from the bottom in TM polarization are given by the vector "**result.TMinc_bottom_reflected.efficiency**".

## 6.2. Rayleigh expansion for propagatives modes

The coefficients of the Rayleigh expansion of Fig. 1 can be obtained from the structure **result**. For instance, when the grating is illuminated from the bottom with a TE polarised mode, we have :

$\mathbf{E}^{m}_{bottom}$=result.TEinc_bottom_reflected.E{m} (3 components in Oxyz)

$\mathbf{H}^{m}_{bottom}$=result.TEinc_bottom_reflected.H{m} (3 components in Oxyz)

$\mathbf{E}^{m}_{top}$ =result.TEinc_bottom_transmitted.E{m} (3 components in Oxyz)

$\mathbf{H}^{m}_{top}$=result.TEinc_bottom_ transmitted.H{m} (3 components in Oxyz)

and the incident plane wave defined in page 4 is given by :

$\mathbf{E}^{inc}_{bottom}$=result.TEinc_bottom.E (3 components in Oxyz)

$\mathbf{H}^{inc}_{bottom}$=result.TEinc_bottom.H (3 components in Oxyz).

## 6.3. Diffracted amplitudes of propagative waves

### 6.3.1 $U_{TE}$, $U_{TM}$, θ, δ and K

Figure 6 defines the geometry of the diffracted order m, for a diffracted wave in the top layer and for a diffracted wave in the bottom layer. The wave vector $\vec{\mathbf{k}}_m$ =(2π/λ) $n_{top}$ $\mathbf{K_m}$ (or (2π/λ)$n_{bottom}\mathbf{K_m}$) of the mth diffracted order is defined by the two angles $\theta_m$ and $\delta_m$. As for the incident wave, the angle $\delta_m$ defines the plane of diffraction. The angle $\theta_m$ is varying between 0° and 90°, and the angle $\delta_m$ is varying between 0° and 360°. The relations linking the Cartesian components of the unitary vector $\mathbf{K_m}$ and the angles $\theta_m$ and $\delta_m$ are the same as the relations defined previously for the incident plane wave (Section 3) :

$\mathbf{K}_m = [\sin(\theta_m)\cos(\delta_m), \sin(\theta_m)\sin(\delta_m), -\cos(\theta_m)]$

The unitary vector $\vec{\mathbf{u}}_{TE}$ is perpendicular to the plane of diffraction and is oriented so that ($\mathbf{K}_m$, $\vec{\mathbf{u}}_{TE}$ , **z**) is direct. The unitary vector $\vec{\mathbf{u}}_{TM}$ is defined by the relation $\vec{\mathbf{u}}_{TM} = \vec{\mathbf{u}}_{TE} \wedge \mathbf{K}_m$. So the base $\vec{\mathbf{u}}_{TM}$ , $\vec{\mathbf{u}}_{TE}$ , $\mathbf{K}_m$ is direct. If the diffracted electric field is parallel to $\vec{\mathbf{u}}_{TE}$ , then the order m is TE polarized, and if the diffracted electric field is parallel to $\vec{\mathbf{u}}_{TM}$ , then the order m is TM polarized. In general, the diffracted electric field of the order m has a non-zero component along both directions.

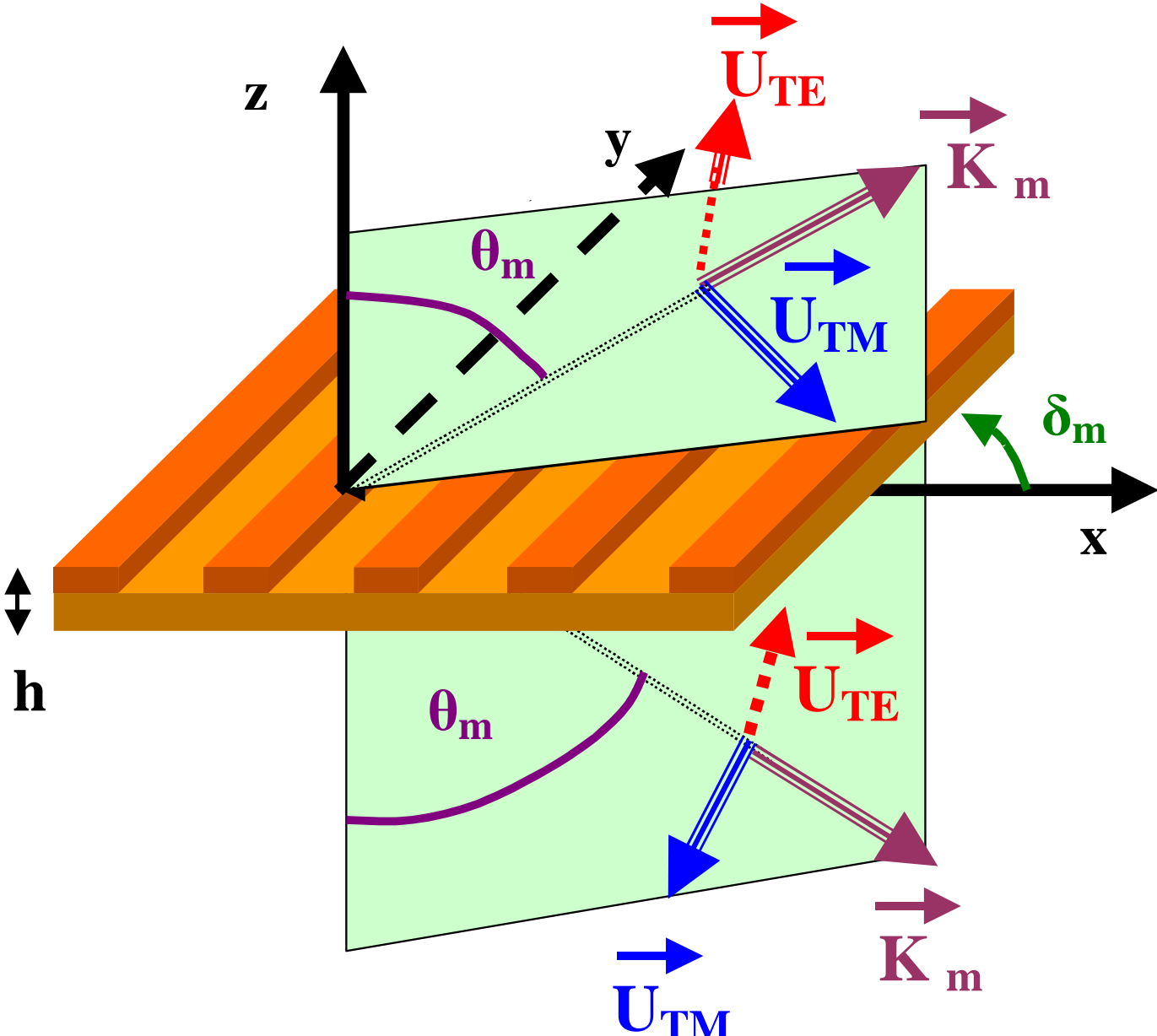

Fig. 6. Definition of the $\theta_m$ and $\delta_m$ for a specific diffracted order m.

### 6.3.2 $O_{top}$ and $O_{bottom}$ points

$O_{top}$ and $O_{top}$ are 2 important points (see Fig. 1). In the Cartesian coordinates system Oxyz , they are defined by : $O_{top}$=(0,0,h) at the top of the grating, and $O_{bottom}$=(0,0,0) at the bottom of the grating.

In addition, let us consider an arbitrary point M=(x,y,z) in the 3D space in Oxyz. Associated to this point, we define the two vectors :

$\mathbf{r}_{top=}\ \overrightarrow{O_{top}M}$, and

$\mathbf{r}_{bottom=}\ \overrightarrow{O_{bottom}M}$ .

### 6.3.3 Jones' matrix

Let us assume that the grating is illuminated from the top layer and let us consider a diffracted order m in the bottom layer. Any other diffraction situation is straightforwardly deduced.

α and β being two given complex numbers, the incident electromagnetic field (6 components of **E** and **H** in every points of the 3D space) can be written :

$$\mathbf{W}^{inc}=\alpha\ \overrightarrow{PW_{TE}}+\beta\ \overrightarrow{PW_{TM}}\ ,$$

where $\overrightarrow{PW_{TE}}$ is a TE-polarized plane wave defined in every point by $\overrightarrow{PW_{TE}}=\mathbf{A}_{TE}\ \exp\left(i\mathbf{k}_{top}^{inc}\ r_{top}\right)$, and $\overrightarrow{PW_{TM}}$ a TM polarized plane wave defined in the same way by $\overrightarrow{PW_{TM}}=\mathbf{A}_{TM}\exp\left(i\mathbf{k}_{top}^{inc}r_{top}\right)$, $\mathbf{A}_{TE}$ and $\mathbf{A}_{TM}$ being the electromagnetic fields (6 components) of the plane wave at M=$O_{top}$. $\mathbf{k}_{top}^{inc}$ is the incident wave vector. $\mathbf{A}_{TE}$ and $\mathbf{A}_{TM}$ and $\mathbf{K}=\mathbf{k}_{top}^{inc}/\left|\mathbf{k}_{top}^{inc}\right|$ are given by the structure "**result**" as will be defined later.

Similarly, the diffracted electromagnetic field in the mth bottom order can be written :

$$\mathbf{W}_m^{dif}=\gamma\ \overrightarrow{PW_{TE}^m}+\mu\overrightarrow{PW_{TM}^m}\ ,$$

where γ and μ are complex numbers, $\overrightarrow{PW_{TE}^m}$ is a TE-polarized plane wave defined in every point by $\overrightarrow{PW_{TE}^m}=\mathbf{A}_{TE}^m\ \exp\left(i\mathbf{k}_{bottom}^m\ \mathbf{r}_{bottom}\right)$, and $\overrightarrow{PW_{TM}^m}$ a TM-polarized plane wave defined in the same way by $\overrightarrow{PW_{TM}^{m}} = \mathbf{A}_{TM}^{m} \exp\left(i\mathbf{k}_{bottom}^{m}\,\mathbf{r}_{bottom}\right)$. $\mathbf{A}_{TE}^{m}$ and $\mathbf{A}_{TM}^{m}$ are the electromagnetic fields (6 components) of the plane wave at M=O$_{bottom}$, and $\mathbf{k}_{bottom}^{m}$ is the wave vector of the m transmitted order. $\mathbf{A}_{TE}^{m}$ and $\mathbf{A}_{TM}^{m}$ and $\mathbf{K}^{m} = \mathbf{k}_{bottom}^{m} / \left|\mathbf{k}_{bottom}^{m}\right|$ are given by the structure "**result**" as will be defined later.

We define the (4x4) Jones' matrix **J**, associated to the order m by :

$$\begin{pmatrix} \gamma \\ \mu \end{pmatrix} = \begin{pmatrix} J_{EE} & J_{ME} \\ J_{EM} & J_{MM} \end{pmatrix} \begin{pmatrix} \alpha \\ \beta \end{pmatrix}.$$

$J_{EE}$, $J_{EM}$,$J_{ME}$, $J_{MM}$ and **J** are all given by the structure "**result**".

The $\mathbf{A}_{TE}$ $\mathbf{A}_{TE}^{m}$ $\mathbf{A}_{TM}$ $\mathbf{A}_{TM}^{m}$ vectors are normalized so that the $|J_{EE}|^2$, $|J_{EM}|^2$,$|J_{ME}|^2$ and $|J_{MM}|^2$ represent diffraction efficiencies. For instance, $|J_{ME}|^2$ =**result.TMinc_top_transmitted.efficency_TE{m}**.

We now define all these data from the "result" structure :
**K** = result.TEinc_top.K or K=result.TMinc_top.K.
$\mathbf{K}^{m}$ = result.TEinc_top_transmitted.K{m} = result.TMinc_top_transmitted.K{m}. Note that if some symmetries are used for the calculation, "result.TEinc_top_transmitted.K{m}" or "result.TMinc_top_transmitted.K{m}" can be an empty vector.

The $\mathbf{A}_{TE}^{m}$ 's coefficients can be obtained either in the Cartesian coordinate system or in the ($\vec{\mathbf{u}}_{TM}$ ,$\vec{\mathbf{u}}_{TE}$ ) basis.
In the Cartesian coordinate system Oxyz :

$$\mathbf{A}_{TE} = \begin{pmatrix} \text{result.TEinc_top.PlaneWave_TE_E} \\ \text{result.TEinc_top.PlaneWave_TE_H} \end{pmatrix}$$

$$\mathbf{A}_{TM} = \begin{pmatrix} \text{result.TMinc_top.PlaneWave_TM_E} \\ \text{result.TMinc_top.PlaneWave_TM_H} \end{pmatrix}$$

$$\mathbf{A}_{TE}^{m} = \begin{pmatrix} \text{result.TEinc_top_transmitted.PlaneWave_TE_E}\{m\} \\ \text{result.TEinc_top_transmitted.PlaneWave_TE_H}\{m\} \end{pmatrix}$$

$$= \begin{pmatrix} \text{result.TMinc_top_transmitted.PlaneWave_TE_E}\{m\} \\ \text{result.TMinc_top_transmitted.PlaneWave_TE_H}\{m\} \end{pmatrix}.$$ (same remark as for $\mathbf{K}^{m}$)

$$\mathbf{A}_{TM}^{m} = \begin{pmatrix} \text{result.TEinc_top_transmitted.PlaneWave_TE_E}\{m\} \\ \text{result.TEinc_top_transmitted.PlaneWave_TE_H}\{m\} \end{pmatrix}$$

$$= \begin{pmatrix} \text{result.TMinc_top_transmitted.PlaneWave_TE_E}\{m\} \\ \text{result.TMinc_top_transmitted.PlaneWave_TE_H}\{m\} \end{pmatrix}.$$ (same remark as for $\mathbf{K}^{m}$)

In the ($\vec{\mathbf{u}}_{TM}$ ,$\vec{\mathbf{u}}_{TE}$ ) basis (with only 2 components for each fields E and H) :

$$\mathbf{A}_{TE} = \begin{pmatrix} \text{result.TEinc_top.PlaneWave_TE_Eu} \\ \text{result.TEinc_top.PlaneWave_TE_Hu} \end{pmatrix}$$

$$\mathbf{A}_{TM} = \begin{pmatrix} \text{result.TMinc_top.PlaneWave_TM_Eu} \\ \text{result.TMinc_top.PlaneWave_TM_Hu} \end{pmatrix}$$

$$\mathbf{A}_{TE}^{m} = \begin{pmatrix} \text{result.TEinc_top_transmitted.PlaneWave_TE_Eu}\{m\} \\ \text{result.TEinc_top_transmitted.PlaneWave_TE_Hu}\{m\} \end{pmatrix}$$

$$= \begin{pmatrix} \text{result.TMinc_top_transmitted.PlaneWave_TE_Eu}\{m\} \\ \text{result.TMinc_top_transmitted.PlaneWave_TE_Hu}\{m\} \end{pmatrix}.$$ (same remark as for $\mathbf{K}^{m}$)

$$\mathbf{A}_{TM}^{m} = \begin{pmatrix} \text{result.TEinc_top_transmitted.PlaneWave_TE_Eu}\{m\} \\ \text{result.TEinc_top_transmitted.PlaneWave_TE_Hu}\{m\} \end{pmatrix}$$

$$= \begin{pmatrix} \text{result.TMinc_top_transmitted.PlaneWave_TE_Eu}\{m\} \\ \text{result.TMinc_top_transmitted.PlaneWave_TE_Hu}\{m\} \end{pmatrix}$$ . (same remark as for $\mathbf{K}^m$)

The Jones' coefficients are :
$J_{EE}$ =result.TEinc_top_transmitted.amplitude_TE{m}
$J_{EM}$ =result.TEinc_top_transmitted.amplitude_TM{m}
$J_{ME}$ =result.TMinc_top_transmitted.amplitude_TE{m}
$J_{MM}$ =result.TMinc_top_transmitted.amplitude_TM{m}

And the Jones' matrix is :

$$\mathbf{J} = \begin{pmatrix} J_{EE} & J_{ME} \\ J_{EM} & J_{MM} \end{pmatrix} = \text{result.Jones.inc_top_transmitted}\{m\}.$$

# 7. Using symmetries to accelerate the computational speed

When the grating possesses some mirror symmetry for the plane $x=x_0$, the user may define "**parm.sym.x=** $x_0$**"** For delta=90 or 270, the x-symmetry will be used.

When angle_delta and parameter k_parallel are compatible with the symmetry, the structure "result" contains only information upon the polarisation selected by parameter **parm.sym.pol**.

Note that the code does not check if the grating symmetry defined by the user is in agreement with the "textures". It is up to the user to define carefully the parameters parm.sym.x.

# 8. Plotting the electromagnetic field and calculating the absorption loss

## 8.1. Computation of the electromagnetic fields

Once the eigenmodes associated to all textures are known, the calculation of the electromagnetic fields everywhere in the grating can be performed. This calculation is done by the subroutine "**res3.m**", following the instruction

**[e,z,index] = res3(x,aa,profile,einc,parm);**

The function"res3.m" can be called without calling "res2.m". This subroutine has 5 input arguments:
-the "**x**" variable is a vector containing the locations where the fields will be calculated in the x-direction. For instance, we may set **x = linspace(-period/2, period/2, 51);** for allocating 51 sampling points in the x-direction,
-the "**aa**" variable contains all the information on the eigenmodes of all textures and is computed by the subroutine res1.m,
-the variable "**profile**" is defined in Section 4.2. Note that it can be redefined,
-the variable "**einc**" defines the complex amplitude of the *incident electric* field at O_top or O_bottom in the basis {$\mathbf{u}_{TM}$, $\mathbf{u}_{TE}$}. For instance, setting einc=[1,0] means that one is looking for TM polarization, and setting einc=[1,1]/sqrt(2) means that one is looking for a 45° polarization.

If one wants to illumine the grating exactly by the TE-polarized incident $\overrightarrow{PW_{TE}}$ defined above, one should set: **einc**= **result.TEinc_top PlaneWave_TE_Eu.**

If symmetry arguments have been used previously, note that the calculation with res1.m is provided only for some specific polarization; it would be a nonsense to specify another polarization for the field plots (in this case the corresponding component of einc is taken as 0).
-the "**parm**" variable, already mentioned is discussed in the following.

There are three possible output arguments for the subroutine "**res3.m**".
-The argument "**e**" contains all the electromagnetic field quantities:

$E_x$=**e**(:,:,1),$E_y$=**e**(:,:,2),$E_z$=**e**(:,:,3),$H_x$=**e**(:,:,4),$H_y$=**e**(:,:,5),$H_z$=**e**(:,:,6).

-The second argument "**z**" is the vector containing the z-coordinate of the sampling points. Note that in the matrix $E_x$=**e**(:,:,1), the first index refer to the z coordinate, and the second to the x-coordinate. Thus $E_x$(i,j) is the $E_x$ field component at the location {z(i), x(j)}.
-The third argument **index**(i,j) is the complex refractive index at the location {**z**(i), **x**(j)}. It can be useful to check the profile of the grating.

Some **important** comments on the **parm**" argument:
1. For calculating precisely the electromagnetics fields, one has to set: "**parm.res1.champ=1"** before calling **res1.m.** This increases the calculation time and memory load but it is hightly recommended. If not, the computation of the field will be correct only in homogenous textures (for example in the top layer and in the bottom layer).
2. Illuminating the grating from the top or the bottom layer : As mentioned earlier, the code compute the diffraction efficiencies of the transmitted and reflected orders for an incident plane wave from the top and for an incident plane wave from the bottom at the same time. When plotting the field, the user must specify the direction of the incident plane wave. This is specified with variable **parm.res3.sens**. For **parm.res3.sens=1**, the grating is illuminated form the top and **parm.res3.sens=-1**, the grating is illuminated form the bottom (default is **parm.res3.sens=1**).
3. Specifying the z locations of the computed fields: This is provided by the variable **parm.res3.npts**. **parm.res3.npts** is a vector whose length is equal to the length of the variable **profile{1}**. For instance let us imagine, a two-layer grating defined by **profile** = {[0.5,1,2,0.6],[1,2,3,4]}. Setting **parm.res3.npts=[2,3,4,5]** implies that the field will be computed in two z=constant plans in the top layer, in three z=constant plans in the first layer (texture 2), in four z=constant plans in the second layer (texture 3), and in five z=constant plans in the bottom layer. Default for **parm.res3.npts** is 10 z=constant plan per layer. **VERY IMPORTANT** : where is the z=0 plan and what are the z-coordinates of the z=constant plan? The z=0 plan is defined at the bottom of the bottom layer. Thus, the field calculation is performed only for z>0 values. For the example **profile** = {[0.5,1,2,0.6],[1,2,3,4]}, and if we refer to texture 4 as the substrate, the z=0 plan is located in the substrate at a distance 0.6 under the grating. The z=constant plans are located by an equidistant sampling in every layer. Always referring to the previous example, it implies that the five z=constant plans in the substrate are located at coordinate z=(p-0.5) 0.6/5, where p=1,2, …5. Note that the z coordinate for z=constant plan are always given by the second output variable of res3.m.
4. How can one specify a given z=constant plan? First, one has to redefine the variable **profile**. For the grating example with the two layers discussed above, let us imagine that one wants to plot the field at z=z0+0.6+0.2 in layer 2. Then one has to set: **profile** = {[0.5,1-z0,0,z0,0.2,0.6],[1,2,2,2,3,4]} and set **parm.res3.npts=[0,0,1,0,0,0]**. Note that it is not necessary to redefine the variable **profile** at the beginning of the program. One just needs to redefine this variable before calling subroutine res3.m.
5. Automatic plots: an automatic plot (showing all the components of the electromagnetic fields and the grating refractive index distribution) is provided by setting **parm.res3.trace**=1. If one wants to plot only some components of the fields, one can set for instance: **parm.res3.champs**=[2,3,6,0], to plot $E_y$, $E_z$, $H_z$ and the object, **parm.res3.champs**=[1] to plot only $E_x$.

## 8.2. Computation of the absorption loss

Once the eigenmodes of all textures are known, the calculation of the absorption loss everywhere in the grating can be performed. The absorption loss in a surface $S$ is given by:

$$L = \frac{\pi}{\lambda} \int_S \Im m\left(\varepsilon\left(M\right)\right)\left(\left|E_X\right|^2 + \left|E_Y\right|^2 + \left|E_Z\right|^2\right) ds$$

The loss calculation is done with the subroutine "**res3.m**", without specifying the sampling points, following the instruction

[**e, Z, index, wZ, loss_per_layer, loss_of_Z, loss_of_Z_X, X, wX**] = res3(**x,aa,profile,einc,parm**);

The important ouput arguments are:
**loss_per_layer**: the loss in every layer defined by **profile**, **loss_per_layer**(1) is the loss in the top layer, **loss_per_layer**(2) the loss in layer 2, ... and **loss_per_layer**(end) the loss in the bottom layer
**loss_of_Z**: the absorption loss density (integrated over **X**) as a function of **Z** (like for **X**, the sampling points **Z** are not equidistant. You may plot this loss density as follows : plot(**Z**, **loss_of_Z**), xlabel('Z'), ylabel('absorption')
**loss_of_Z_X**(**Z**,**X**) = π/λ Im(**index**(**Z**,**X**).^2) (|**e**(**Z**,**X**,**1**)|$^2$+|**e**(**Z**,**X**,**2**)|$^2$+|**e**(**Z**,**X**,**3**)|$^2$)

**index**: **index**(i,j) is the complex refractive index at the location {**z**(i), **x**(j)}.

To compute integrals like the loss or the electromagnetic energy, we use a Gauss-Legendre integration method. This method, which is very powerful for 'regular' functions, becomes inaccurate for discontinuous functions. Thus, the integration domain should be divided into subdomains where the electric field **E** is continuous. For the integration in **X**, this difficult task is performed by the program, so that the user should only define the limits of integration: the input "**x**" argument is now a vector of length 2, which represent the limits of the x interval (to compute the loss over the entire period, we may take **x**(2)=**x**(1)+**period**. The integration domain is then divided into subintervals where the permittivity is continuous, each subinterval having a length less than $\lambda/(2\pi)$. For every subinterval, a Gauss-Legendre integration method of degree 10 is used. This default value can be changed by setting **parm.res3.gauss_x**=.... The actual points of computation of the field are returned in the output argument **X**.

For the z integration, the discontinuity points are more easily determined by the variable **'profile'**. The user may choose the number of subintervals and the degree in every layer using the parameter **parm.res3.npts**, which is now an array with two lines (in subsection 8.1 this variable is a line vector): the first line defines the degree and the second line the numbers of subintervals of every layer. For example: **parm.res3.npts** = [ [10 , 0 , 12 ] ; [3 ,  1 ,  5 ] ]; means that 3 subintervals with 10-degree points are used in the first layer, 1 subintervals with 0 point in the second layer, 5 subintervals with 12degree points in the third layer.

The actual z-points of computation of the field are returned in the output variable **Z**, and the vector **wZ** represents the weights and we have sum(**loss_of_Z**.***wZ**)=sum(**loss_per_layer**). Although the maximum degree that can be handled by reticolo is 468, it is recommended to limit the degree values to modest numbers (10-30 maximum) and to increase the number of subintervals (the larger the degree, the denser the sampling points in the vicinity of the subinterval boundaries).

Note that if **einc**= **result.TEinc_top PlaneWave_TE_Eu**, the energie conservation test for a TE incident plane wave from the top is

sum(**result.TEinc_top_reflected.efficiency**)+
sum(**result.TEinc_top_transmitted.efficiency**)+
sum(**loss_per_layer**) / (.5***period**) = 1.

Usually, this equality is achieved with an absolute error of $<10^{-5}$.

For specialists:
-**loss_of_Z_X** =pi/ **wavelength***imag(**index**.^2).*sum(abs(**e**(:,:,1:3)).^2,3);
-**loss_of_Z** =(**loss_of_Z_X*****wX**(:)).';
-by setting **index**(**index** ~= index_chosen)=0 in the previous formulas, one may calculate the absorption loss in the medium of refractive index index_chosen.

# 9. Bloch-mode effective indices

RETICOLO gives access to an other output: the Bloch-modes effective indices of all textures and the diffracted propagative waves.

Bloch mode number k of the texture number l can be written

$$\left|\Phi_k^{\,l}\right\rangle = \sum_m a_m^{k,l} \exp\left[i\left(k_{incx}+mK_x\right)x\right]\exp\left[i\,k_{incy}\,y\right]\exp\left[i\frac{2\pi}{\lambda}n_{eff}^{k,l}\,z\right],$$

where $n_{eff}^{k,l}$ is the effective index of the Bloch mode k of the texture l.

Instruction :
**[aa,n_eff] = res1(wavelength,period,textures,nn,kparallel,delta0,parm)**;
Note that the "n_eff" variable is a Matlab cell array: "**n_eff{ii}**" is a column vector containing all the Bloch-mode effective indices associated to the texture "**textures{ii}**". The element number 5 of this vector, for example, is called by the instruction "**n_eff{ii}(5)**;". An attenuated Bloch-mode has a complex effective index.

# 10. Annex

## 10.1. Checking that the textures are correctly set up

Setting “**parm.res1.trace = 1**;” generates a Matlab figure which represents the refractive-index distribution of all the textures.

## 10.2. The “retio” instruction

RETICOLO automatically creates temporary files in order to save memory. These temporary files are of the form “abcd0.mat”, “abcd1.mat” … with abcd randomly chosen) .They are created in the current directory. In general RETICOLO automatically erases these files when they are no longer needed, but it is recommended to finish all programs by the instruction “**retio**;”, which erases all temporary files. Also, if a program anormally stopsone may execute the instruction “**retio**” before restarting the program.
The “retefface” instruction allows to know all the “abcd0.mat” files and to erase them if wanted.

## 10.3. How to save and to reload the “aa” variable

To save the “**aa**” variable in a “.mat” file, the user has to define a new parameter containing the name of the file he or she wants to create : “**parm.res1.fperm = 'file_name'**;”. field_name is a char string with at least one letter. The program will automatically save “**aa**” in the file “**file_name.mat**”. In a new utilisation it is sufficient to write aa = **'file_name'**;.

<u>Example of a program which calculates and saves the “aa” variable</u>
[...] % Definition of the input parameters, see Section 3
**parm.res1.fperm = 'toto'**;
[...] % Definition of the textures, see Section 4.1
**aa = res1(wavelength,period,textures,nn,k_parallel,angle_delta,parm)**;
<u>Example of a program which uses the “aa” variable and then calculates the diffracted waves</u>
[...] % Definition of the profile, see Section 4.2. Note that the textures used to define the profile argument have to correspond to the textures defined in the program which has previously calculated the “aa” variable.
**aa=’toto’**;
**result = res2(aa,profile)**;
**retio**;

## 10.4. Asymmetry of the Fourier harmonics retained in the computation

**nn = [-15;20]**; % this defines the set of non-symmetric Fourier harmonics retained for the computation. In this case, the Fourier harmonics from –15 to +20 are retained.
The instructions “**nn = 10**;” and “**nn = [-10;10]**;” are equivalent.
Take care that the use of symmetry imposes symmetric Fourier harmonics, if not the computation will be done without any symmetry consideration.

# 11. Summary

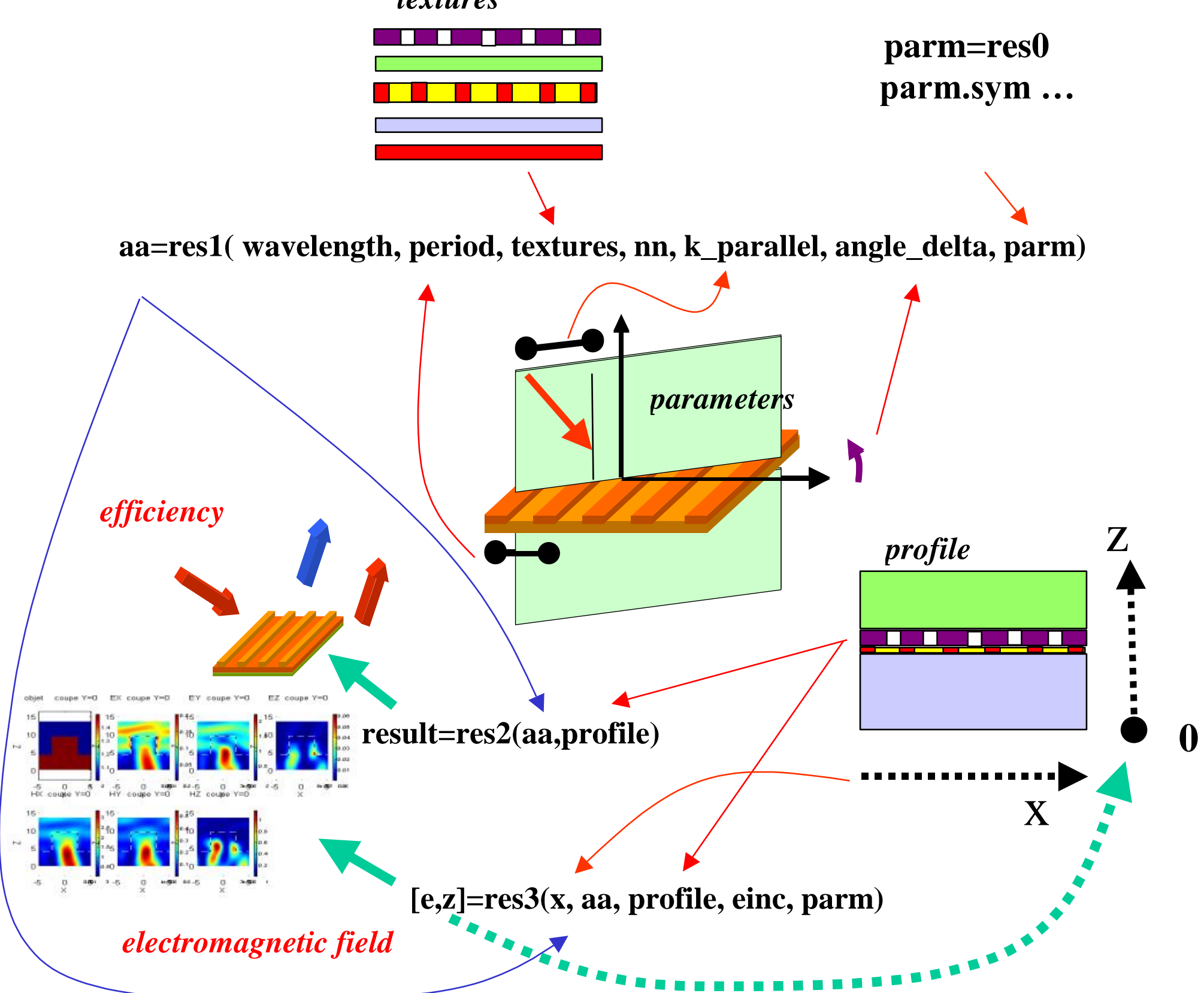

Fig. 7 Summary.

# 12. Examples

The following examples can be copied and executed in Matlab.

```
%%%%%%%%%%%%%%%%%%%%%%%%%%%%%
% SIMPLE EXAMPLE 1D CONICAL %
%%%%%%%%%%%%%%%%%%%%%%%%%%%%%
wavelength=8;
period=10;% same unit as wavelength
n_incident_medium=1;%refractive index of the top layer
n_transmitted_medium=1.5;% refractive index of the bottom layer
angle_theta0=10;k_parallel=n_incident_medium*sin(angle_theta0*pi/180);
angle_delta=-20;

parm=res0;          % default parameters for "parm"
parm.res1.champ=1; % the electromagnetic field is calculated accurately
nn=5;   %  Fourier harmonics run from [-5,5]

% textures for all layers including the top and bottom layers
texture=cell(1,3);
textures{1}= n_incident_medium;                     % uniform texture
textures{2}= n_transmitted_medium;                  % uniform texture
textures{3}={[-2.5,2.5],[n_incident_medium,n_transmitted_medium] };

aa=res1(wavelength,period,textures,nn,k_parallel,angle_delta,parm);
```

```
profile={[4.1,5.2,4.1],[1,3,2]};

conical=res2(aa,profile)

eff_TETM=conical.TEinc_top_reflected.efficiency{-1}
% -1 order efficiency (TE+TM) for a TE-illumination from the top layer
eff_TE=conical.TEinc_bottom_transmitted.efficiency_TE{-1}
% -1 order TE efficiency for a TE-illumination from the top layer
J=conical.Jones.inc_bottom_transmitted{-1};% Jones'matrix
abs(J).^2 % -1 order efficiencies for an illumination from the top layer

% field calculation
x=linspace(-period/2,period/2,51);% x coordinates(z-coordinates are
determined by res3.m)
einc=[0,1]; % E-field components in the (u, v) basis (default is
illumination from the top layer)
parm.res3.trace=1; % plotting automatically
parm.res3.npts=[50,50,50];
[e,z,index]=res3(x,aa,profile,einc,parm);
figure;pcolor(x,z,real(squeeze(e(:,:,3)))); % user plotting
shading flat;xlabel('x');ylabel('y');axis equal;title('Real(Ez)');

% Loss calculation
textures{3}={[-2.5,2.5],[n_incident_medium,.1+5i] };
aa_loss=res1(wavelength,period,textures,nn,k_parallel,angle_delta,parm);
conical_loss=res2(aa_loss,profile)
parm.res3.npts=[[0,10,0];[1,3,1]];
einc=conical_loss.TEinc_top.PlaneWave_TE_Eu;

[e,z,index,wZ,loss_per_layer,loss_of_Z,loss_of_Z_X,X,wX]=res3([-
period/2,period/2],aa_loss,profile,einc,parm);

Energie_conservation=sum(conical_loss.TEinc_top_reflected.efficiency)+sum(c
onical_loss.TEinc_top_transmitted.efficiency)+sum(loss_per_layer)/(.5*
period)-1

retio % erase temporary files

%%%%%%%%%%%%%%%%%%%%%%%%%
% EXAMPLE 1D (TE or TM) %
%%%%%%%%%%%%%%%%%%%%%%%%%

wavelength=8;
period=10;% same unit as wavelength
n_incident_medium=1;% refractive index of the top layer
n_transmitted_medium=1.5;% refractive index of the bottom layer

angle_theta0=-10;k_parallel=n_incident_medium*sin(angle_theta0*pi/180);

parm=res0(1);% TE polarization. For TM : parm=res0(-1)
parm.res1.champ=1;% the electromagnetic field is calculated accurately

nn=40;% Fourier harmonics run from [-40,40]

% textures for all layers including the top and bottom layers
texture=cell(1,3);
textures{1}= n_incident_medium;                    % uniform texture
textures{2}= n_transmitted_medium;                 % uniform texture
textures{3}={[-2.5,2.5],[n_incident_medium,n_transmitted_medium] };
```

```
aa=res1(wavelength,period,textures,nn,k_parallel,parm);

profile={[4.1,5.2,4.1],[1,3,2]};

one_D_TE=res2(aa,profile)
eff=one_D_TE.inc_top_reflected.efficiency{-1}
J=one_D_TE.Jones.inc_top_reflected{-1};% Jones'coefficients
abs(J)^2 % first order efficiency for an illumination from the top layer

% field calculation
x=linspace(-period/2,period/2,51);% x coordinates(z-coordinates are
determined by res3.m)
einc=1;
parm.res3.trace=1; % plotting automatically
parm.res3.npts=[50,50,50];
[e,z,index]=res3(x,aa,profile,einc,parm);
figure;pcolor(x,z,real(squeeze(e(:,:,1)))); % user plotting
shading flat;xlabel('x');ylabel('y');axis equal;title('Real(Ey)');

% Loss calculation
textures{3}={[-2.5,2.5],[n_incident_medium,.1+5i] };
aa_loss=res1(wavelength,period,textures,nn,k_parallel,parm);
one_D_loss=res2(aa_loss,profile)
parm.res3.npts=[[0,10,0];[1,3,1]];
einc=one_D_loss.inc_top.PlaneWave_E(2);
[e,z,index,wZ,loss_per_layer,loss_of_Z,loss_of_Z_X,X,wX]=res3([-
period/2,period/2],aa_loss,profile,einc,parm);

Energie_conservation=sum(one_D_loss.inc_top_reflected.efficiency)+sum(one_D
_loss.inc_top_transmitted.efficiency)+sum(loss_per_layer)/(.5* period)-1

retio % erase temporary files
```

# III. RETICOLO CODE 2D for the analysis of the diffraction by stacks of lamellar 2D gratings

**Reticolo code 2D is a free software for analyzing 2D crossed grating. It operates under Matlab. To install it, copy the companion folder "reticolo_allege" and add the folder in the Matlab path. <u>The code may also be used to analyze thin-film stacks with homogeneous and anisotropic materials</u>, see the end of Section 4.1.**

## Outline

# 1. Generality

RETICOLO is a code written in the language MATLAB 9.0. It computes the diffraction efficiencies and the diffracted amplitudes of gratings composed of stacks of lamellar structures. It incorporates routines for the calculation and visualisation of the electromagnetic fields inside and outside the grating. With this version, 2D periodic (crossed) gratings cannot be analysed.

As free alternative to MATLAB, RETICOLO can also be run in GNU Octave with minimal code changes. For further information, please contact tina.mitteramskogler@profactor.at.

In brief, RETICOLO implements a frequency-domain modal method (known as the Rigorous Coupled wave Analysis/RCWA). To get an overview of the RCWA, the interested readers may refer to the following articles:
1D-classical and conical diffraction
M.G. Moharam et al., JOSAA **12**, 1068 (1995),
M.G. Moharam et al, JOSAA **12**, 1077 (1995),
P. Lalanne and G.M. Morris, JOSAA **13**, 779 (1996),
G. Granet and B. Guizal, JOSAA **13**, 1019 (1996),
L. Li, JOSAA **13**, 1870 (1996), see also C. Sauvan et al., Opt. Quantum Electronics **36**, 271-284 (2004) which simply explains the raison of the convergence-rate improvement of the Fourier-Factorization rules without requiring advanced mathematics on Fourier series and generalizes to other kinds of expansions.
2D-crossed gratings
L. Li, JOSAA **14**, 2758-2767 (1997),
E. Popov and M. Nevière, JOSAA **17**, 1773 (2000),
which describe the up-to-date formulation of the approach used in RETICOLO. Note that the formulation used in the last article (which proposes an improvement for analysing metallic gratings with continuous profiles like sinusoidal gratings) is not available in the RETICOLO version of the web. The RCWA relies on the computation of the eigenmodes in all the layers of the grating structure in a Fourier basis (plane-wave basis) and on a scattering matrix approach to recursively relate the mode amplitudes in the different layers.

**Eigenmode solver:** For conical diffraction analysis of 1D gratings, the Bloch eigenmode solver used in Reticolo is based on the article "P. Lalanne and G.M. Morris, JOSAA **13**, 779 (1996)".

**Scattering matrix approach:** The code incorporates many refinements that we have not published and that we do not plan to publish. For instance, although it is generally admitted that the S-matrix is inconditionnally stable, it is not always the case. We have developed an in-house transfer matrix method which is more stable and accurate. The new transfer matrix approach is also more general and can handle perfect metals. The essence of the method has been rapidly published in "J.-P. Hugonin, M. Besbes and P. Lalanne, Op. Lett. **33**, 1590 (2008)".

**Field calculation:** The calculation of the near-field electromagnetic fields everywhere in the grating is performed according to the method described in "P. Lalanne, M.P. Jurek, JMO **45**, 1357 (1998)" and to its generalization to crossed gratings (unpublished). Basically, no Gibbs phenomenon will be visible in the plots of the discontinuous electromagnetic quantities, but field singularities at corners will be correctly handled.

**Acknowledging the use of RETICOLO**: In publications and reports, acknowledgments have to be provided by referencing to J.P. Hugonin and P. Lalanne, RETICOLO software for grating analysis, Institut d'Optique, Orsay, France (2005), arXiv:2101:00901.

**In journal publications and in addition, one may fairly quote the following references:**
-L. Li, "New formulation of the Fourier modal method for crossed surface-relief gratings", J. Opt. Soc. Am. A **14**, 2758-2767 (1997),
-P. Lalanne and M.P. Jurek, "Computation of the near-field pattern with the coupled-wave method for TM polarization", J. Mod. Opt. **45**, 1357-1374 (1998), if near-field electromagnetic-field distributions are shown.

# 2. The diffraction problem considered

In general terms, RETICOLO-2D solves the diffraction problem by a grating defined by a stack of layers which have all identical periods in the x- and y-directions. In the following, the (x,y) plane and the z-direction will be referred to as the transverse plane and the longitudinal direction, respectively. To define the grating structure, first we must define a top layer and a bottom layer. This is rather arbitrary since the top or the bottom layers can be the substrate or the cover of a real structure. It is up to the user. Once the top and the bottom layers have been defined, the user can choose to illuminate the structure from the top or from the bottom. The z-axis is oriented from bottom to top.
RETICOLO-2D is written with the $exp(-i\omega t)$ convention for the complex notation of the fields. So, if the materials are absorbant, one expects that all indices have a positive imaginary part. The Maxwell's equations are of the form

$$\boldsymbol{\nabla} \times \mathbf{E} = \frac{2i\pi}{\lambda}\mathbf{H} \ (\varepsilon_0 = \mu_0 = c = 1)$$
$$\boldsymbol{\nabla} \times \mathbf{H} = -\frac{2i\pi}{\lambda}\varepsilon\mathbf{E},$$

where $\varepsilon = n^2$ is the relative permittivity, a complex number, and $\lambda$ is the wavelength in a vacuum.

RETICOLO-2D returns the diffraction efficiencies of the transmitted and reflected orders for an incident plane wave from the top and for an incident plane wave from the bottom, both for TM and TE polarizations. The four results are obtained by the same calculation (incident TE wave from the top, incident TM wave from the top, incident TE wave from the bottom and incident TM wave from the bottom). Of course, the two incident plane waves must have identical parallel wave-vector in the transverse plane [ $\mathrm{k}_{\mathrm{x}}^{\mathrm{inc}}$, $\mathrm{k}_{\mathrm{y}}^{\mathrm{inc}}$ ]. This possibility which is not mentioned in the literature to our knowledge is important in practice since the user may get, for the same computational loads, the grating diffraction efficiencies for an illumination from the substrate or from the cover.

RETICOLO-2D calculates the electric and magnetic fields diffracted by the grating for the following incident plane wave :

$$\boldsymbol{E}_{top}^{inc}\, exp\left(i\left(k_x^{inc}x + k_y^{inc}y + k_{z\,top}^{inc}(z-h)\right)\right)$$
$$\boldsymbol{H}_{top}^{inc}\, exp\left(i\left(k_x^{inc}x + k_y^{inc}y + k_{z\,top}^{inc}(z-h)\right)\right), \text{ if incident from the top layer,}$$
$$\text{where } k_{z\,top}^{inc} = -\sqrt{\left(2\pi n_{top}/\lambda\right)^2 - (k_x^{inc})^2 - \left(k_y^{inc}\right)^2}.$$
$$\boldsymbol{E}_{bottom}^{inc}\, exp\left(i\left(k_x^{inc}x + k_y^{inc}y + k_{z\,bottom}^{inc}(z-h)\right)\right)$$
$$\boldsymbol{H}_{bottom}^{inc}\, exp\left(i\left(k_x^{inc}x + k_y^{inc}y + k_{z\,bottom}^{inc}(z-h)\right)\right), \text{ if incident from the bottom layer,}$$
$$\text{where } k_{z\,bottom}^{inc} = \sqrt{(2\pi n_{bottom}/\lambda)^2 - (k_x^{inc})^2 - \left(k_y^{inc}\right)^2}.$$

The z-component of the Poynting vector of the incident plane wave is ±0.5.

The Rayleigh-expansion of the diffracted electric fields are shown in the following figure.

$$\boldsymbol{E}_{top}^{dif} = \textstyle\sum_m \boldsymbol{E}_{top}^{m}\, exp\left[i((k_x^{inc} + mK_x)x + \left(k_y^{inc} + nK_y\right)y + k_{z\,top}^{m}(z-h)\right]$$
$$\boldsymbol{H}_{top}^{dif} = \textstyle\sum_m \boldsymbol{H}_{top}^{m}\, exp\left[i((k_x^{inc} + mK_x)x + \left(k_y^{inc} + nK_y\right)y + k_{z\,top}^{m}(z-h)\right]$$
$$\text{where } k_{z\,top}^{m} = \sqrt{\left(2\pi n_{top}/\lambda\right)^2 - (k_x^{inc} + nK_x)^2 - \left(k_y^{inc} + nK_y\right)^2}.$$

$$\mathbf{E}_{bottom}^{dif} = \textstyle\sum_{m,n} \mathbf{E}_{bottom}^{m,n}\, exp\left[i((k_x^{inc} + mK_x)x + \left(k_y^{\mathrm{inc}} + nK_y y\right) + k_{z\,bottom}^{\mathrm{m,n}} z\right]$$
$$\mathbf{H}_{bottom}^{dif} = \textstyle\sum_{m,n} \mathbf{H}_{bottom}^{m,n}\, exp\left[i((k_x^{inc} + mK_x)x + \left(k_y^{\mathrm{inc}} + mK_y y\right) + k_{z\,bottom}^{\mathrm{m,n}} z\right]$$
$$\text{where } \mathrm{k}_{z\,bottom}^{\mathrm{m,n}} = -\sqrt{(2\pi n_{bottom}/\lambda)^2 - (k_x^{inc} + mK_x)^2 - \left(k_y^{\mathrm{inc}} + nK_y\right)^2}$$

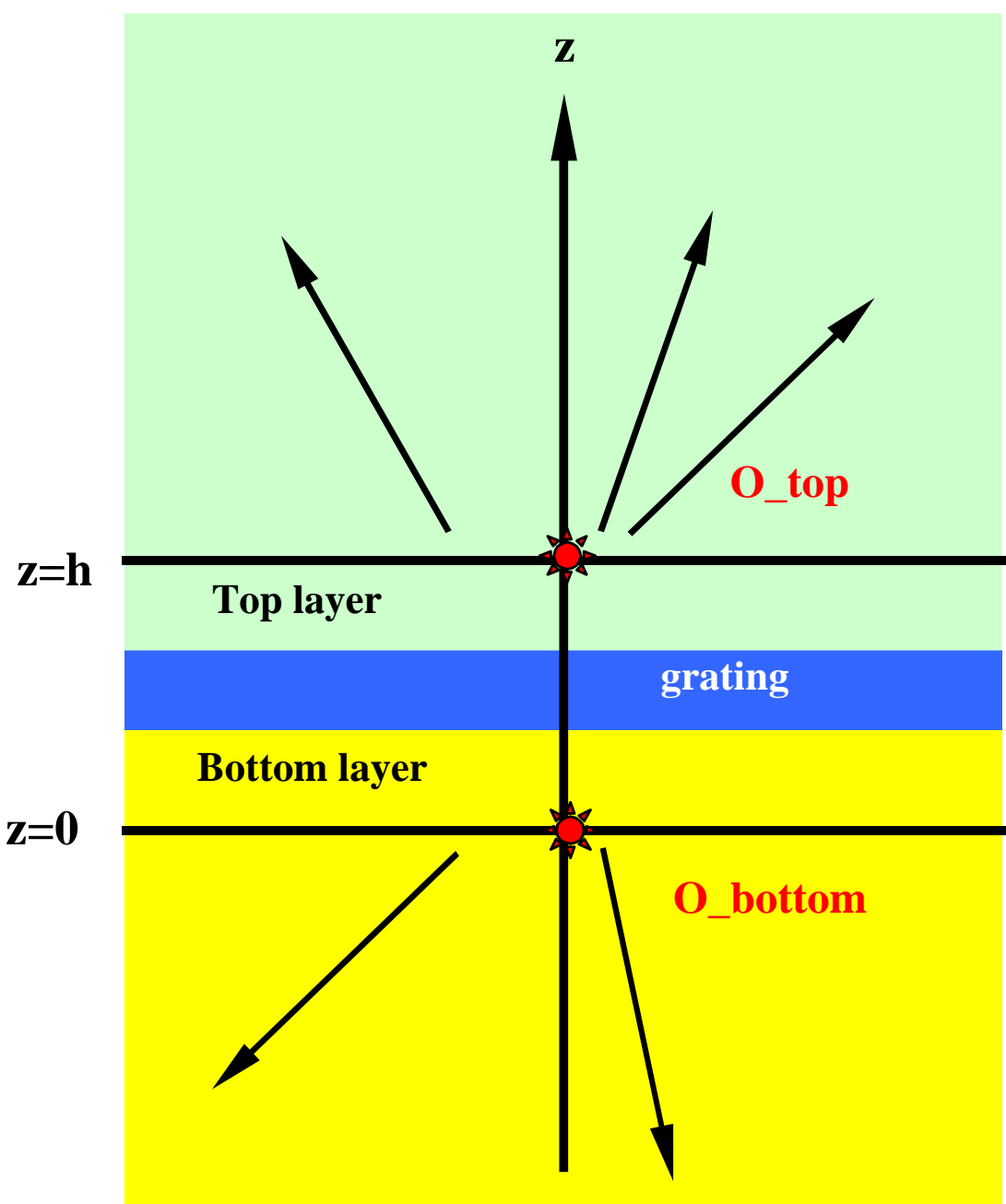

Fig. 1. Rayleigh expansion for the diffracted fields. $K_x = (2\pi)/period_x$, $K_y = (2\pi)/period_y$. The (m,n)th order has a parallel momentum equal to $(k_x^{inc} + mK_x)\,\vec{x} + (k_y^{inc} + mK_y)\,\vec{y}$. We define two points $O_{top}$= (0,0,h) at the top of the grating, and $O_{bottom}$= (0,0,0) at the bottom of the grating.

The following is organized so that one can straightforwardly write a code using the software.

# 3. Preliminary input parameters

The name of the following parameters are given as examples. The user may define his own parameter vocabulary.

**wavelength** = **3**; % wavelength (λ) in a vacuum. It might be 3 nm or 3 µm. You do not need to specify the unit but all other dimensions are of course in the same unit as the wavelength.

**period = [period_x , period_y]**; % the first variable is always related to the x-direction.

**nn** = **[3,2]**; This defines the set of Fourier harmonics retained for the computation. More specifically, 2×nn(1)+1 represent the number of Fourier harmonics retained in the x-direction from -nn(1) to nn(1), and 2×nn(2)+1 represent the number of Fourier harmonics retained in the y-direction from -nn(2) to nn(2). Note that the x-direction is always set up first.

If all the textures are homogeneous (case of a thin-film stack), we may set nn=0 and the period may be arbitrarily set to any value, [1,1] for example. NB: Because of our normalization choice (Poynting vector equal to 1), the computed reflected and transmitted amplitude coefficients are not identical to those provided by the Fresnel formulas.

**angle_delta** = **30**; % in degrees, see the following figure for a definition of "angle_delta" for the incident plane wave. This angle is varying between 0° and 360°. This angle has to be defined in the incident medium.

**k_parallel = n_incident_medium*sin(angle_theta*pi/180)**;
The parameters "**angle_delta**" and "**angle_theta**" which are used to specify **the plane of incidence and the angle of incidence** are denoted by δ and $\theta_{inc}$ in Fig. 2. The angle δ defines the plane of incidence. This plane allows to define the polarization of the incident plane wave: if the electric field of the incident plane wave is perpendicular to this plane, the incident wave is TE polarized, and if it is parallel to this plane, the incident wave is TM polarized. The incident wave vector is

$\mathbf{k}_{inc}=(2\pi/\lambda)\, n_{inc}\, \mathbf{K}_{inc}$

with $\mathbf{K}_{inc}$ = [sin(θ)cos(δ), sin(θ)sin(δ), -cos(θ)].
$n_{inc}$ is the refractive index of the top (or bottom) layer. One expects that it is a positive real number and that the texture (see Section 4.1) associated to the top (or the bottom) layer has a background with a uniform refractive index “$n_{inc}$”.
(Note that the “k_parallel” variable is defined **without** the factor 2π/λ.)

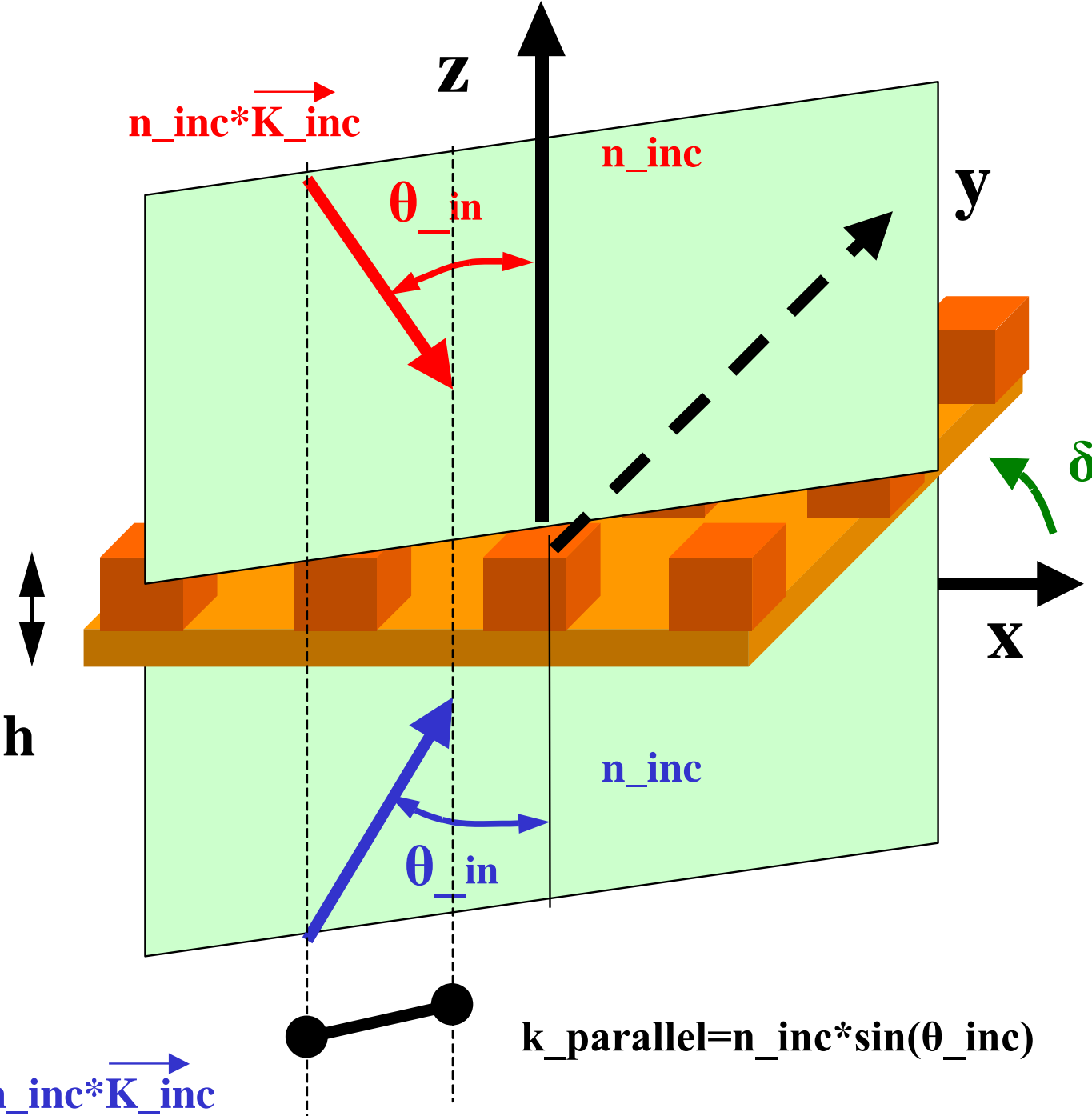

Fig. 2. Definition of θ_inc, δ, k_parallel.

In general, the user has in mind to illuminate the grating from the substrate or from the upperstrate (air in general). “n_incident_medium” (denoted also $n_{inc}$) is the refractive index of the incident medium. One expects that it is a positive real number and that the texture (see Section 4.1) associated to the top or the bottom layer has a background with a refractive index “n_incident_medium”.

It is very important to keep in mind that whether the user defines the incident plane wave in the top layer or in the bottom layer, the calculation will be done for both an incident wave from the top and an incident wave from the bottom, with an identical parallel wave vector, i.e. for a specified $\left[k_x^{inc}, k_y^{inc}\right]$ which is the same in the bottom and top layers.

# 4. Structure definition (grating parameters)

The grating encompasses a uniform upperstrate, called the top in the following, a uniform substrate, called the bottom in the following, and many layers which define the grating, which is defined by a stack of layers. Every layer is defined by a “texture” and by its thickness. Two different layers may be identical (identical texture and thickness), may have different thicknesses with identical texture, may have different thicknesses and textures. To define the diffraction geometry, one needs to define the different textures and then the different layers.

## 4.1. How to define a texture?

We have first to define a background by its refractive index “nbackground”. Then in this uniform background, we successively incorporate inclusions with a refractive index “ninclusion”. The geometry of this inclusion can be an ellipse or a rectangle, defined by the position (c_x,c_y) of its center and its dimensions Lx and Ly along the x and the y direction respectively. Note that the ellipse axes are parallel to the x and y directions

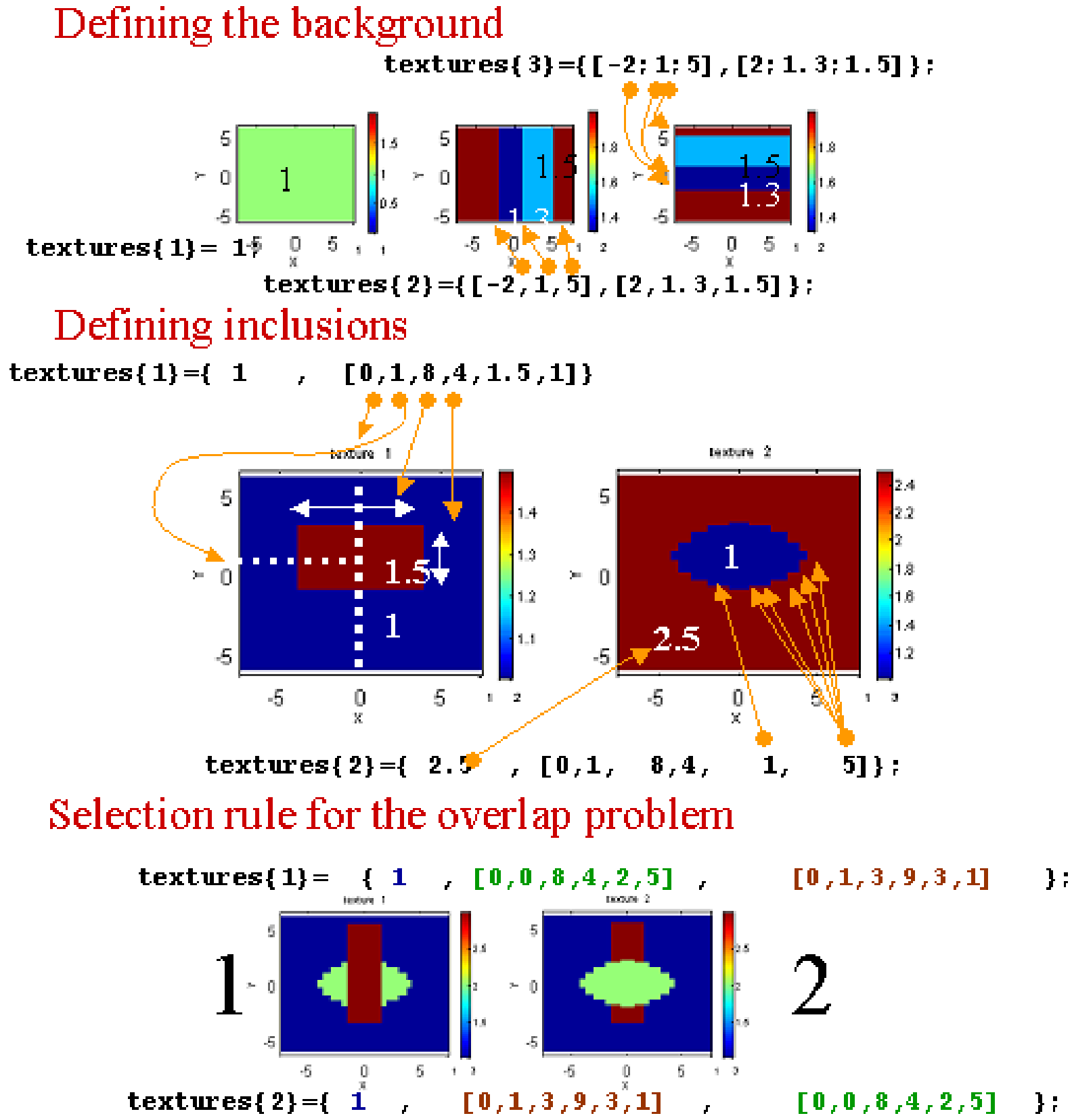

Fig. 3. Principle of textures definition.

<u>Some examples</u>

Definition of a uniform texture 1 with a refractive index “nbackground” :
textures{1} = { nbackground };

Definition of a texture 2 composed of a rectangle of refractive index “ninclusion” in an uniform background with a refractive index “nbackground”:
textures{2} = { nbackground,[cx,cy,Lx,Ly,ninclusion,1]};
Note that the last number “1” indicates that the inclusion is a rectangle. Of course, if Lx = Ly, we define a square.

Definition of a texture 3 composed of an ellipse of refractive index “ninclusion” in a uniform background with a refractive index “nbackground”:
textures{3} = { nbackground,[cx,cy,Lx,Ly,ninclusion,N]};
Note that the last number “N” indicates that the inclusion is an ellipse. If Lx = Ly, we define a circle. The ellipse is in fact coded by a staircase approximation, and 4xN represents the number of edges of the staircase pattern used to represent the continuous smooth profile. As N increases, the staircase approximation becomes more and more accurate. We recommend to use N ≥ 5.

Definition of an intricate texture 4 (selection rule for the overlap problem):
In the following example, we have two rectangular pillars in the period.
textures{4} = {1, [0,0,5,2,  2,  1], [0,0,1,10,  3,  1]};
There is an overlap between the two rectangles, and the refractive index of the overlap region is fixed by the last inclusion, “3” in the example.

The following figure shows the refractive indices of the 4 generated textures.

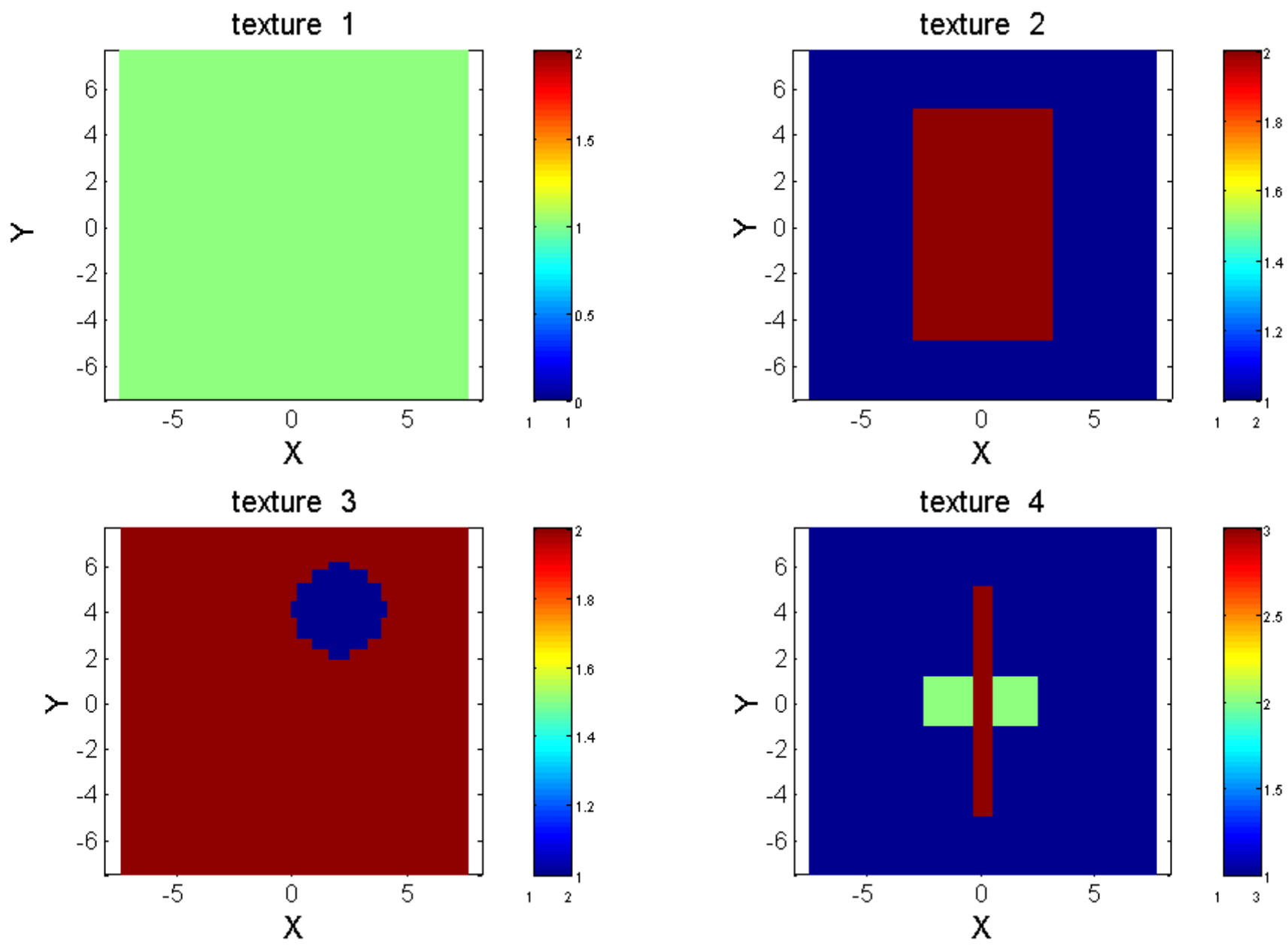

Fig. 4. Different textures, see the text for their generation.

Dielectric rectangles in a perfectly-conducting metallic background:
The background can be an infinitly conducting metal. In this uniform background, inclusions with a complex or real refractive index “ninclusion” can be incorporated. In this case, the geometry of this inclusion can be only rectangular, defined by the position (c_x, c_y) of its center and its dimensions Lx and Ly along the x and the y direction respectively. The inclusions *cannot* overlap
For example:
textures {5}= { inf, [c_x1,c_y1,Lx1,Ly1,ninclusion1],[c_x2,c_y2,Lx2,Ly2, ninclusion2]}

Anisotropic layers:
Grating layers (not the substrate nor the superstrate) can be anisotropic with diagonal tensors ($\varepsilon_{xy} = \varepsilon_{xz} \ldots = 0$). To implement diagonal anisotropy

parm.res1.change_index={[$n_{prov}^1$, $n_x^1$, $n_y^1$, $n_z^1$] , [$n_{prov}^2$, $n_x^2$, $n_y^2$, $n_z^2$]}; % NB: $n_{prov}^1 \neq n_{prov}^2$

The refractive index $n_{prov}^1$ is then replaced **in all textures** by epsilon=diag([$(n_x^1)^2$, $(n_y^1)^2$, $(n_z^1)^2$]). Beware if the superstate (or substrate) has a refractive index $n_{prov}^1$, it will also be replaced and this is not allowed. Thus we recommend using an unusual value for $n_{prov}^1$ (e.g. 89.99999 or rand(1)).
The user may also diagonal permeability tensors

parm.res1.change_index={ [$n_{prov}^1$, $n_x^1$, $n_y^1$, $n_z^1$ , $m_x^1$, $m_y^1$, $m_z^1$ ] , [$n_{prov}^2$, $n_x^2$, $n_y^2$, $n_z^2$] };

The refractive index $n_{prov}^1$ is then replaced **in all textures** by epsilon=diag( [$(n_x^1)^2$, $(n_y^1)^2$, $(n_z^1)^2$] ), mu=diag( [$(m_x^1)^2$, $(m_y^1)^2$, $(m_z^1)^2$] ).
For slits in perfectly-conducting metallic textures, anisotropy cannot be implemented.

Fully-anisotropic homogeneous layers and thin-film-stack modeling:
Homogeneous layers (with permittivity and permeability independent of x and z) can be simulated for arbitrary anisotropies (not necessarily diagonal) textures {4} = {epsilon};

with epsilon an arbitrary 3×3 matrix. The user may also implement magnetic anisotropy textures {4} = {epsilon, mu};

with epsilon and mu arbitrary 3×3 matrices.

Note that the substrate and superstrates should be uniform and isotropic materials. If all layers are uniform, a thin-film stack can be computed for arbitrary epsilon and mu 3×3 matrices by retaining a single Fourier component, nn = 0.

To check if the set of textures is correctly set up, the user can set the variable parm.res1.trace equal to 1: “parm.res1.trace = 1;”. Then a MatLab figure will show the refractive-index distributions of all textures. Every texture is represented with the coordinate x varying from - period(1)/2 to +period(1)/2 and with the coordinate y varying from - period(2)/2 to +period(2)/2.

## 4.2. How to define the layers?

This is performed by defining the “Profile” variable which contains, starting from the top layer and finishing by the bottom layer, the successive information (thickness and texture-label) relative to every layer. Here is an example that illustrates how to set up the “Profile” variable:

**Profile** = {[0,1,0.5,0.5,1,0.5,0.5,2,0],[1,3,2,4,3,2,4,6,2]}; (1)

It means that from the top to the bottom we have: the top layer is formed by a thickness 0 of texture 1, then we have twice textures 3, 2 and 4 with depth 1, 0.5 and 0.5 respectively, texture 6 with depth 2, and finally the bottom layer (formed by texture 2) with null thickness. Since textures 1 and 2 correspond to the top and bottom layers, they must be uniform. In this example, the top and bottom layers have a null thickness. However, one may set an arbitrary thickness. Especially, if one needs to plot the electromagnetic fields in the bottom and top layers, the thicknesses $h_b$ and $h_h$ (see Fig. 4) over which the fields have to be visualized has to be specified. For $h_b$=$h_h$=0, the Rayleigh expansions of the fields in the top and bottom layers are not plotted.

In this particular profile, the structure formed by texture 3 with thickness 1, texture 2 with thickness 0.5 and texture 4 with thickness 0.5 is repeated twice. It is possible to simplify the instruction defining the “Profile” variable in order to take into account the repetitions:

**Profile** = {{0,1},{[1,0.5,0.5], [3,2,4], 2},{[2,0],[6,2]}}; (2)

If a structure is repeated many times, the above “factorized” instruction of Eq. 2 is better than the “expanded” one of Eq. 1, in terms of computational speed, because the calculation will take into account the repetitions.

The profile is shown below.

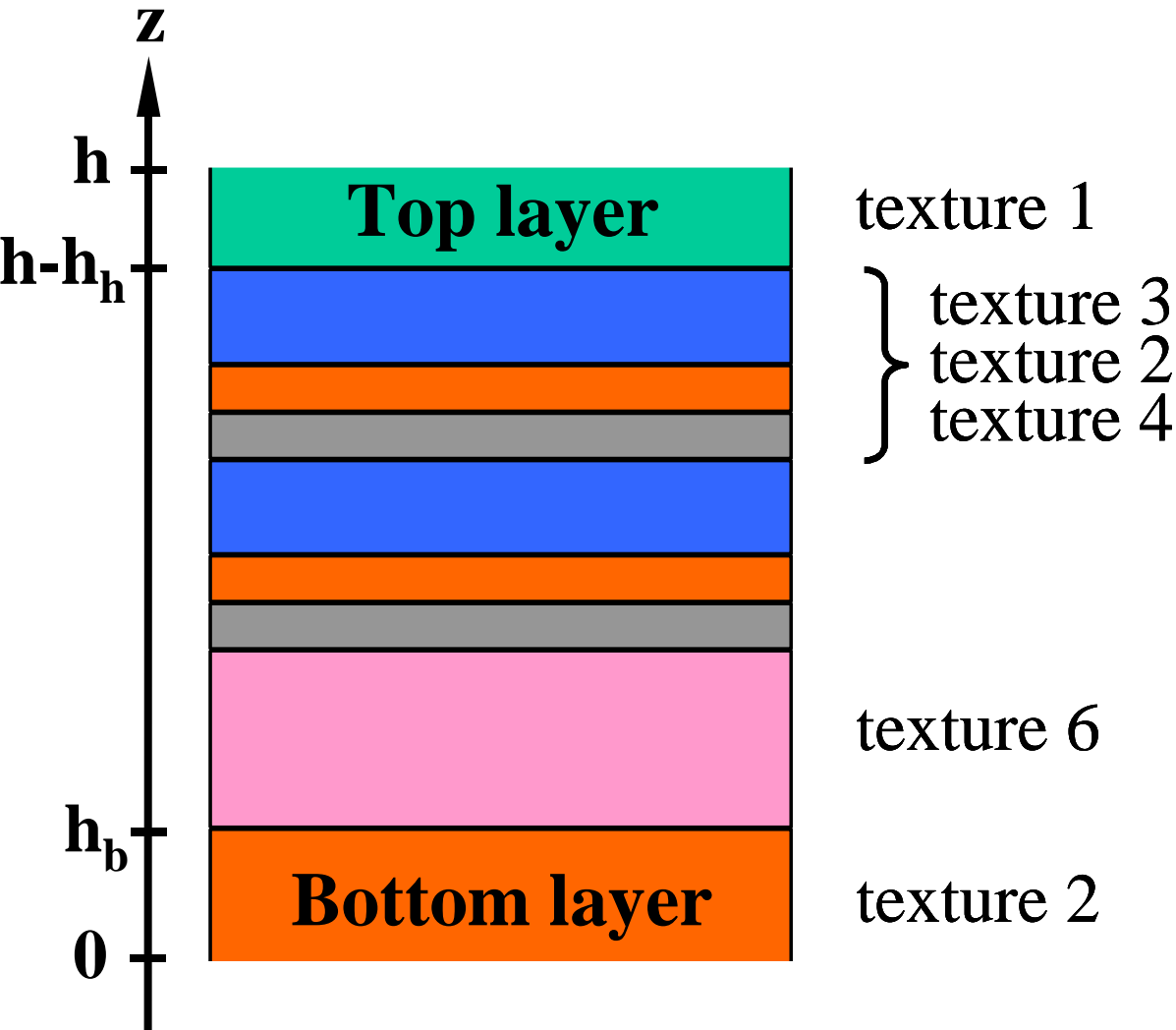

Fig. 5. Texture stacks. The example corresponds to a profile defined by **Profile** = {[$h_h$,1,0.5,0.5,1,0.5,0.5,2, $h_b$],[1,3,2,4,3,2,4,6,2]}; . The top and bottom layers have uniform and isotropic textures.

# 5. Solving the eigenmode problem for every texture

The first computation with the RCWA consists in calculating the eigenmodes associated to all textures. This is done by the subroutine “res1.m”, following the instruction:

**aa = res1(wavelength,period,textures,nn,k_parallel,angle_delta,parm)**;

The first-six input parameters are absolutely required by the code : the wavelength “**wavelength**”, the period of the grating “**period**”, the “**textures**” variable, the number of Fourier harmonics “**nn**”, the norm of the parallel incident wave vector “**k_parallel**”, the angle that defines the plane of incidence “**angle_delta**.
Some other additional parameters can be defined. For example, the default parameters do not take the symmetry of the problem into account. So if the user wants to use symmetries, new parameters have to be defined : “**parm.sym.x**”, “**parm.sym.y**”, and “**parm.sym.pol**”. These parameters are defined in Section 7.

**parm = res0**;
res0.m is a function that changes the default values. This instruction has to be executed before res1.m, if one wants to modify the default values (for instance to use symmetry).

It is very important to note that if one has to study the diffraction by many different gratings composed of the same textures, one needs to compute only once the eigenmodes. It is possible to save the “aa” variable in a “.mat” file and to reload it for the computation of the diffracted waves, see an example in Annex 9.3.

# 6. Computing the diffracted waves

This is the second step of the computation. This is done by the subroutine “res2.m”, following the instruction:

**result = res2(aa,Profile)**;

This subroutine has 2 input arguments: the output “**aa**” of the subroutine “res1.m” and the “**Profile**” variable. The output argument “**result**” contains all the information on the diffracted fields. “**result**” is an object of class ‘reticolo’ that can be indexed as an usual structure with parentheses, or with the labels of the considered orders between curly braces. Examples will be given in the following.
This information is divided into the following sub-structures fields :

- “**result.TEinc_top**”
- “**result.TEinc_top_reflected**”
- “**result.TEinc_top_transmitted**”

- “**result.TEinc_bottom**”
- “**result.TEinc_bottom_reflected**”
- “**result.TEinc_bottom_transmitted**”

- “**result.TMinc_top**”
- “**result.TMinc_top_reflected**”
- “**result.TMinc_top_transmitted**”

- “**result.TMinc_bottom**”
- “**result.TMinc_bottom_reflected**”
- “**result.TMinc_bottom_transmitted**”

The sub-structure “**result.TEinc_top_reflected**” contains all the information concerning the propagative *reflected* waves for *the incident wave from the top* of the grating in *TE polarization* which is described in the sub-structure **“result.TEinc_top”**
The sub-structure “**result.TMinc_bottom_transmitted**” contains all the information concerning the propagative *transmitted* waves for *the incident wave from the bottom* of the grating *in TM polarization*, which is described in the sub-structure “**result.TMinc_bottom”**. And so on.

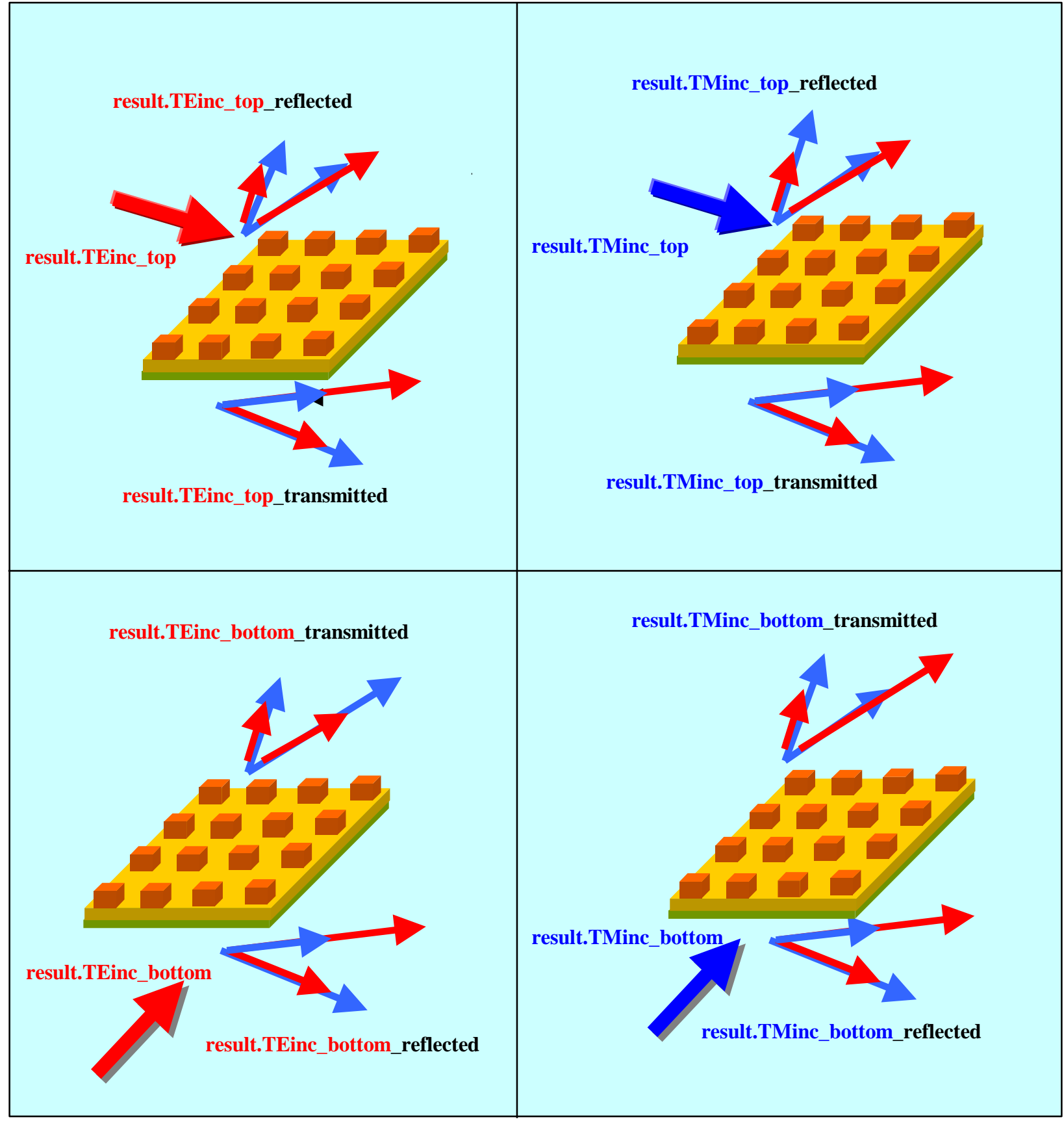

Fig. 6. The 4 solutions obtained.

Each sub-structure of **result** is composed of the following fields. Each field is a Matlab column vector or matrix having the same number N of lines. N is the number of propagative orders considered and can be 0.

| **Field name** | **signification** | size |
|---|---|---|
| order | orders of the diffracted propagative plane waves | N, 1 |
| theta | angle $\theta_{m,n}$ of every diffracted order | N, 1 |
| delta | angle $\delta_{m,n}$ of every diffracted order | N, 1 |
| **K** | normalized wave vector | N, 3 |
| efficiency | efficiency in each order | N, 1 |
| efficiency_TE | efficiency in TE polarization in every order | N, 1 |
| efficiency_TM | efficiency in TM polarization in every order | N, 1 |
| amplitude_TE | complexe amplitude in TE polarization in every order | N, 1 |
| amplitude_TM | complexe amplitude in TM polarization in every order | N, 1 |
| **E** | electric field $(E_x,E_y,E_z)$ of the diffracted orders at O_top or O_bottom when the amplitude of the incident plane wave is one. | N, 3 |
| **H** | magnetic field $(H_x,H_y,H_z)$ of the diffracted orders at O_top or O_bottom when the amplitude of the incident plane wave is one. | N, 3 |
| **PlaneWave_TE_E** | **E**-vector components of the TE-polarized $\overrightarrow{PW}$ 's (in the Oxyz basis) | N, 3 |
| **PlaneWave_TE_H** | **H**-vector components of the TE-polarized $\overrightarrow{PW}$ 's (in the Oxyz basis) | N, 3 |
| **PlaneWave_TE_Eu** | **E**-vector components of the TE-polarized $\overrightarrow{PW}$ 's (in the $\mathbf{u}_{TM}$ $\mathbf{u}_{TE}$ basis) | N, 2 |
| **PlaneWave_TE_Hu** | **H**-vector components of the TE-polarized $\overrightarrow{PW}$ 's (in the $\mathbf{u}_{TM}$ $\mathbf{u}_{TE}$ basis) | N, 2 |
| **PlaneWave_TM_E** | **E**-vector components of the TM-polarized $\overrightarrow{PW}$ 's (in the Oxyz basis) | N, 3 |
| **PlaneWave_TM_H** | **H**-vector components of the TM-polarized $\overrightarrow{PW}$ 's (in the Oxyz basis) | N, 3 |
| **PlaneWave_TM_Eu** | **E**-vector components of the TM-polarized $\overrightarrow{PW}$ 's (in the $\mathbf{u}_{TM}$ $\mathbf{u}_{TE}$ basis) | N, 2 |
| **PlaneWave_TM_Hu** | **H**-vector components of the TM-polarized $\overrightarrow{PW}$ 's (in the $\mathbf{u}_{TM}$ $\mathbf{u}_{TE}$ basis) | N, 2 |

## 6.1. Efficiency

For a given diffraction order (m,n), the diffraction efficiency is defined as the ratio between the flux of the diffracted Poynting vector and the flux of the incident Poynting vector (flux through a period of the grating). The total diffraction efficiency is equal to efficiency = efficiency_TE + efficiency_TM.

The efficiencies of all propagative reflected and transmitted plane waves (for a TE-polarized plane wave incident from the top of the grating) are given by the two vectors "**result.TEinc_top_reflected.efficiency**" and "**result.TEinc_top_transmitted.efficiency**". If all refractive indices are real, the sum of all elements of these two vectors is equal to one because of the energy conservation. The label "(m,n)" of the corresponding orders are found in "**result.TEinc_top_reflected.order**" (see below for a description of the other fields of this sub-structure).

For example, if the desired diffracted order is evanescent for the wavelength or the incidence angle considered, the result returned is 0.

Some examples

1) The TE-efficiency of the reflected order (m=-3, n=4) for an illumination from the top under TM polarisation is equal to **efficiency_TE**=**result.TMinc_top_reflected.efficency_TE{-3,4}**. If this order is evanescent, the efficiency is equal to zero.
The total efficiency (TE+TM) in this order is **result.TMinc_top_reflected.efficiency{-3,4}**,

2) The N propagative orders of the transmitted plane waves for an incident wave from the top of the grating in TE polarization are given by the vector of size (N,2) "**result.TEinc_top_transmitted.order**".

3) The efficiencies of all propagative reflected waves for an incident wave from the bottom in TM polarization are given by the vector of size (N,2) "**result.TMinc_bottom_reflected.efficiency**".

3) The efficiencies of all propagative reflected and transmitted waves for an incident wave from the top of the grating in TE polarization are given by the two vectors "**result.TEinc_top_reflected.efficiency**" and "**result.TEinc_top_transmitted.efficiency**". If all refractive indices are real, the sum of all the elements of these two vectors is equal to one because of energy conservation.

## 6.2. Rayleigh expansion for propagatives modes

The coefficients of the Rayleigh expansion of Fig. 1 can be obtained from the structure **result**. For instance, when the grating is illuminated from the bottom with a TE polarised mode, we have :

$\mathbf{E}^{m}_{\mathrm{bottom}}$=result.TEinc_bottom_reflected.E{m} (3 components in Oxyz)

$\mathbf{H}^{m}_{\mathrm{bottom}}$=result.TEinc_bottom_reflected.H{m} (3 components in Oxyz)

$\mathbf{E}^{m}_{\mathrm{top}}$=result.TEinc_bottom_transmitted.E{m} (3 components in Oxyz)

$\mathbf{H}^{m}_{\mathrm{top}}$=result.TEinc_bottom_ transmitted.H{m} (3 components in Oxyz)

and the incident plane wave defined in page 4 is given by :

$\mathbf{E}^{\mathrm{inc}}_{\mathrm{bottom}}$=result.TEinc_bottom.E (3 components in Oxyz)

$\mathbf{H}^{\mathrm{inc}}_{\mathrm{bottom}}$=result.TEinc_bottom.H (3 components in Oxyz).

## 6.3. Diffracted amplitudes of propagative waves

### 6.3.1 $U_{TE}$, $U_{TM}$, θ, δ and K

Figure 7 defines the geometry of the diffracted order m, for a diffracted wave in the top layer and for a diffracted wave in the bottom layer. The wave vector $k_{m,n} = (2\pi/\lambda)\, n_{top}\, \mathbf{K}_{m,n}$ (or $(2\pi/\lambda)\, n_{bottom}\, \mathbf{K}_{m,n}$) of the (m,n)th diffracted order is defined by the two angles $\theta_{m,n}$ and $\delta_{m,n}$. As for the incident wave, the angle $\delta_{m,n}$ defines the plane of diffraction. The angle $\theta_{m,n}$ varies between 0° and 90°, and the angle $\delta_{m,n}$ varies between 0° and 360°. The relations linking the Cartesian components of the unitary vector $\mathbf{K}_{m,n}$ and the angles $\theta_{m,n}$ and $\delta_{m,n}$ are the same as the relations defined previously for the incident plane wave (Section 3) :

$\mathbf{K}_{m,n} = [\sin(\theta_{m,n})\cos(\delta_{m,n}), \sin(\theta_{m,n})\sin(\delta_{m,n}), -\cos(\theta_{m})]$

The unitary vector $\vec{\mathbf{u}}_{TE}$ is perpendicular to the plane of diffraction and is oriented such that ($\mathbf{K}_{m,n}$, $\vec{\mathbf{u}}_{TE}$, $\mathbf{z}$) is direct. The unitary vector $\vec{\mathbf{u}}_{TM}$ is defined by $\vec{\mathbf{u}}_{TM} = \vec{\mathbf{u}}_{TE} \wedge \mathbf{K}_{m,n}$. So the base ($\vec{\mathbf{u}}_{TM}$, $\vec{\mathbf{u}}_{TE}$, $\mathbf{K}_{m,n}$) is direct. If the diffracted electric field is parallel to $\vec{\mathbf{u}}_{TE}$, then the order (m,n) is TE polarized, and if the diffracted electric field is parallel to $\vec{\mathbf{u}}_{TM}$, then it is TM polarized. In general, the diffracted electric field of order (m,n) has a non-zero component along both directions.

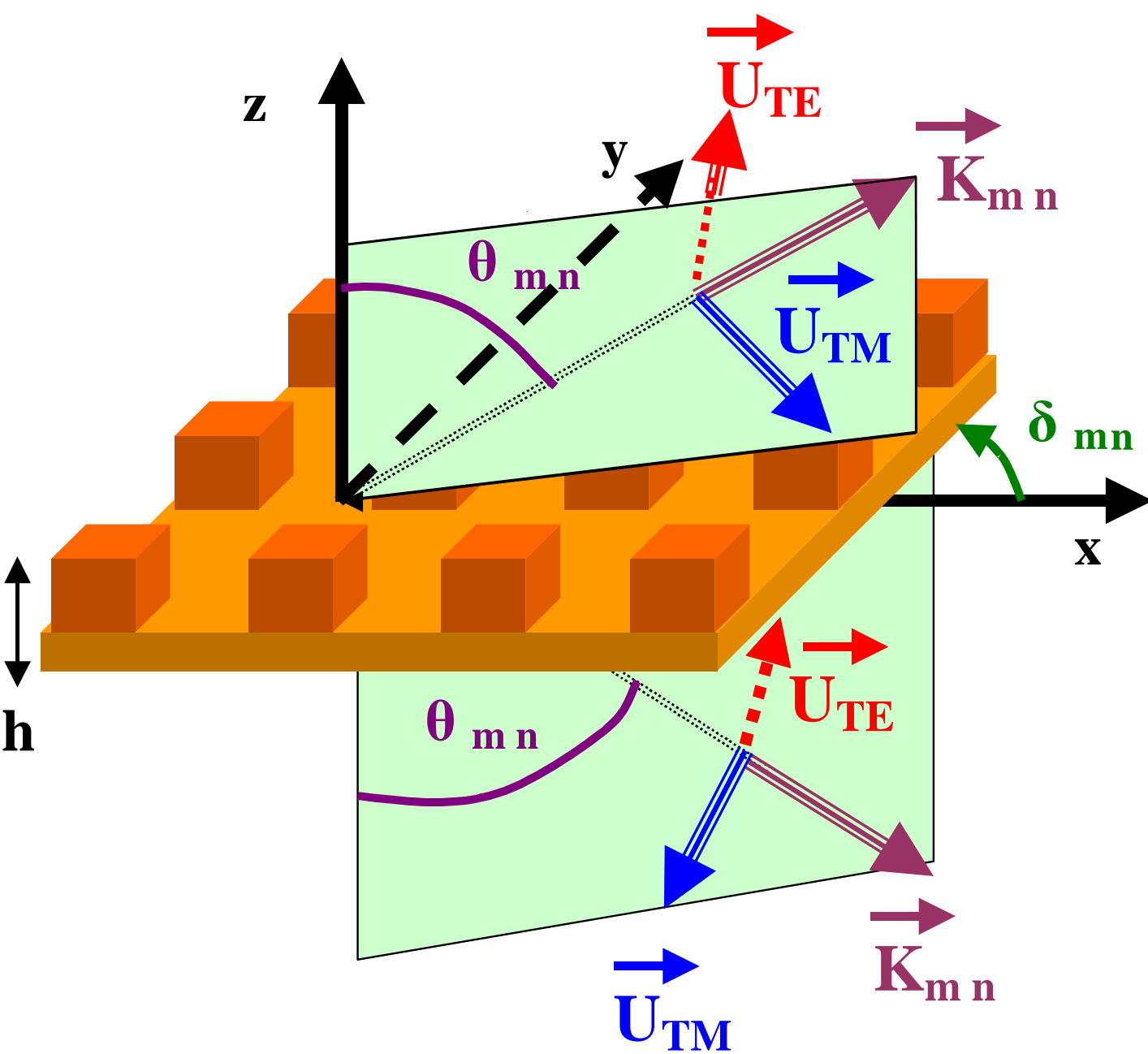

Fig. 7. Definition of the $\mathbf{U}_{TE}$,$\mathbf{U}_{TM}$,$\mathbf{K}_{m,n}$,$\theta_{m,n}$ and $\delta_{m,n}$ for a specific diffracted order (m,n).

### 6.3.2 $O_{top}$ and $O_{bottom}$ points

$O_{top}$ and $O_{top}$ are 2 important points (see Fig. 1). In the Cartesian coordinates system Oxyz , they are defined by : $O_{top}$=(0,0,h) at the top of the grating, and $O_{bottom}$=(0,0,0) at the bottom of the grating.

In addition, let us consider an arbitrary point M=(x,y,z) in the 3D space in Oxyz. Associated to this point, we define the two vectors :

$\mathbf{r}_{top} = \overrightarrow{O_{top}M}$, and

$\mathbf{r}_{bottom} = \overrightarrow{O_{bottom}M}$ .

### 6.3.3 Jones' matrix

Let us assume that the grating is illuminated from the top layer and let us consider a diffracted order m in the bottom layer. Any other diffraction situation is straightforwardly deduced.

α and β being two given complex numbers, the incident electromagnetic field (6 components of **E** and **H** in every points of the 3D space) can be written :

$$\mathbf{W}^{inc} = \alpha\, \overrightarrow{PW_{TE}} + \beta\, \overrightarrow{PW_{TM}} \;,$$

where $\overrightarrow{PW_{TE}}$ is a TE-polarized plane wave defined in every point by $\overrightarrow{PW_{TE}} = \mathbf{A}_{TE} \exp\left(i\mathbf{k}_{top}^{inc}\, \mathbf{r}_{top}\right)$, and $\overrightarrow{PW_{TM}}$ a TM polarized plane wave defined in the same way by $\overrightarrow{PW_{TM}} = \mathbf{A}_{TM} \exp\left(i\mathbf{k}_{top}^{inc}\, \mathbf{r}_{top}\right)$, $\mathbf{A}_{TE}$ and $\mathbf{A}_{TM}$ being the electromagnetic fields (6 components) of the plane wave at M=$O_{top}$. $\mathbf{k}_{top}^{inc}$ is the incident wave vector. $\mathbf{A}_{TE}$ and $\mathbf{A}_{TM}$ and $\mathbf{K} = \mathbf{k}_{top}^{inc} / \left|\mathbf{k}_{top}^{inc}\right|$ are given by the structure "**result**" as will be defined later.

Similarly, the diffracted electromagnetic field in the mth bottom order can be written :

$$\mathbf{W}_{m,n}^{dif} = \gamma\, \overrightarrow{PW_{TE}^{m,n}} + \mu \overrightarrow{PW_{TM}^{m,n}} \;,$$

where $\gamma$ and $\mu$ are complex numbers, $\overrightarrow{PW_{TE}^{m,n}}$ is a TE-polarized plane wave defined in every point by $\overrightarrow{PW_{TE}^{m,n}} = \mathbf{A}_{TE}^{m,n} \exp\left(i\mathbf{k}_{bottom}^{m,n}\, \mathbf{r}_{bottom}\right)$, and $\overrightarrow{PW_{TM}^{m,n}}$ a TM-polarized plane wave defined in the same way by $\overrightarrow{PW_{TM}^{m,n}} = \mathbf{A}_{TM}^{m,n} \exp\left(i\mathbf{k}_{bottom}^{m,n}\, \mathbf{r}_{bottom}\right)$. $\mathbf{A}_{TE}^{m,n}$ and $\mathbf{A}_{TM}^{m,n}$ are the electromagnetic fields (6 components) of the plane wave at $M=O_{bottom}$, and $\mathbf{k}_{bottom}^{m}$ is the wave vector of the m transmitted order. $\mathbf{A}_{TE}^{m,n}$ and $\mathbf{A}_{TM}^{m,n}$ and $\mathbf{K}^{m,n} = \mathbf{k}_{bottom}^{m,n} / \left|\mathbf{k}_{bottom}^{m,n}\right|$ are given by the structure "**result**" as will be defined later.

We define the (4x4) Jones' matrix **J**, associated to the order m by :

$$\begin{pmatrix} \gamma \\ \mu \end{pmatrix} = \begin{pmatrix} J_{EE} & J_{ME} \\ J_{EM} & J_{MM} \end{pmatrix} \begin{pmatrix} \alpha \\ \beta \end{pmatrix}.$$

$J_{EE}$, $J_{EM}$,$J_{ME}$, $J_{MM}$ and **J** are all given by the structure "**result**".
The $\mathbf{A}_{TE}$ $\mathbf{A}_{TE}^{m,n}$ $\mathbf{A}_{TM}$ $\mathbf{A}_{TM}^{m,n}$ vectors are normalized so that the $|J_{EE}|^2$, $|J_{EM}|^2$,$|J_{ME}|^2$ and $|J_{MM}|^2$ represent diffraction efficiencies. For instance, $|J_{ME}|^2$ =**result.TMinc_top_transmitted.efficency_TE{m,n}**.

We now define all these data from the "**result**" structure :
**K** = result.TEinc_top.K or K=result.TMinc_top.K.
$\mathbf{K}^{m,n}$ = result.TEinc_top_transmitted.K{m,n} = result.TMinc_top_transmitted.K{m,n}. Note that if some symmetries are used for the calculation, "result.TEinc_top_transmitted.K{m,n}" or "result.TMinc_top_transmitted.K{m,n}" can be an empty vector.

The $\mathbf{A}_{TE}^{m,n}$ 's coefficients can be obtained either in the Cartesian coordinate system or in the ($\vec{\mathbf{u}}_{TM}$ ,$\vec{\mathbf{u}}_{TE}$ ) basis.
In the Cartesian coordinate system Oxyz :

$$\mathbf{A}_{TE} = \begin{pmatrix} \text{result.TEinc_top.PlaneWave_TE_E} \\ \text{result.TEinc_top.PlaneWave_TE_H} \end{pmatrix}$$

$$\mathbf{A}_{TM} = \begin{pmatrix} \text{result.TMinc_top.PlaneWave_TM_E} \\ \text{result.TMinc_top.PlaneWave_TM_H} \end{pmatrix}$$

$$\mathbf{A}_{TE}^{m,n} = \begin{pmatrix} \text{result.TEinc_top_transmitted.PlaneWave_TE_E}\{m,n\} \\ \text{result.TEinc_top_transmitted.PlaneWave_TE_H}\{m,n\} \end{pmatrix}$$

$$= \begin{pmatrix} \text{result.TMinc_top_transmitted.PlaneWave_TE_E}\{m,n\} \\ \text{result.TMinc_top_transmitted.PlaneWave_TE_H}\{m,n\} \end{pmatrix}.$$ (same remark as for $\mathbf{K}^{m,n}$)

$$\mathbf{A}_{TM}^{m,n} = \begin{pmatrix} \text{result.TEinc_top_transmitted.PlaneWave_TE_E}\{m,n\} \\ \text{result.TEinc_top_transmitted.PlaneWave_TE_H}\{m,n\} \end{pmatrix}$$

$$= \begin{pmatrix} \text{result.TMinc_top_transmitted.PlaneWave_TE_E}\{m,n\} \\ \text{result.TMinc_top_transmitted.PlaneWave_TE_H}\{m,n\} \end{pmatrix}.$$ (same remark as for $\mathbf{K}^{m,n}$)

In the ($\vec{\mathbf{u}}_{TM}$ ,$\vec{\mathbf{u}}_{TE}$ ) basis (with only 2 components for each fields E and H) :

$$\mathbf{A}_{TE} = \begin{pmatrix} \text{result.TEinc_top.PlaneWave_TE_Eu} \\ \text{result.TEinc_top.PlaneWave_TE_Hu} \end{pmatrix}$$

$$\mathbf{A}_{TM} = \begin{pmatrix} \text{result.TMinc_top.PlaneWave_TM_Eu} \\ \text{result.TMinc_top.PlaneWave_TM_Hu} \end{pmatrix}$$

$$\mathbf{A}_{TE}^{m,n} = \begin{pmatrix} \text{result.TEinc_top_transmitted.PlaneWave_TE_Eu}\{m,n\} \\ \text{result.TEinc_top_transmitted.PlaneWave_TE_Hu}\{m,n\} \end{pmatrix}$$

$$= \begin{pmatrix} \text{result.TMinc_top_transmitted.PlaneWave_TE_Eu}\{m,n\} \\ \text{result.TMinc_top_transmitted.PlaneWave_TE_Hu}\{m,n\} \end{pmatrix}. \text{ (same remark as for } \mathbf{K}^{m,n}\text{)}$$

$$\mathbf{A}_{TM}^{m,n} = \begin{pmatrix} \text{result.TEinc_top_transmitted.PlaneWave_TE_Eu}\{m,n\} \\ \text{result.TEinc_top_transmitted.PlaneWave_TE_Hu}\{m,n\} \end{pmatrix}$$

$$= \begin{pmatrix} \text{result.TMinc_top_transmitted.PlaneWave_TE_Eu}\{m,n\} \\ \text{result.TMinc_top_transmitted.PlaneWave_TE_Hu}\{m,n\} \end{pmatrix}. \text{ (same remark as for } \mathbf{K}^{m,n}\text{)}$$

The Jones' coefficients are :
$J_{EE}$ =result.TEinc_top_transmitted.amplitude_TE{m,n}
$J_{EM}$ =result.TEinc_top_transmitted.amplitude_TM{m,n}
$J_{ME}$ =result.TMinc_top_transmitted.amplitude_TE{m,n}
$J_{MM}$ =result.TMinc_top_transmitted.amplitude_TM{m,n}

And the Jones' matrix is :

$$\mathbf{J} = \begin{pmatrix} J_{EE} & J_{ME} \\ J_{EM} & J_{MM} \end{pmatrix} = \text{result.Jones.inc_top_transmitted}\{m,n\}.$$

# 7. Using symmetries to accelerate the computational speed

This is very important for 2D gratings : one has to use symmetry as much as possible. Typical acceleration rate improvements by use of symmetry can be found in the beginning of Section 4 in "Ph. Lalanne, J. Opt. Soc. Am. A **14**, 1592-1598 (1997)". It has to be understood that symmetries can be used only when the illumination **and** the grating structure possess some mirror symmetries for the plane $x = x_0$ and/or $y = y_0$.

To use the symmetry, the user needs to define three new parameters: "**parm.sym.x**", "**parm.sym.y**" and "**parm.sym.pol**".
"**parm.sym.x**" defines the position of the mirror symmetry plane in the x-direction (if exist)
"**parm.sym.y**" defines the position of the mirror symmetry plane in the y-direction (if exist)

**If** the illumination possesses the same symmetry as the grating (for details see the table below),. RETICOLO-2D will perform the calculation for a single polarisation, according to :
**parm.sym.pol = 1** ; % TE polarization
**parm.sym.pol = -1** ; % TM polarization
In this case, the fields or structure "**result**" corresponding to the other polarization will be empty. If one wants to obtain the result for both polarizations, two independent calculations have to be executed. This is preferable to performing the calculation for both polarizations at the same time without using symmetries.

**If** the illumination does not possess the same symmetry as the grating, RETICOLO-2D will perform the calculation for both polarisations without using the symmetry.

Note that the code does not verify if the grating symmetries defined by the user are in agreement with all the texture symmetries. It is up to the user to define carefully the parameters parm.sym.x and parm.sym.y.

Using symmetries is not difficult but, in order to check, it is recommended to first execute the code with a small number of retained Fourier harmonics without using symmetry, then to re-execute the code with the same number of retained harmonics using symmetries. The calculated efficiencies must be identical.

Depending on the values of the angles $\theta$ and $\delta$, the computation will be done using full symmetry (two mirror plans), one symmetry (one mirror plane) or no symmetry. In particular, it is important to remember that **the angle $\delta$ can take only 4 values** if one wants to use symmetry : $\delta = 0°$, 90°, 180° and 270°. The following table recapitulates the conditions on the angles $\theta$ and $\delta$ for using symmetries, and the associated values of "**parm.sym.x**" and "**parm.sym.y**".

| **parm.sym.x** | **parm.sym.y** | **$\delta$** | **$\theta$** | **symmetries used by RETICOLO** |
|---|---|---|---|---|
| $x_0$ | $y_0$ | 0° | 0° | **full symmetry** (plane $x = x_0$ and plane $y = y_0$) |
| | | 90° | | |
| | | 180° | | |
| | | 270° | | |
| $x_0$ | $y_0$ | 0° | $\neq 0°$ | **one symmetry** (plane $y = y_0$) |
| | | 90° | | **one symmetry** (plane $x = x_0$) |
| | | 180° | | **one symmetry** (plane $y = y_0$) |
| | | 270° | | **one symmetry** (plane $x = x_0$) |
| $x_0$ | [] | 0° | any value | **no symmetry** |
| | | 90° | | **one symmetry** (plane $x = x_0$) |
| | | 180° | | **no symmetry** |
| | | 270° | | **one symmetry** (plane $x = x_0$) |
| [] | $y_0$ | 0° | any value | **one symmetry** (plane $y = y_0$) |
| | | 90° | | **no symmetry** |
| | | 180° | | **one symmetry** (plane $y = y_0$) |
| | | 270° | | **no symmetry** |
| [] | [] | any value | any value | **no symmetry** |

Remember that the direction of the incident electric field is defined by the polarization and the angle $\delta$. The next table recapitulates the directions of the incident electric field depending on the polarization and the angle $\delta$.

| $\delta$ | 0° | 90° | 180° | 270° |
|---|---|---|---|---|
| TE (parm.sym.pol = 1) | E parallel to y | E parallel to x | E parallel to y | E parallel to x |
| TM (parm.sym.pol = -1) | E parallel to the (x,z) plane | E parallel to the (y,z) plane | E parallel to the (x,z) plane | E parallel to the (y,z) plane |

# 8. Plotting the electromagnetic field and calculating the absorption loss

## 8.1. Computation of the electromagnetic fields

Once the eigenmodes associated to all textures are known, the calculation of the electromagnetic fields everywhere in the grating can be performed. This calculation is done by the subroutine "**res3.m**", following the instruction

**[e,z,index] = res3(x,y,aa,Profile,einc,parm);**

The function"res3.m" can be called without calling "res2.m". This subroutine has 6 input arguments:
-the "**x**" variable is a vector containing the locations where the fields will be calculated in the x-direction. For instance, we may set **x = linspace(-period_x/2, period_x/2, 51);** for allocating 51 sampling points in the x-direction,
-the "**y**" variable is a vector containing the locations where the fields will be calculated in the y-direction. For instance, we may set **y = linspace(-period_y/2, period_y/2, 51);** for allocating 51 sampling points in the y-direction,
-the "**aa**" variable contains all the information on the eigenmodes of all textures and is computed by the subroutine res1.m,

-the variable “**Profile**” is defined in Section 4.2. Note that it can be redefined, note also that the “repetition” trick of Eq. (2) cannot be used,
-the variable “**einc**” defines the complex amplitude of the *incident electric* field at O_top or O_bottom in the basis {$\mathbf{u}_{TM}$, $\mathbf{u}_{TE}$}. For instance, setting einc=[1,0] means that one is looking for TM polarization, and setting einc=[1,1]/sqrt(2) means that one is looking for a 45° polarization.

For illuminating the grating exactly by the TE-polarized incident $\overrightarrow{PW_{TE}}$ defined above, one should set: einc= **result.TEinc_top.PlaneWave_TE_Eu.**

If symmetry arguments have been used previously, note that the calculation with res1.m is provided only for some specific polarization ; it would be a nonsense to specify another polarization for the field plots (in this case the corresponding component of einc is taken as 0).
-the “**parm**” variable, already mentioned is discussed hereafter.

There are three possible output arguments for the subroutine “**res3.m**” :
-The “**e**” argument contains all the electromagnetic field quantities:

$E_x$=**e**(:,:,1),$E_y$=**e**(:,:,2),$E_z$=**e**(:,:,3),$H_x$=**e**(:,:,4),$H_y$=**e**(:,:,5),$H_z$=**e**(:,:,6).

-The second argument “**z**” is the vector containing the z-coordinate of the sampling points. Note that in the matrix $E_x$=e(:,:,:,1), the first index refer to the z-coordinate, the second to the x-coordinate and the third to the y-coordinate. Thus $E_x$(i,j,k) is the $E_x$ field-component at the location {z(i), x(j), y(k)}.
-The third argument **index**(i,j,k) is the complex refractive index at the location {**z**(i), **x**(j), **y**(k)}. It can be useful to check the profile of the grating.

Some **important** comments on the **parm**” argument:
1. For calculating precisely the electromagnetics fields, one has to set: ”**parm.res1.champ=1**” before calling **res1.m**. This increases the calculation time and memory load but it is highly recommended. If not, the computation of the field will be correct only in homogenous textures (for example in the top layer and in the bottom layer).

2. Illuminating the grating from the top or the bottom layer : As mentioned earlier, the code compute the diffraction efficiencies of the transmitted and reflected orders for an incident plane wave from the top and for an incident plane wave from the bottom at the same time. When plotting the field, the user must specify the direction of the incident plane wave. This is specified with variable **parm.res3.sens**. For **parm.res3.sens=1**, the grating is illuminated form the top and **parm.res3.sens=-1**, the grating is illuminated form the bottom (default is **parm.res3.sens=1**).

3. Specifying the z locations of the computed fields: This is provided by the variable **parm.res3.npts**. **parm.res3.npts** is a vector whose length is equal to the length of the variable **Profile{1}**. For instance let us imagine, a two-layer grating defined by **Profile** = {[0.5,1,2,0.6],[1,2,3,4]}. Setting **parm.res3.npts=[2,3,4,5]** implies that the field will be computed in two z=constant plans in the top layer, in three z=constant plans in the first layer (texture 2), in four z=constant plans in the second layer (texture 3), and in five z=constant plans in the bottom layer. Default for **parm.res3.npts** is 10 z=constant plan per layer.

**VERY IMPORTANT**: where is the z=0 plan and what are the z-coordinates of the z=constant plan? The z=0 plan is defined at the bottom of the bottom layer. Thus, the field calculation is performed only for z>0 values. For the example **Profile** = {[0.5,1,2,0.6],[1,2,3,4]}, and if we refer to texture 4 as the substrate, the z=0 plan is located in the substrate at a distance 0.6 under the grating. The z=constant plans are located by an equidistant sampling in every layer. Always referring to the previous example, it implies that the five z=constant plans in the substrate are located at coordinate z=(p−0.5) 0.6/5, where p=1,2, …5. Note that the z coordinate for z=constant plan are always given by the second output variable of res3.m.

4. How can one specify a given z=constant plan? First, one has to redefine the variable **Profile**. For the grating example with the two layers discussed above, let us imagine that one wants to plot the field at z=z0+0.6+0.2 in layer 2. Then one has to set: **Profile** = {[0.5,1-z0,0,z0,0.2,0.6],[1,2,2,2,3,4]} and set **parm.res3.npts=[0,0,1,0,0,0]**. Note that it is not necessary to redefine the variable **Profile** at the beginning of the program. One just needs to redefine this variable before calling subroutine res3.m.

5. Automatic plots: an automatic plot (showing all the components of the electromagnetic fields and the grating refractive index distribution) is provided by setting **parm.res3.trace**=1. If one wants to plot only some components of the fields, one can set for instance: **parm.res3.champs**=[2,3,6,0], to plot $E_y$, $E_z$, $H_z$ and the object, **parm.res3.champs**=[1] to plot only $E_x$. Take care that automatic plots are only available when one of the variables x, y or z is of length 1 (field-distribution plots are available in a plane, not in a volume).

## 8.2. Computation of the absorption loss

The loss calculation is done with the subroutine "**res3.m**".

First approach based on integrals (not valid for homogeneous layers with non-diagonal anisotropy):
The absorption loss in a surface $S$ is given by:
$L = \frac{\pi}{\lambda}\int_S Im\ (\varepsilon_{XX}(M)|E_X(M)|^2 + \varepsilon_{YY}(M)|E_Y(M)|^2 + \varepsilon_{ZZ}(M)|E_z(M)|^2)\ dV.$
The integral can be computed with the following instruction

[**e, Z, index, wZ, loss_per_layer, loss_of_Z, loss_of_Z_X, X, wX**] = res3(**x,aa,Profile,einc,parm**);

The important ouput arguments are:
**loss_per_layer**: the loss in every layer defined by **Profile**, **loss_per_layer**(1) is the loss in the top layer, **loss_per_layer**(2) the loss in layer 2, ... and **loss_per_layer**(end) the loss in the bottom layer
**loss_of_Z**: the absorption loss density (integrated over **X**) as a function of **Z** (like for **X**, the sampling points **Z** are not equidistant. You may plot this loss density as follows : plot(**Z**, **loss_of_Z**), xlabel('Z'), ylabel('absorption')
**loss_of_Z_X**(**Z**,**X**) = π/λ Im(**index**(**Z**,**X**).^2) (|**e**(**Z**,**X**,**1**)|$^2$+|**e**(**Z**,**X**,**2**)|$^2$+|**e**(**Z**,**X**,**3**)|$^2$)
**index**: **index**(i,j) is the complex refractive index at the location {**z**(i), **x**(j)}.
An alternative to calculate the losses in the slices is to calculate the difference in the flux of the incoming and outgoing Poynting vector. This method is faster, but in some cases the calculation of the integral can be more precise. In homogeneous layers with non-diagonal anisotropy, only this method is possible

Second approach based on Poynting theorem (always valid even for homogeneous layers with non-diagonal anisotropy):
An alternative approach to compute the losses in the layers consists in calculating the difference in the flux of the incoming and outgoing Poynting vectors. This approach is faster, but in some cases, the computation of the integral can be more accurate. In homogeneous layers with non-diagonal anisotropy, only this approach is possible.

To specify which approach used per layer, we define a vector

**parm.res3.pertes_poynting** = [0,0,0,1,0]; % for instance for a 5-layer grating with "0", the integral approach is used (default option) and with "1", the Poynting approach is used. The length of **parm.res3.pertes_poynting** is equal to the number of layers. We may set **parm.res3.pertes_poynting** = 0 or 1; the scalar is then repeated for all layers.

We may then compute the flux of the Poynting vector in the layer-boundary planes

[**e, Z, index, wZ,loss_per_layer,loss_of_Z,loss_of_Z_X,X,wX,Flux_Poynting**] = res3(**x,aa,Profile,einc,parm**);

**Flux_Poynting** is a vector. **Flux_Poynting(1)** corresponds to the upper interface of the top layer. The flux is computed for a normal vector equal to the $\hat{\mathbf{z}}$ vector. If **Flux_Poynting(p)** > 0, the energy flows toward the top and, if it is negative, the energy flows toward the bottom.

For an illumination from the top and a lossy substrate, the substrate absorption is **–Flux_Poynting (end)/(0.5*prod(period))**. For an illumination from the bottom and a lossy superstrate, the superstrate absorption is **Flux_Poynting (1)/(0.5*prod(period))**

Note on the computation accuracy of the integral approach:
To compute volume integrals like the loss or the electromagnetic energy, we use a Gauss-Legendre integration method. This method, which is very powerful for 'regular' functions, becomes inaccurate for discontinuous functions. Thus, the integration domain should be divided into subdomains where the electric field **E** is continuous. For the integration in **X** and **Y**, this difficult task is performed by the program, so that the user should only define the limits of integration: the input "**x**" and "**y**" arguments are now vectors of length 2, which represent the limits of the x and y intervals (to compute the loss over the entire period, we may take **x**(2)=**x**(1)+**period_x**, **y**(2)=**y**(1)+**period_y**. The integration domain is then divided into subintervals where the permittivity is continuous, each subinterval having a length less than λ/(2π). For every subinterval, a Gauss-Legendre integration method of degree 10 is used. This default value can be changed by setting **parm.res3.gauss_x**=..., **parm.res3.gauss_y=...**. The actual points of computation of the field are returned in the output arguments **X** and **Y**.

For the z integration, the discontinuity points are more easily determined by the variable **'Profile'**. The user may choose the number of subintervals and the degree in every layer using the parameter **parm.res3.npts**, which is now an array with two lines (in subsection 8.1 this variable is a line vector): the first line defines the degree and the second line the numbers of subintervals of every layer. For example: **parm.res3.npts** = [ [10 , 0 , 12 ] ; [3 , 1 , 5 ] ]; means that 3 subintervals with 10-degree points are used in the first layer, 1 subintervals with 0 point in the second layer, 5 subintervals with 12degree points in the third layer.

The actual z-points of computation of the field are returned in the output variable **Z**, and the vector **wZ** represents the weights and we have sum(**loss_of_Z**.***wZ**)=sum(**loss_per_layer**). Although the maximum degree that can be handled by reticolo is 1000, it is recommended to limit the degree values to modest numbers (10-30 maximum) and to increase the number of subintervals (the larger the degree, the denser the sampling points in the vicinity of the subinterval boundaries).

Note that if **einc**= **result.TEinc_top PlaneWave_TE_Eu**, the energie conservation test for a TE incident plane wave from the top is

sum(**result.TEinc_top_reflected.efficiency**)+
sum(**result.TEinc_top_transmitted.efficiency**)+
sum(**loss_per_layer**) / (.5***period_x*** **period_y** ) = 1.

Usually, this equality is achieved with an absolute error $< 10^{-5}$.

For specialists:
-**loss_of_Z_X_Y** =pi/ **wavelength***imag(**index**.^2).*sum(abs(**e**(:,:,:,1:3)).^2,4);
-**loss_of_Z** =(reshape(**loss_of_Z_X_Y**,length(**Z**),[])***wXY**(:)).';
-by setting **index**(**index** ~= index_chosen)=0 in the previous formulas, one may calculate the absorption loss in the medium of refractive index index_chosen.

# 9. Bloch-mode effective indices

RETICOLO gives access to another output: the Bloch mode associated to all textures. The Bloch mode *k* of the texture *l* can be written

$$|\Phi_k{}^l\rangle = \sum_{m,n} a_{m,n}^{k,l}\, exp[i(k_x^{inc} + mK_x)x]\, exp\left(i\left(k_y^{inc} + nK_y\right)y\right)\, exp\left(i\frac{2\pi}{\lambda} n_{eff}^{k,l} z\right),$$

where $n_{eff}^{k,l}$ is the effective index of the Bloch mode *k* of the texture *l*.

Instruction:
**[aa, n_eff] = res1(wavelength, period, textures, nn, kparallel, delta0, parm)**;
Note that the "n_eff" variable is a Matlab cell array: "**n_eff{ii}**" is a column vector containing all the Bloch-mode effective indices associated to the texture "**textures{ii}**". The element number 5 of this vector, for example, is called by the instruction "**n_eff{ii}(5)**;". An attenuated Bloch-mode has a complex effective index.

Bloch mode profile visualization:
To plot the profile of Bloch mode Num_mode of the texture Num_texture:

**res1(aa, neff, Num_texture, Num_mode)**;

To obtain the profile datas in the format given by res3:

**[e,o] = res1(aa, neff, Num_texture, Num_mode)**; % by default, for |x| < period(1)/2 and |y| < period(2)/2

**[e,o] = res1(aa, neff, Num_texture, Num_mode, x, y)**; % by specifying the x,y vectors, x=linspace(0,3*period(1),100) and y=linspace(3,3+period(2),100) for example.

# 10. Annex

## 10.1. Checking that the textures are correctly set up

Setting "**parm.res1.trace = 1**;" generates a Matlab figure which represents the refractive-index distribution of all the textures.

## 10.2. The "retio" instruction

RETICOLO automatically creates temporary files in order to save memory. These temporary files are of the form "abcd0.mat", "abcd1.mat" … with abcd randomly chosen). They are created in the current directory. In general RETICOLO automatically erases these files when they are no longer needed, but it is recommended to finish all programs by the instruction "**retio**;", which erases all temporary files. Also, if a program anormally stopsone may execute the instruction "**retio**" before restarting the program.

The “retefface” instruction allows to know all the “abcd0.mat” files and to erase them if wanted.

If we are not limited by memory (this is often the case with modern computers), we can prevent the writing of intermediate files on the hard disk by the setting

parm.not_io = 1;

before the call to res1. Then it is no longer necessary to use the retio instruction at the end of the programs to erase the files.
IMPORTANT: to use parfor loops, it is imperative to take the option parm.not_io = 1.

## 10.3. How to save and to reload the “aa” variable

To save the “**aa**” variable in a “.mat” file, the user has to define a new parameter containing the name of the file he or she wants to create : “**parm.res1.fperm = 'file_name'**;”. field_name is a char string with at least one letter. The program will automatically save “**aa**” in the file “**file_name.mat**”. In a new utilisation it is sufficient to write aa = **'file_name'**;.

Example of a program which calculates and saves the “aa” variable
[...] % Definition of the input parameters, see Section 3
**parm.res1.fperm = 'toto'**;
[...] % Definition of the textures, see Section 4.1
**aa = res1(wavelength,period,textures,nn,k_parallel,angle_delta,parm)**;
Example of a program which uses the “aa” variable and then calculates the diffracted waves
[...] % Definition of the Profile, see Section 4.2. Note that the textures used to define the Profile argument have to correspond to the textures defined in the program which has previously calculated the “aa” variable.
**aa='toto'**;
**result = res2(aa,Profile)**;
**retio**;

## 10.4. Asymmetry of the Fourier harmonics retained in the computation

**nn = [[-3,-2];[2,4]]**;
This defines the set of non-symmetric Fourier harmonics retained for the computation. In this case, the Fourier harmonics from –3 to +2 are retained in the x-direction, and the Fourier harmonics from –2 to +4 are retained in the y-direction.
The instructions “**nn = [3,2]**;” and “**nn = [[-3,-2];[3,2]]**;” are equivalent.
Take care that the use of symmetry imposes symmetric Fourier harmonics, if not the computation will be done without any symmetry consideration.

# 11. Summary

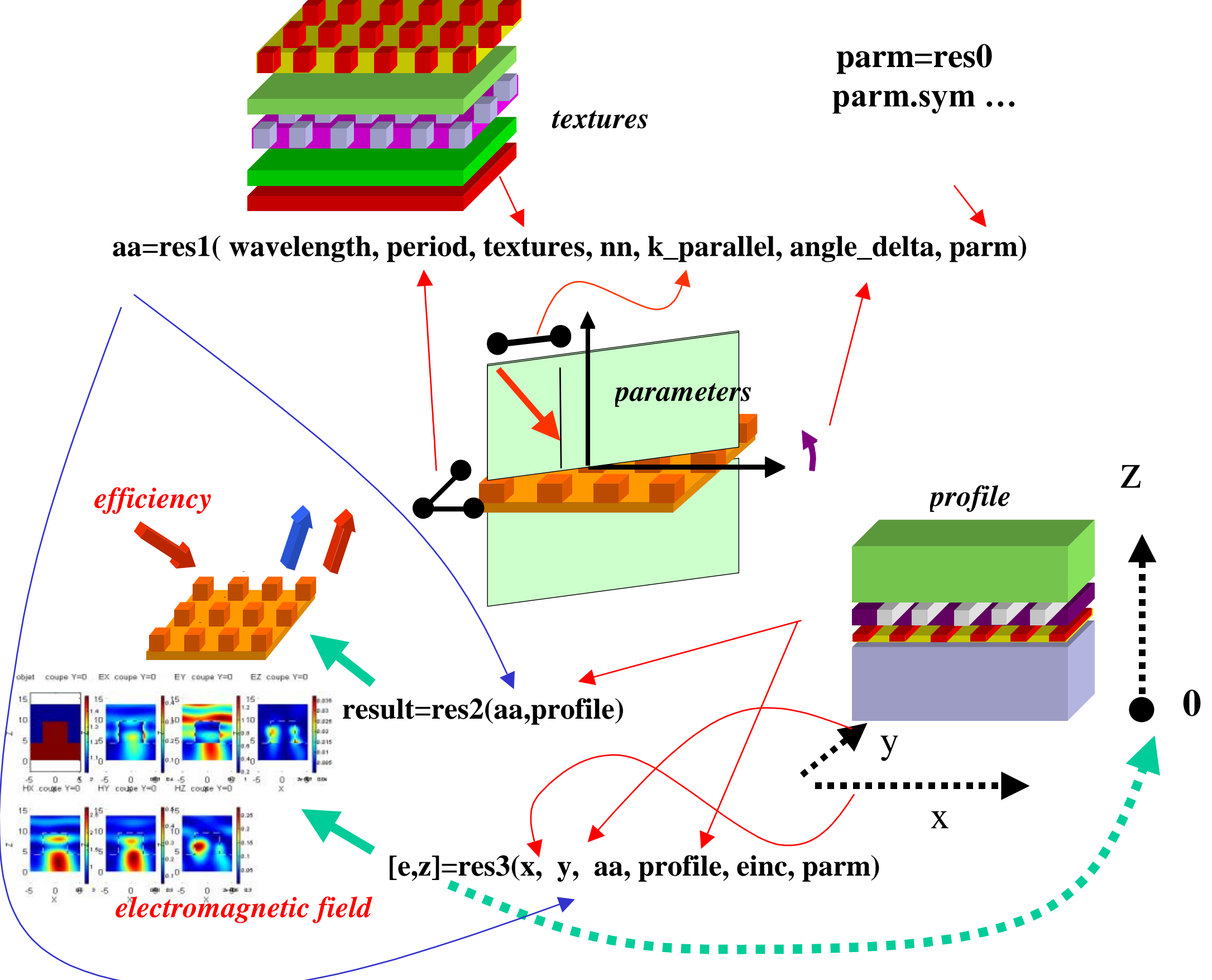

Fig. 8 Summary.

# 12. Examples

The following examples can be copied and executed in Matlab.

```
%%%%%%%%%%%%%%%%%%%%%%%
% SIMPLE EXAMPLE 2D   %
%%%%%%%%%%%%%%%%%%%%%%%
wavelength=8
period=[10,15];% same unit as wavelength
n_incident_medium=1;% refractive index of the top layer
n_transmitted_medium=1.5;% refractive index of the bottom layer
angle_theta=10;k_parallel=n_incident_medium*sin(angle_theta*pi/180);
angle_delta=-20;

parm=res0;         % default parameters for "parm"
parm.res1.champ=1; % the eletromagnetic field is calculated accurately
nn=[3,2]; % Fourier harmonics run from [-3,3]in x and [-2,2] in y
% textures for all layers including the top and bottom
texture=cell(1,3);
textures{1}= n_incident_medium;                % uniform texture
textures{2}= n_transmitted_medium;             % uniform texture
textures{3}={n_incident_medium,[0,0,5,2,n_transmitted_medium,1] };

aa=res1(wavelength,period,textures,nn,k_parallel,angle_delta,parm);
```

```
Profile={[4.1,5.2,4.1],[1,3,2]};
two_D=res2(aa,Profile)

eff_TETM=two_D.TEinc_top_reflected.efficiency{-1,1}
% (-1,1) order efficiency (TE+TM) for a TE-illumination from the top layer
eff_TE=two_D.TEinc_bottom_transmitted.efficiency_TE{-1,1}
% (-1,1) TE efficiency for a TE-illumination from the top layer
J=two_D.Jones.inc_bottom_transmitted{-1,1};% Jones'matrix
abs(J).^2 % (-1,1) order efficiency for an illumination from the bottom layer

% field calculation in plane y=0
x=linspace(-period(1)/2,period(1)/2,51);y=0;%(x,y) coordinates (z-coordinates are
determined by res3.m)
einc=[0,1];% E-field components in the (u, v) basis (default is illumination from
the top layer)
parm.res3.trace=1; % plotting automatically
parm.res3.npts=[50,50,50];
[e,z,index]=res3(x,y,aa,Profile,einc,parm);
figure;pcolor(x,z,real(squeeze(e(:,:,:,2)))); % user plotting
shading flat;xlabel('x');ylabel('y');axis equal;title('Real(Ey)');

% Loss calculation
textures{3}={.1+5i,[0,0,5,2,1,1] };
aa_loss=res1(wavelength,period,textures,nn,k_parallel,angle_delta,parm);
two_D_loss=res2(aa_loss,Profile)
parm.res3.npts=[[0,10,0];[1,3,1]];
einc= two_D_loss.TEinc_top.PlaneWave_TE_Eu;
parm.res3.trace=0;
[e,z,index,wZ,loss_per_layer,loss_of_Z,loss_of_Z_X_Y,X,Y,wXY]=res3([-
period(1)/2,period(1)/2],[-period(2)/2,period(2)/2],aa_loss,Profile,einc,parm);

Energie_conservation=sum(two_D_loss.TEinc_top_reflected.efficiency)+sum(two_D_loss.
TEinc_top_transmitted.efficiency)+sum(loss_per_layer)/(.5*prod(period))-1
retio % erase temporary files

%%%%%%%%%%%%%%%%%%%%%%%%%%
% THIN FILM STACK VITH FULL ANISOTROPY %
%%%%%%%%%%%%%%%%%%%%%%%%%%
wavelength=8;
period=10; % same unit as wavelength
n_incident_medium=1; %refractive index of the top layer
n_transmitted_medium=1.5; % refractive index of the bottom layer
angle_theta0=10;k_parallel=n_incident_medium*sin(angle_theta0*pi/180);
angle_delta=-20;
parm=res0;parm.not_io=1; % default parameters for "parm"
parm.res1.champ=1; % the electromagnetic field is calculated accurately
nn=0; % Fourier harmonics only 0

% textures for all layers including the top and bottom layers
textures=cell(1,3);
textures{1}= n_incident_medium; % uniform textures
textures{2}= n_transmitted_medium; % uniform textures
epsilon=[[2.1160 0 0.7165];[0 1.3995 0]; [0.7165 0 2.1160]];
textures{3}={epsilon} ;
[aa,neff]=res1(wavelength,period,textures,nn,k_parallel,angle_delta,parm);

Profile={[4.1,5.2,4.1],[1,3,2]};
conical=res2(aa,Profile);

% field calculation
x=linspace(-period/2,period/2,51); % x coordinates(z-coordinates are determined by
res3.m)
einc=[0,1]; % E-field components in the (u, v) basis (default is illumination from
the top layer)
parm.res3.trace=1; % plotting automatically
parm.res3.npts=[50,50,50];
[e,z,index]=res3(x,aa,Profile,einc,parm);
```

```
figure;pcolor(x,z,real(squeeze(e(:,:,3)))); % user plotting
shading flat;xlabel('x');ylabel('y');axis equal;title('Real(Ez)');

% Loss calculation
epsilon=randn(3)+1i*randn(3);epsilon=epsilon+epsilon';H=randn(3,1)+1i*randn(3,1);ep
silon=1i*H*H'+epsilon';
% integral method: general non-diagonal anisotropy without amplification
textures{3}={epsilon};
aa_loss=res1(wavelength,period,textures,nn,k_parallel,angle_delta,parm);
conical_loss=res2(aa_loss,Profile);
einc=conical_loss.TEinc_top.PlaneWave_TE_Eu;
parm.res3.npts=[[5,10,5];[4,10,4]];
% Poynting method: diagonal anisotropy only
parm.res3.trace=0;
parm.res3.pertes_poynting=1;
[e,z,index,wZ,loss_per_layer]=res3([-period/2,period/2],aa_loss,Profile,einc,parm);
Energie_conservation_Poynting=sum(conical_loss.TEinc_top_reflected.efficiency)+sum(
conical_loss.TEinc_top_transmitted.efficiency)+sum(loss_per_layer)/(.5* period)-1
%%%%%%%%%%%%%%%%%%%%%%%%%%%%
```